\documentclass{article}
\usepackage{amsmath}
\usepackage{amssymb}
\usepackage[dvips]{epsfig}
\usepackage{color}
\usepackage{cite}

 \advance \textwidth by 0.1cm
 \advance \textheight by 0.06cm
\def\##1{{\bf#1}}
\def\=#1{\underline{\underline{#1}}}
\def\r#1{(\ref{#1})}

\def\le{\left(}
\def\ri{\right)}
\def\les{\left[}
\def\ris{\right]}

\def\.{\mbox{ \tiny{$^\bullet$} }}

\def\eps{\epsilon}

\def\curl{\nabla\times}
\def\nablap{\nabla^\prime}
\def\nablapsq{\nabla^{\prime2}}
\def\curlp{\nablap\times}

\def\rt{(\#r)}

\def\uz{\hat{\#z}^\prime}

\begin{document}

\noindent {\bf Debye--H\"uckel solution for steady
electro-osmotic flow of a micropolar fluid in a cylindrical
microcapillary} \vspace{0.3cm}

\noindent {Abuzar A. Siddiqui$^{1}$ and Akhlesh
Lakhtakia$^2$\footnote{Corresponding author. E-mail: akhlesh@psu.edu }\\


\noindent $^1$Department of Basic Sciences, Bahauddin Zakariya University,
Multan, Pakistan.\\

\noindent$^2$Nanoengineered Metamaterials Group, Department of Engineering Science and Mechanics,
Pennsylvania State University,
University Park, PA 16802, USA.

\label{firstpage}

\vspace{0.4cm}

\noindent\unskip\textbf{Abstract} Analytic expressions for the speed, flux, microrotation, stress, and
couple stress in a micropolar fluid exhibiting steady, symmetric and
one-dimensional electro-osmotic flow   in a uniform cylindrical
microcapillary were derived under the constraint of the
Debye--H\"uckel approximation, which is applicable when the
cross-sectional radius of the microcapillary  exceeds the Debye
length, provided that the zeta potential is sufficiently small in
magnitude.  As the aciculate particles in a micro\-polar fluid can
rotate without translation, micropolarity  influences fluid speed,
fluid flux, and one of the two non-zero components of the stress
tensor.
 The axial speed  in a micropolar fluid
 intensifies  as the radius increases. The stress tensor is confined to the region near the wall of the microcapillary
 but the couple stress tensor is uniform across the cross-section.

\vspace{0.5cm}


\section{Introduction}\label{Intro}

Electro-osmotic flows on the micrometer scale occur in many
technoscientific settings, including: microchannels in chips for
protein analysis and intravenous drug delivery \cite{ACPI}; microchannels in
biological and chemical instruments \cite{FFGH}; micropumps, microturbines,
and micromachines \cite{DSG,FH,HMG}; injectors of detoxification agents
\cite{Keane}; and water desalinators \cite{MRT}. These flows occur by virtue of
the movement of free ions in the liquid in the electric double layer
(EDL) when an electric field is applied, and, in turn, cause bulk
motion of the entire fluid. The EDL is an infinitesimal region
between a layer of charges of one polarity on the
electrolytic-liquid side of the solid-liquid interface (i.e., the
wall of the channel) and a layer of charges of the opposite polarity
on the solid side.

The discovery of electro-osmosis is just about two centuries old
\cite{Probstein}. Fifty years after that event, electro-osmotic flow in a
channel was experimentally found to be proportional to the applied
current, which prompted Helmholtz to develop the EDL theory in 1879.
Two decades later, the EDL thickness was measured as
 much smaller than the cross-sectional dimensions of the channel. In 1923,
Debye and H\"uckel published their analysis on the distribution of
ions in a low-ionic-energy solution by using the Boltzmann
distribution for the ionic energy \cite{DH}, a seminal work that is widely
used even nowadays.

Many fluids are not simple Newtonian fluids---wherein stress is a
symmetric tensor of the second rank.
  Instead, these fluids are micropolar as they contain aciculate particles
that can rotate about axes passing through their centroids \cite{ATS,Eringen1973}.
 Therefore, not only is stress
asymmetric in micropolar fluids but they also sustain body couples.

Micropolar fluids are exemplified by colloidal suspensions, liquid
crystals, and epoxies \cite{Eringen2001}. Blood too is micropolar \cite{MG,TSA}, along
with other body fluids containing particulate materials. As body
fluids are subjected to electric fields in labs-on-a-chip \cite{OV},
electro-osmotic flows of micropolar fluids in microchannels  \cite{SL1,SL2}
are of emerging technoscientific importance in nano\-medicine \cite{EV,RSL}.

Our present interest lies in the spatial characteristics of
 fluid speed, stress, microrotation, and couple
stress in a micropolar fluid flowing steadily in a microcapillary of
circular cross-section with a  cross-sectional radius exceeding the
Debye length. With the assumption that zeta potential due to the EDL
is  sufficiently small,
 the Debye--H\"uckel approximation can be applied to obtain analytical results. Analysis
of these results would clearly show how the micropolarity influences
fluid flow.

This paper is organized as follows: The formulation of the relevant
boundary--value problem is presented in Sec.~\ref{baf}, while
Sec.~\ref{DBAsol} contains the description of analytical solution of
the problem based on the Debye--H\"uckel approximation.
Section~\ref{CRD} contains numerical illustrations of the obtained
analytical results and discussions thereon. The main conclusions are
summarized in Sec.~\ref{CR}.

 \section{Basic Analysis}\label{baf}
In a micropolar fluid, the stress tensor $\=\sigma^\prime$ is
defined as a function of position $\#r^\prime$ as
\begin{eqnarray}
&&\=\sigma^\prime(\#r^\prime) =
-p^\prime(\#r^\prime)\=I+\lambda\=I\nabla^\prime\.\#V^\prime(\#r^\prime)
+ (\mu+\chi/2)\left\{\nabla^\prime\#V^\prime(\#r^\prime)
+\left[\nabla^\prime\#V^\prime(\#r^\prime)\right]^T\right\}
\nonumber\\
&&\qquad\qquad
-(\chi/2)\=I\times\left[\nabla^\prime\times\#V^\prime(\#r^\prime)\right]
+\chi\=I\times\#v^\prime(\#r^\prime)\,, \label{stress-def}
\end{eqnarray}
and the couple stress tensor $\=m^\prime$ as
\begin{eqnarray}
&&\=m^\prime(\#r^\prime)
 = \alpha\=I\nabla^\prime\.\#v^\prime(\#r^\prime) +
\beta\nabla^\prime\#v^\prime(\#r^\prime)+\gamma\left[\nabla^\prime\#v^\prime(\#r^\prime)\right]^T\,.
\label{couplestress-def}
\end{eqnarray}
 Here,  {$\=I=\nabla^\prime\#r^\prime$} is the idempotent; $p^\prime$ is the hydrostatic
pressure; $\#V^\prime$ and $\#v^\prime$, respectively, are the fluid
velocity and the microrotation; $\alpha$, $\beta$, and $\gamma$ are
the three spin-gradient viscosity coefficients; and $\mu$ and
$\chi$, respectively, are the Newtonian shear viscosity coefficient
and the vortex viscosity coefficient  related by the inequality
 $2\mu+\chi\geq 0$, where $\chi\geq 0$ \cite[p.~14]{Eringen2001}.
The parameters $\#v^\prime$,  $\chi$, $\alpha$, $\beta$, and
$\gamma$ are null-valued in a simple Newtonian fluid.  {The
superscript $T$ denotes the transpose.}

We examine here the steady flow of a micropolar fluid in an
infinitely extended microcapillary of cross-sectional radius $R$,
such that its axis coincides with the $z^\prime$ axis of the
cylindrical polar coordinate system ${\bf
r}^\prime\equiv(\rho^\prime,\theta^\prime,z^\prime)$. The study is
undertaken under the following assumptions: (i) the zeta potential
is uniform in the microcapillary; (ii) the surface of the
microcapillary is perfectly insulated and impermeable; (iii) the
applied electric field is spatiotemporally uniform electric field
and aligned parallel to the axis of the microcapillary; (iv) the
fluid is ionized, incompressible, and viscous; (v) the flow is fully
developed, steady, laminar, {axial}, and radially symmetric; (vi)
the effect of gravity is negligible; (vii) neither a body couple nor
a pressure gradient is present; (viii) the Joule heating effects are
small enough to be ignored; and (ix) $R$ is much greater than the
Debye length $\lambda_D$.

Under these conditions, our starting point comprises the following
three  equations of micropolar-fluid flow \cite{Eringen2001}:
\begin{eqnarray} \label{2.1.0}
&&\nabla^\prime\.\#V^\prime(\#r^\prime)=0\,,
\\[8pt]
\label{2.1.1} &&\nabla^\prime\.\=\sigma^\prime(\#r^\prime)
+\rho^\prime_e(\#r^\prime)\,\#E^\prime_{app}=\rho_m
[\#V^\prime(\#r^\prime)\.\nabla^\prime]\#V^\prime(\#r^\prime)\,,\\[8pt]
\label{2.1.2} &&\nabla^\prime\.\=m^\prime(\#r^\prime) +\=I
\stackrel{\.}{\times}\=\sigma^\prime(\#r^\prime) =\rho_m j_o
[\#V^\prime(\#r^\prime)\.\nabla^\prime]\#v^\prime(\#r^\prime)\,.
\end{eqnarray}
Introducing Eqs.~(\ref{stress-def}) and (\ref{couplestress-def}) in
these three equations, we get
\begin{eqnarray}
\label{2.1} &&\nabla^\prime\.\#V^\prime(\#r^\prime)=0\,,
\\[8pt]
\nonumber && -(\mu+\chi)\curlp\les\curlp\#V^\prime(\#r^\prime)\ris
+\chi\curlp\#v^\prime(\#r^\prime)+\rho^\prime_e(\#r^\prime)\,\#E^\prime_{app}
\\
\label{2.2} &&\qquad\qquad =\rho_m
[\#V^\prime(\#r^\prime)\.\nabla^\prime]\#V^\prime(\#r^\prime)\,,
\\[8pt]
\nonumber
&&(\alpha+\beta+\gamma)\nablap\les\nablap\.\#v^\prime(\#r^\prime)\ris
-\gamma \curlp\les\curlp\#v^\prime(\#r^\prime)\ris
-2\chi\#v^\prime(\#r^\prime)
\\
\label{2.3} &&\qquad\qquad +\chi\curlp\#V^\prime(\#r^\prime)
   =\rho_m j_o
[\#V^\prime(\#r^\prime)\.\nabla^\prime]\#v^\prime(\#r^\prime)\,.
\end{eqnarray}
Here,  $\#E^\prime_{app}$ is the applied electric field, whereas
$\rho_m$ and $j_o$, respectively, are the mass density and the
microinertia. The microinertia is null-valued in a simple Newtonian
fluid.

In the {absence} of a significant convective or electrophoretic
disturbance to the EDL, the charge density
$\rho_e^\prime(\#r^\prime)$ is described by a Boltzmann
distribution, and takes the following form for a symmetric, dilute,
and univalent electrolyte \cite{Li}:
\begin{equation}
\label{Boltz1} \rho_e^\prime(\#r^\prime) = -2 z_o e n_o
\sinh\les{z_oe\psi^\prime (\#r^\prime)}/{k_B T}\ris\,.
\end{equation}
Here, $z_o$ is the absolute value of the ionic valence,
$\psi^\prime$ is the electric potential, $e$ is the charge of an
electron, $n_o$ is the number density of ions in the fluid far away
from any charged surface, $k_B$ is the Boltzmann constant, and $T$
is the temperature. With $\epsilon$ denoting the static permittivity
of the fluid, the charge density and the electric potential are also
related by the Gauss law
\begin{equation}
\nablapsq\psi^\prime (\#r^\prime) =- \rho_e^\prime
(\#r^\prime)/\epsilon\,; \label{Gauss1}
\end{equation}
thus,
\begin{equation}
\nablapsq\psi^\prime (\#r^\prime) = \left({2 z_o e
n_o}/{\epsilon}\right) \sinh\les{z_oe\psi^\prime (\#r^\prime)}/{k_B
T}\ris\,. \label{Gauss2}
\end{equation}
The Debye length $\lambda_D = \left({z_o e}\right)^{-1}\,
\left({\epsilon k_B T}/{2  n_{o}}\right)^{1/2}$ is assumed in this
paper to be much smaller than the cross-sectional radius $R$.
Furthermore, we set $\#E^\prime_{app}=\uz\,E_o$ as the applied
electric field.

Before proceeding, let us define the non-dimensionalized quantities
\begin{equation}\label{1-dim}
\left.\begin{array}{llll} \#r= \#r^\prime/R\,, &\,\,\#V =
\#V^\prime/U\,, &\,\,\#v=({R}/{U})\,\#v^\prime\,  \\[6pt]
\=\sigma=(R/U)(\mu+\chi)^{-1}\=\sigma^\prime\,,&\,\, \=m=(R^2
/\gamma U)\=m^\prime
\\[6pt]
\#\psi = \psi^\prime/{\psi_o}\,, &
\#\rho_e=({R^2}/{\epsilon\psi_o})\,\rho_e^\prime\,
\end{array}\right\}\,.
\end{equation}
Here, ${\bf r}\equiv (\rho,\theta,z)$ with $\rho=\rho^\prime/R$,
$\theta=\theta^\prime$, and $z=z^\prime/R$;
\begin{equation}
\label{2.5} U =  -\frac{\epsilon\psi_o  E_o}{\mu+\chi}\,
\end{equation}
is a characteristic speed \cite{SL1};  and $\psi_o$ is called the zeta
potential \cite{Li} which is assumed to be temporally constant and
spatially uniform at the wall $\rho=1$ of the microcapillary. With
these quantities, Eqs.\r{2.1}--\r{2.3} and \r{Gauss2}, respectively,
simplify to
\begin{eqnarray}
\label{2.6a} && \nabla\.\#V\rt = 0\,,
\\[6pt]
\label{2.6b} && \nonumber
-\curl\les\curl\#V\rt\ris + k_1\curl\#v\rt -\rho_e\left(\#r\right)\uz\\[6pt]
&&\qquad= R_e[\#V(\#r)\.\nabla]\#V(\#r)\,,
\\[6pt]
\nonumber &&-\curl\les\curl\#v\rt\ris -2k_2\#v\rt
+k_3\nabla\les\nabla\.\#v\rt\ris+k_2\curl\#V\rt
\\
\label{2.6c} &&\qquad =R_o[\#V(\#r)\.\nabla]\#v(\#r)\,,
\\[6pt]
\label{2.6d} &&\nabla^2\psi\le\#r\ri=\left({m_o
^2}/{\alpha_o}\right)\sinh\les\alpha_o\psi\le\#r\ri\ris\,,
\end{eqnarray}
where
\begin{equation}
\label{k-defs} \left.\begin{array}{llll} k_1
=\frac{\chi}{\mu+\chi}\,,& \quad k_2=\frac{\chi
R^2}{\gamma}\,,&\quad
k_3 =\frac{\alpha+\beta+\gamma}{\gamma}\\[6pt]
R_e=\frac{\rho_m U R }{\mu+\chi} \,, &\quad R_o =\frac{\rho_m j_o U
R}{\gamma}\,,&\quad m_o=\frac{R}{\lambda_D}\,,&\quad \alpha_o =
\frac{z_o e \psi_o}{k_B T}
\end{array}\right\}\,.
\end{equation}
Here,  $k_1$ couples  the two viscosity coefficients, $k_2$ and
$k_3$ are normalized micropolar parameters, $R_e$ may be called the
Reynolds number, $R_o$ may be called the microrotation Reynolds
number \cite{Eringen2001}, and $\alpha_o$ is the ionic-energy parameter \cite{Li}. The
Gauss law \r{Gauss1} can now be written as
\begin{equation}
\nabla^2\psi (\#r) =- \rho_e (\#r)\,. \label{Gauss3}
\end{equation}
The general condition of radial symmetry implies that
$\partial/\partial\theta\equiv 0$. As the length of the
microcapillary is infinite while its cross-sectional radius is
finite, we set $\partial/\partial z\equiv 0$; thus,  ${\bf
V}(\#r)\equiv\#V(\rho)$ and ${\bf v}(\#r)\equiv\#v(\rho)$ Since the
flow is assumed to be axial and laminar as well, it follows that
${\bf V}(\rho) \simeq V_z(\rho)\,\uz$. Equation~(\ref{2.6a}) is then
automatically satisfied, whereas Eq.~(\ref{2.6d}) reduces to
\begin{equation}
\label{3.1}
\frac{d}{d\rho}\left[\rho\,\frac{d\psi(\rho)}{d\rho}\right]
=\frac{\rho m_o ^2}{\alpha_o}\,\sinh\les\alpha_o\psi\le\rho\ri\ris\,
\end{equation}
and Eq.~(\ref{Gauss3}) to
\begin{equation}
\rho_e(\rho)=-\rho^{-1}\,\frac{d}{d\rho}\left[\rho\,\frac{d\psi(\rho)}{d\rho}\right]
\,. \label{Gauss3a}
\end{equation}

Using Eq.~(\ref{Gauss3a}), we obtain
\begin{equation}
\label{2.16a} \frac{d}{d \rho}\left\{\rho\left[
\frac{dV_z(\rho)}{d\rho}+ \frac{d\psi(\rho)}{d\rho} +k_1
v_\theta(\rho)\right]\right\}=0
\end{equation}
and
\begin{equation}
\label{2.16b} \frac{dv_z(\rho)}{d \rho}=0\,
\end{equation}
from Eq.~(\ref{2.6b}). After using Eq.~(\ref{2.16b}), we get the
following three equations from Eq.~(\ref{2.6c}):
\begin{eqnarray}
\label{new1} &&\frac{d}{d\rho}\left\{
\rho^{-1}\,\frac{d}{d\rho}\left[\rho\,v_\rho(\rho)\right]\right\}
-2k_2v_\rho(\rho)=0\,,
\\
\label{2.17a} &&\frac{d}{d\rho}\left\{
\rho^{-1}\,\frac{d}{d\rho}\left[\rho\,v_\theta(\rho)\right]\right\}
-2k_2v_\theta(\rho)-k_2\frac{dV_z(\rho)}{d\rho}=0\,,
\\
\label{new2} &&v_z(\rho)=0\,.
\end{eqnarray}
According to Eq.~(\ref{new1}), $v_\rho$ is not influenced by the
axial flow. Therefore, we set
\begin{equation}
v_\rho(\rho)\equiv 0\,,
\end{equation}
and focus our attention on Eqs.~(\ref{3.1}), (\ref{2.16a}) and
(\ref{2.17a}) involving $\psi(\rho)$, $V_z(\rho)$ and
$v_\theta(\rho)$.

We need to provide appropriate conditions for these three
quantities. First, by virtue of the definition of the zeta
potential,
\begin{equation}
\label{new4} \psi(1)=1\,;
\end{equation}
next, the condition
\begin{equation}
\label{2.1.3c-1} \psi(0)\ll1
\end{equation}
is engendered by the assumption $R \gg\lambda_D$ \cite[p.~187]{Probstein};
finally, flow symmetry dictates that
\begin{equation}
\label{2.19-2} \frac{d\psi(\rho)}{d\rho} \Big\vert_{\rho=0} = 0\,.
\end{equation}
The no-slip boundary condition \cite[pp.~100-101]{Li} on the surface
$\rho=1$ of the capillary is
\begin{equation}
\label{2.18} V_z(1) = 0\,
\end{equation}
and the symmetric flow requires that
\begin{equation}
\label{2.19-1} \frac{dV_z(\rho)}{d\rho}\Big\vert_{\rho=0} = 0\,.
\end{equation}
 Finally, the boundary conditions imposed on microrotation
are:
\begin{eqnarray}
\label{2.22} &&v_\theta(1)=\beta_o\,\frac{dV_z(\rho)}{d\rho}
\Big\vert_{\rho=1}   \,,\qquad v_\theta(0) = 0\,
\end{eqnarray}
where $\beta_o\in\les-1,0\ris$  is some constant. Although some
researchers have ignored microrotation effects near a solid wall by
setting $\beta_o = 0$ \cite{Eringen2001,PBA}, others have held that
 that $\beta_o<0$ because the existence of the boundary layer requires
 that   the shear and couple
stresses on a wall must be high in magnitude in comparison to
locations elsewhere \cite{HL}. Furthermore, the limiting case of $\beta_o
= -1$ accounts for turbulence near the wall \cite{RB}. As we
think that $\beta_o<0$ may be reasonable if electro-osmosis occurs
because of the presence of the EDL near the wall $\rho=1$, results
are provided in this paper for $\beta_o\ne0$.

\section{Debye--H\"uckel Approximation}\label{DBAsol}
If the electric potential energy is small as compared to the thermal
energy of the ions, i.e., $\vert z_oe\psi_o\psi\vert\ll\vert
k_BT\vert$, then $\vert \alpha_o\psi\vert\ll1$; accordingly,
\begin{equation}
\label{DHa} \sinh(\alpha_o\psi) \approx \alpha_o\psi\,,
\end{equation}
which is called the Debye--H\"uckel approximation \cite{Li}. In light of
this approximation Eq.~\r{3.1} simplifies to
\begin{equation}
\label{3.3} \rho \frac{d^2\psi(\rho)}{d\rho^2}+
\frac{d\psi(\rho)}{d\rho} \approx m_o^2\,\rho \,\psi(\rho)\,,
\end{equation}
whose solution
\begin{equation}
\label{3.4} \psi(\rho) \approx \frac{I_0(m_o \rho)}{I_0(m_o)}\,.
\end{equation}
 satisfies the boundary conditions
\r{new4} and \r{2.19-2}; here and hereafter, $I_n(\.)$ is the
modified Bessel function of the first kind and order $n$ \cite{AS}.  As
$m_o=R/\lambda_D\to\infty$, we see that
$\psi(0)\approx1/I_0(m_o)\to0$, thereby fulfilling the requirement
\r{2.1.3c-1}.

Integrating both sides of Eq.~\r{2.16a} with respect to $\rho$ and
using the boundary conditions \r{2.19-2}, \r{2.19-1} and
\r{2.22}$_2$, we obtain
\begin{equation}
\label{3.5} \frac{dV_z(\rho)}{d\rho}=-k_1v_\theta(\rho)-\frac{d\psi
(\rho)}{d\rho}\,.
\end{equation}
Eliminating $dV_z/d\rho$ from Eq. \r{2.17a}, we next get
\begin{equation}
\label{3.6}
\rho^2\frac{d^2v_\theta(\rho)}{d\rho^2}+\rho\frac{dv_\theta(\rho)}{d\rho}
-[1+(2-k_1)k_2\rho^2)]
v_\theta(\rho)+k_2\rho^2\frac{d\psi(\rho)}{d\rho}=0\,.
\end{equation}
Finally, using the Debye--H\"uckel solution \r{3.4} for
$\psi(\rho)$,   we reduce Eq.~\r{3.6} to
\begin{equation}
\label{3.7}
\rho^2\frac{d^2v_\theta(\rho)}{d\rho^2}+\rho\frac{dv_\theta(\rho)}{d\rho}-
\left(1+k_0^{2}\rho^2\right)v_\theta
(\rho)=-\frac{I_1(m_o\rho)}{I_0(m_o)}k_2m_o\rho^2 \,,
\end{equation}
where
\begin{equation}
\label{k0def} k_0=\sqrt{(2-k_1)k_2}\,.
\end{equation}

The solution of the homogeneous counterpart of Eq.~\r{3.7} is
$c_1I_1(k_0\rho)+c_2K_1(k_0\rho)$, where $K_n(\.)$ is the modified
Bessel function of the second kind and order $n$ \cite{AS}, while $c_1$
and $c_2$ are constants to be determined later. The method of
variation of parameters then yields the particular solution of
Eq.~\r{3.7} as
${k_2m_o}\left({k_0^2-m_o^2}\right)^{-1}{I_1(m_o\rho)}/{I_0(m_o)}$,
after the identity
$I_n(\xi)K_{n+1}(\xi)+I_{n+1}(\xi)K_{n}(\xi)=\xi^{-1}$ [1, Eq.
9.6.15] has been exploited. Since $K_1(\xi)\to\infty$ as
$\xi\to\infty$ but $I_1(0)=0$, satisfaction of the boundary
condition \r{2.22}$_2$ requires that $c_2=0$. Therefore, the
complete solution of Eq.~\r{3.7} is
\begin{equation}
\label{3.8} v_\theta(\rho) = c_1I_1(k_0\rho) +
\frac{k_2m_o}{k_0^2-m_o^2}\,\frac{I_1(m_o\rho)}{I_0(m_o)}\,.
\end{equation}
Substitution of
 Eqs.~\r{3.5} and \r{3.8}
in the boundary condition \r{2.22}$_1$ leads to
\begin{equation}
\label{3.10} c_1=-m_o\left(\frac{k_2 }{k_0^2-m_o^2}+\frac{
\beta_o}{1+k_1\beta_o}\right) \frac{I_1(m_o)}{I_0(m_o)\,I_1(k_0)}\,.
\end{equation}
Finally, using Eqs.~\r{3.4} and \r{3.8} in \r{3.5}, and exploiting
the boundary condition \r{2.18}, we get
\begin{equation}
\label{Vzrho}
V_z(\rho)=c_1\frac{k_1}{k_0}\left[I_0(k_0)-I_0(k_0\rho)\right]
+\left(1 +\frac{k_1k_2}{k_0^2-m_o^2}\right)
\left[1-\frac{I_0(m_o\rho)}{I_0(m_o)}\right]\,,
\end{equation}
which automatically satisfies the requirement \r{2.19-1}.

Thus, Eqs.~\r{3.8}--\r{Vzrho} constitute the solution of the
boundary-value problem for steady flow of a micropolar fluid when
the Debye--H\"uckel approximation (\ref{DHa}) holds, for all values
of $\beta_o$.

On setting $\chi=0$ for a simple Newtonian fluid, we get
$k_1=k_2=k_0=0$. Accordingly, Eqs.~\r{3.8} and Eq.~\r{Vzrho}
simplify to
\begin{equation}\label{3.11}
v_\theta(\rho)\equiv 0\,,\qquad V_z(\rho)=1-\frac{
I_0(m_o\rho)}{I_0(m_o)}\,,
\end{equation}
which is a known result  \cite[p.~102]{Li}.

\section{RESULTS AND DISCUSSION}\label{CRD}

All calculations were made with the following material properties
fixed: $n_o=6.02\times 10^{22}$~m$^{-3}$,   $z_o=1$,
$\epsilon=10\epsilon_o$, $\epsilon_o=8.854\times
10^{-12}$~F~m$^{-1}$,
 $\mu=3\times10^{-2}$~Pa~s, and $\gamma=10^{-4}$~kg~m~s$^{-1}$.
 The temperature was fixed at $T=290$~K. Since the Boltzmann constant
 $k_B=1.38\times10^{-23}$~J~K$^{-1}$ and the electron charge $e=1.6\times 10^{-19}$~C,
 the Debye length $\lambda_D=10.72$~nm. Consistently with the assumption that $R \gg\lambda_D$,
  we chose $R\in\les107.2, 5360\ris$~nm
 so that $m_o\in\les10, 500\ris$.
{We  fixed} $\psi_o=-25\times10^{-3}$~V, which is about the upper
limit for the Debye--H\"uckel approximation to be valid at about the
room temperature \cite[p.~25]{Hunter}.  The parameters
$\beta\in\les-\gamma,\gamma\ris$ and $k_1\in\les0,0.95\ris$ were
kept as variables, after noting that $k_1\to 1$ as $\chi\to\infty$.
The magnitude of the applied electric field was fixed at
$E_o=10^4$~V~m$^{-1}$, which is a reasonable practical value. The
dependences of the relevant components of the fluid velocity,
microrotation, stress tensor, and couple stress tensor on $m_o$,
$k_1$, and $\beta_o$ were investigated for steady flow.

Besides the speed $V^\prime_z(\rho^\prime)$ and the microrotation
$v^\prime_\theta(\rho^\prime)$, the only non-zero components of the
stress tensor and the couple stress tensor, according to
Eqs.~(\ref{stress-def}) and (\ref{couplestress-def}) are
\begin{equation} \label{4.1.3.0} \left.\begin{array}{ll}
\sigma^\prime_{\rho{z}}(\rho^\prime)=(\mu+\chi)\frac{dV^\prime_z(\rho^\prime)}{d\rho^\prime}+\chi
v^\prime_\theta(\rho^\prime)
\\[5pt]
\sigma^\prime_{z\rho}(\rho^\prime)=\mu\frac{dV^\prime_z(\rho^\prime)}{d\rho^\prime}-\chi
v^\prime_\theta(\rho^\prime)\\[5pt]
m^\prime_{\rho\theta}(\rho^\prime)=\beta\frac{dv^\prime_\theta(\rho^\prime)}{d\rho^\prime}-
\gamma\frac{v_\theta(\rho^\prime)}{\rho^\prime}\\[5pt]
m^\prime_{\theta\rho}(\rho^\prime)=-\beta\frac{v_\theta(\rho^\prime)}{\rho^\prime}+
\gamma\frac{dv^\prime_\theta(\rho^\prime)}{d\rho^\prime}
\end{array}\right\}\,.
\end{equation}
The non-dimensionalized form of these equations are
\begin{equation} \label{4.1.3.1} \left.\begin{array}{ll}
\sigma_{\rho{z}}(\rho)=\frac{dV_z(\rho)}{d\rho}+k_1 v_\theta(\rho)
\\[5pt]
\sigma_{z\rho}(\rho)=(1-k_1)\frac{dV_z(\rho)}{d\rho}-k_1
v_\theta(\rho)\\[5pt]
m_{\rho\theta}(\rho)=k_7\frac{dv_\theta(\rho)}{d\rho}-
\frac{v_\theta(\rho)}{\rho}\\[5pt]
m_{\theta\rho}(\rho)=-k_7\frac{v_\theta(\rho)}{\rho}+
\frac{dv_\theta(\rho)}{d\rho}
\end{array}\right\}\,.
\end{equation}
where $k_7=\beta/\gamma \in\left[-1,1\right]$.

\subsection{Fluid speed}\label{fs}

In order to examine the influence of parameters $m_o$, $k_1$, and
$\beta_o$ on the fluid speed, we computed $V_z^\prime(\rho)$ using
its Debye--H\"uckel expression (\ref{Vzrho}). Some representative
plots of $V_z^\prime(\rho)$ vs. $\rho$ are provided in Figure
\ref{Vz_rfig1} for $m_o\in\left\{50,500\right\}$,
$k_1\in\left\{0,0.5,0.95\right\}$, and $\beta_o\in\left\{-0.01,
-0.1, -0.5,-1\right\}$.

Figure \ref{Vz_rfig1} indicates that  the speed of a simple
Newtonian fluid ($k_1=0$) depends on $m_o$ but not on $\beta_o$, in
conformity with Eq.~(\ref{3.11})$_2$.  Moreover, since $0\leq
\frac{I_0(m_o\rho)}{I_0(m_o)}\leq 1$ for all $m_o\in [50, 500]$ and
$\rho\in [0,1]$,
 $0\leq V_z^\prime(\rho)\leq U$ by virtue of the
 same  equation. In addition, for any $\rho\in[0,1)$,
$V_z^\prime(\rho)$ increases with $m_o$ and the maximum fluid speed
exists in the center ($\rho=0$) of the microcapillary.

\begin{figure}[!htb]
\centering \psfull
\epsfig{file=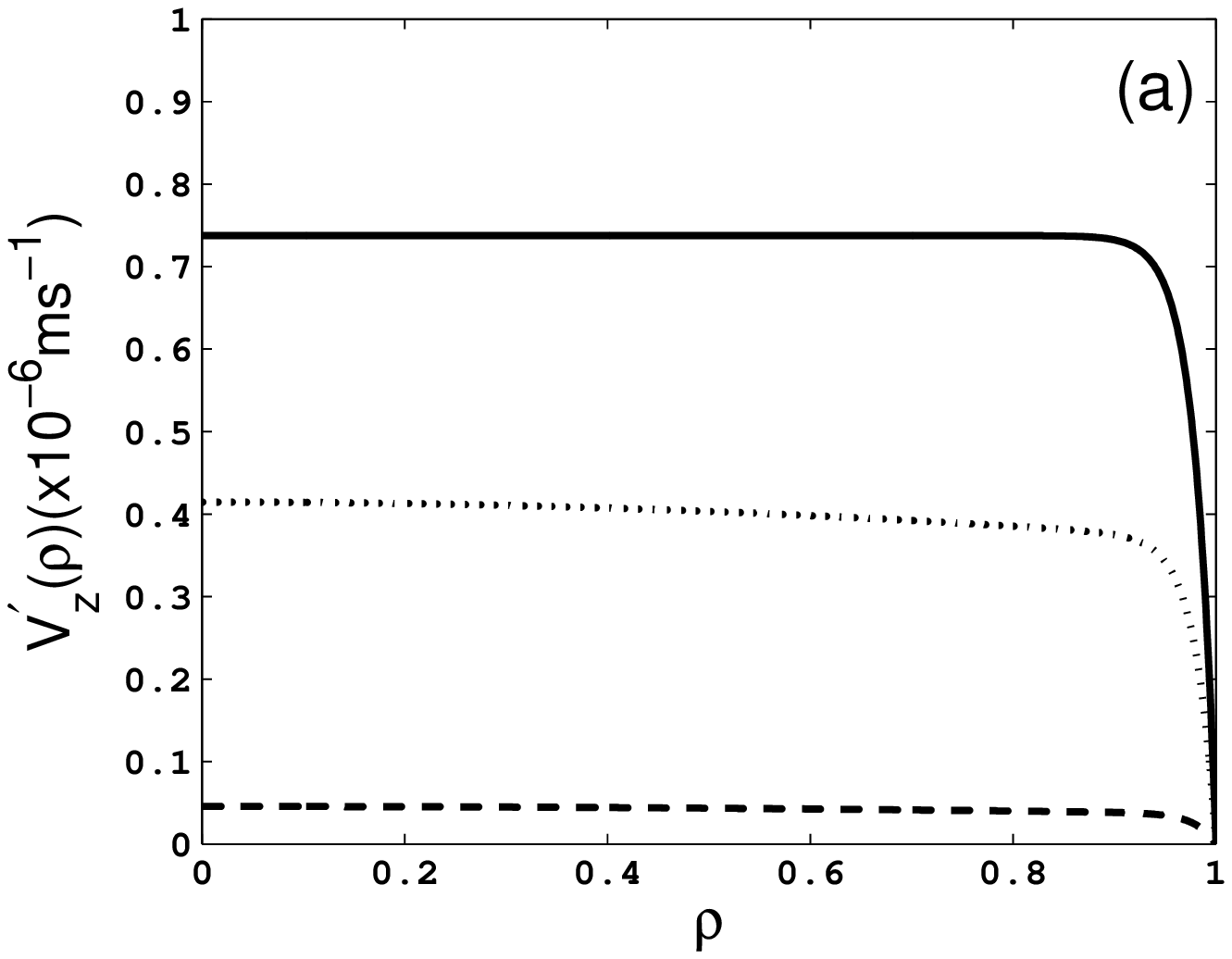,height=3.5cm}
\epsfig{file=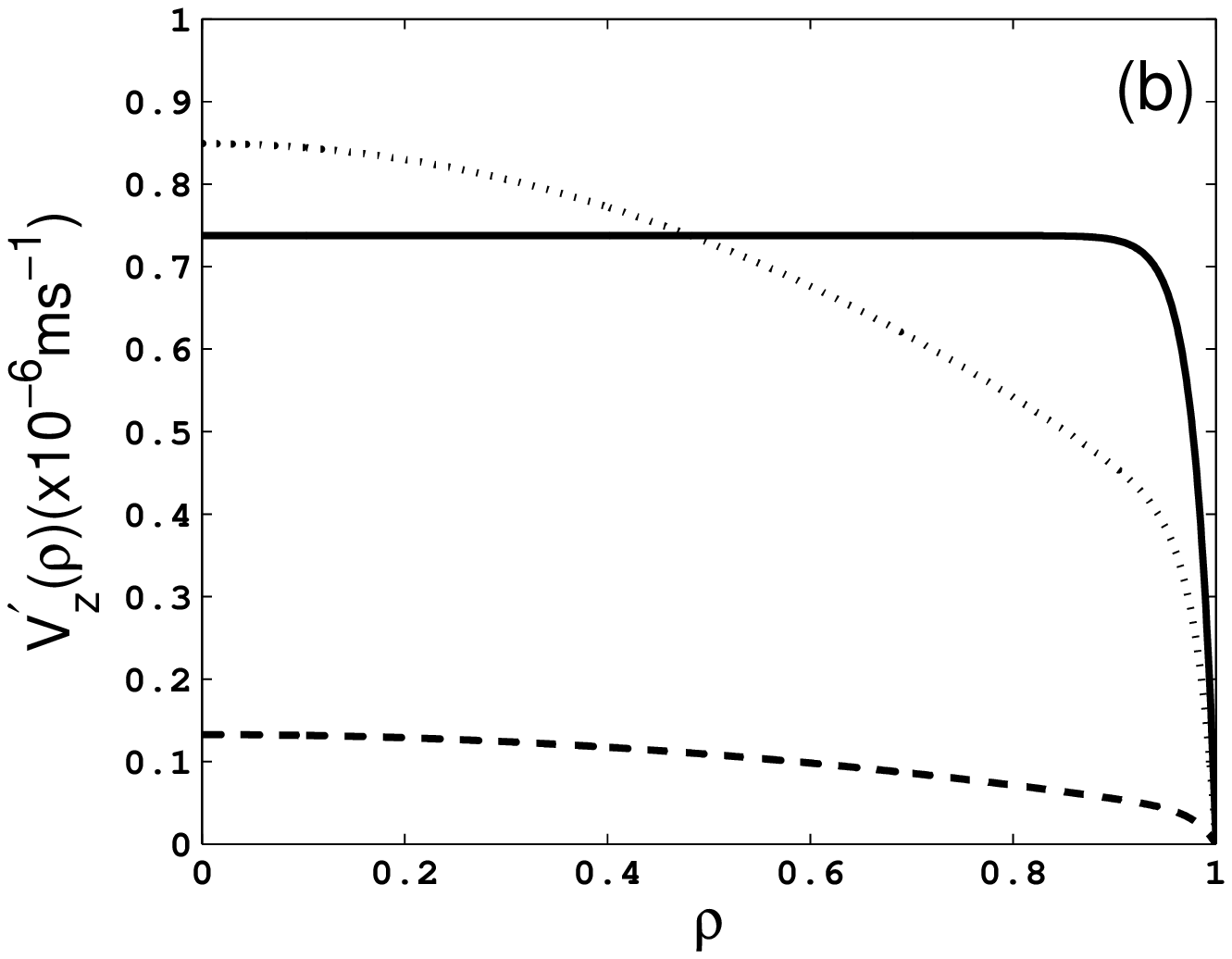,height=3.5cm}\\
\epsfig{file=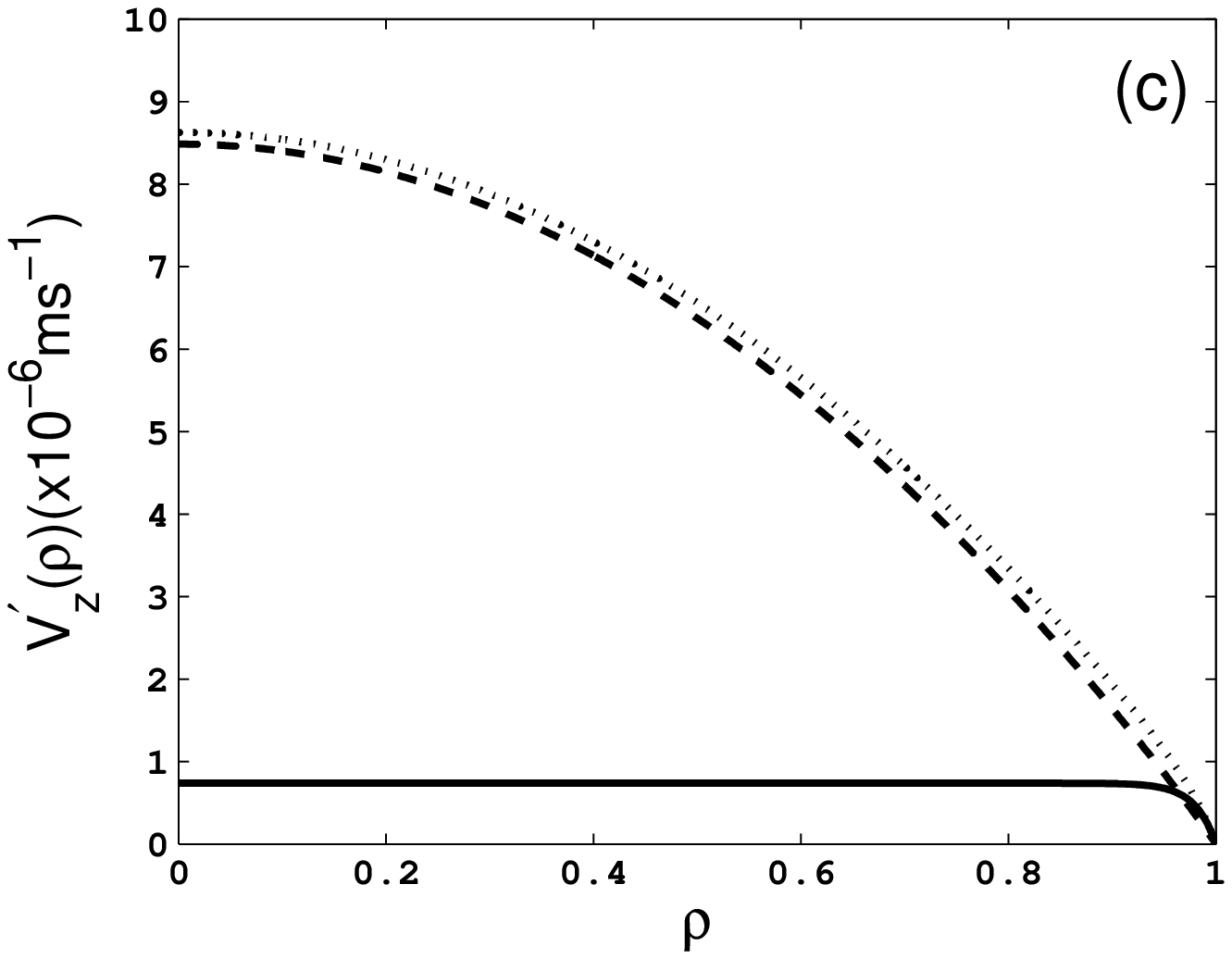,height=3.5cm}
\epsfig{file=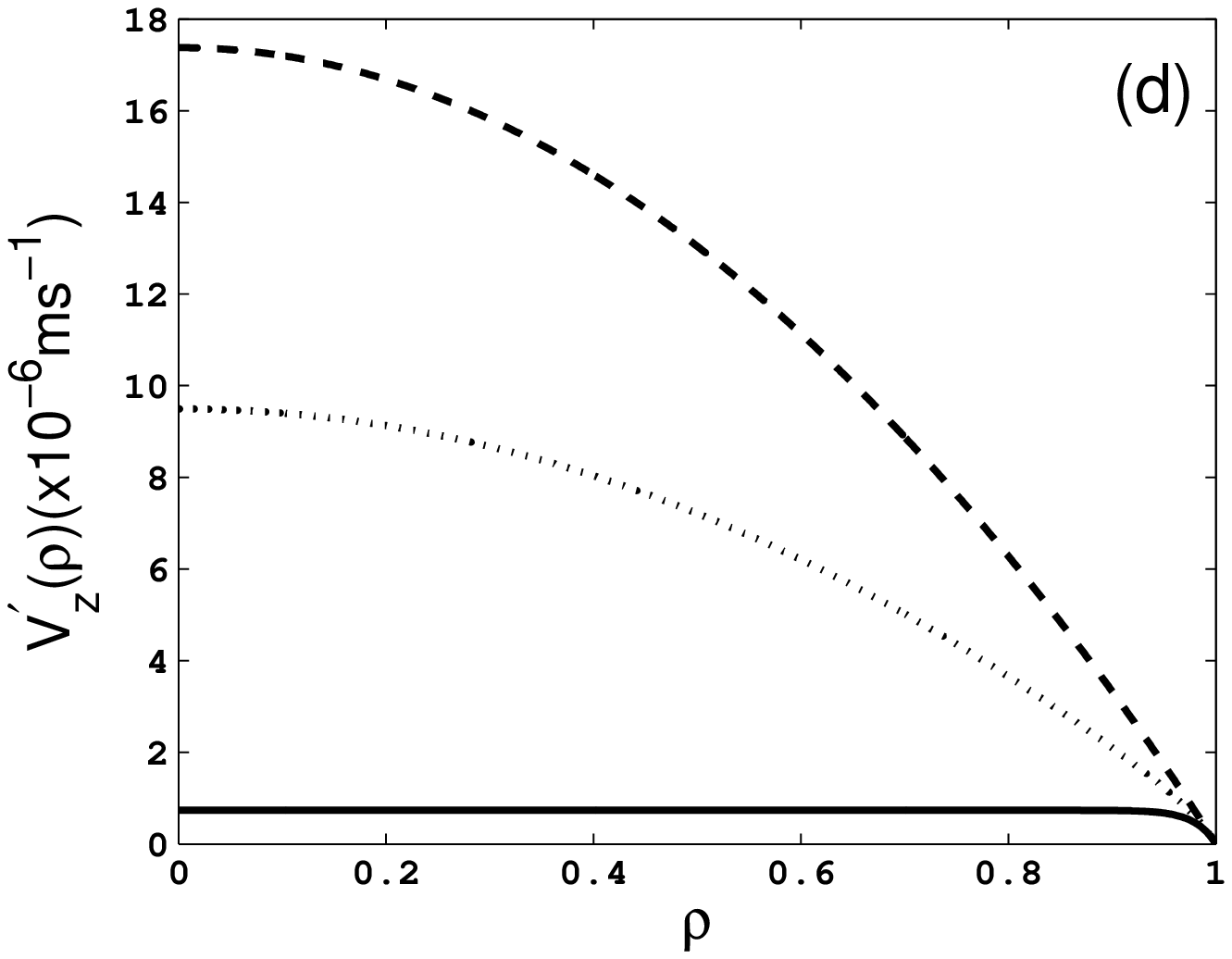,height=3.5cm}\\
\epsfig{file=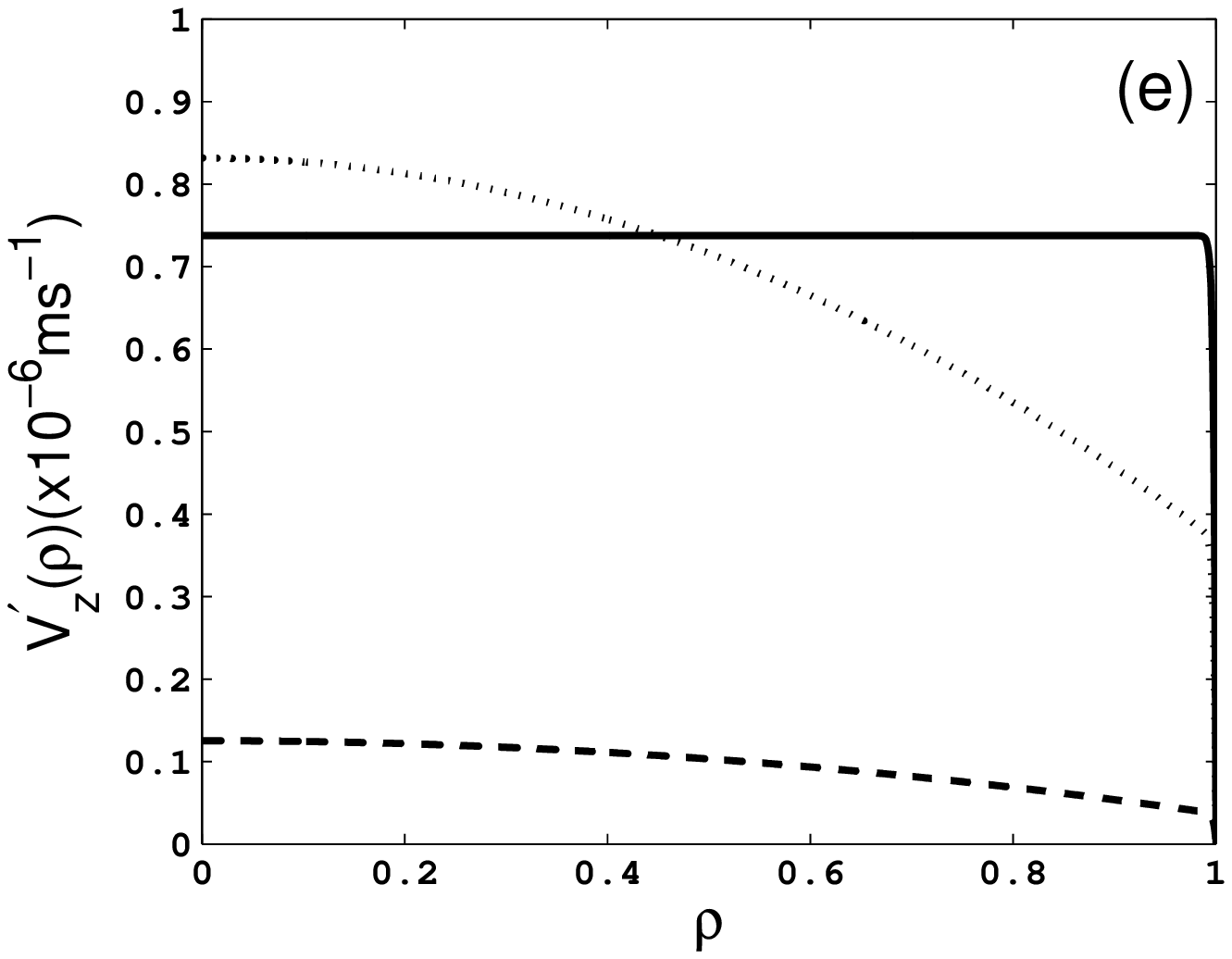,height=3.5cm}
\epsfig{file=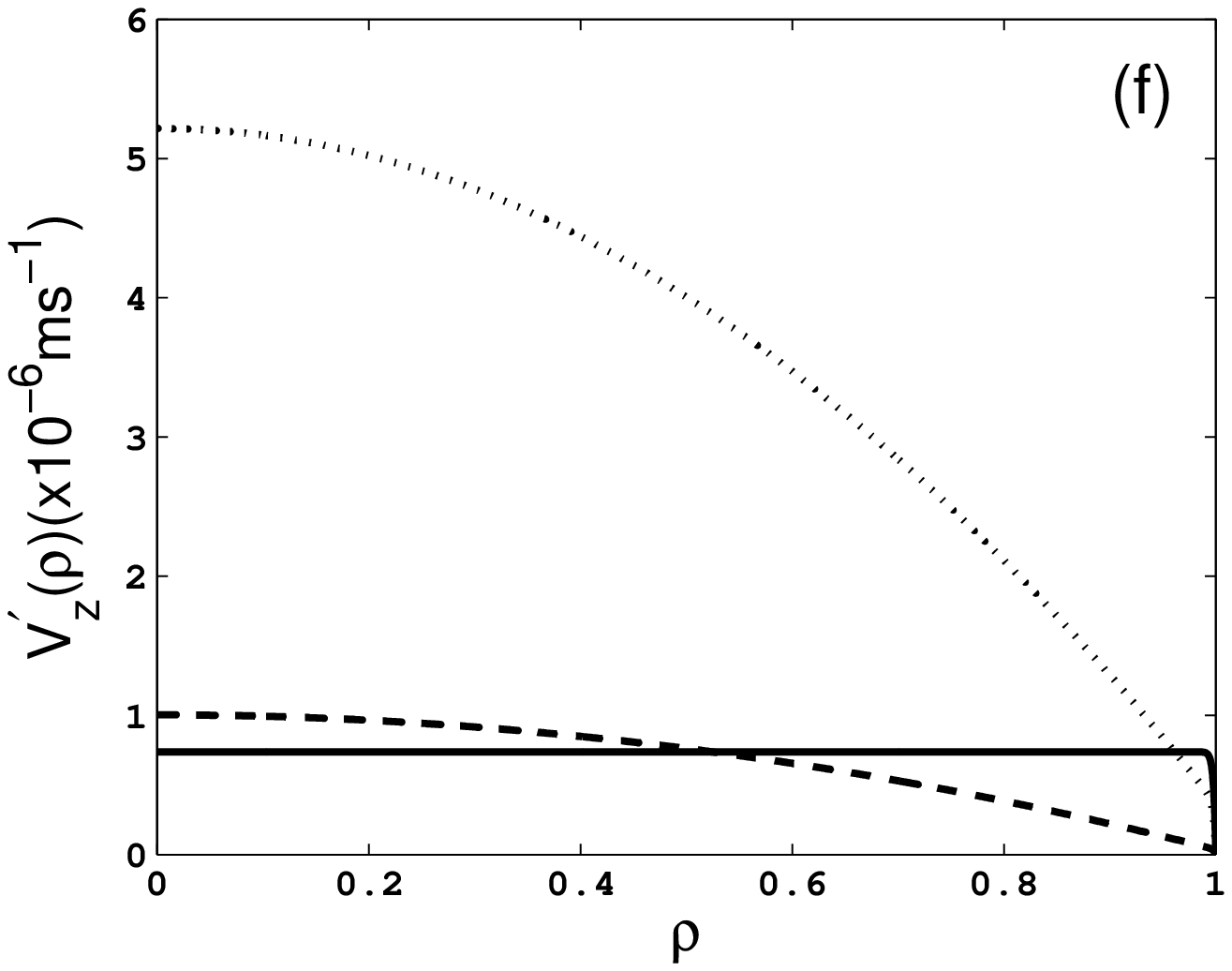,height=3.5cm}\\
\epsfig{file=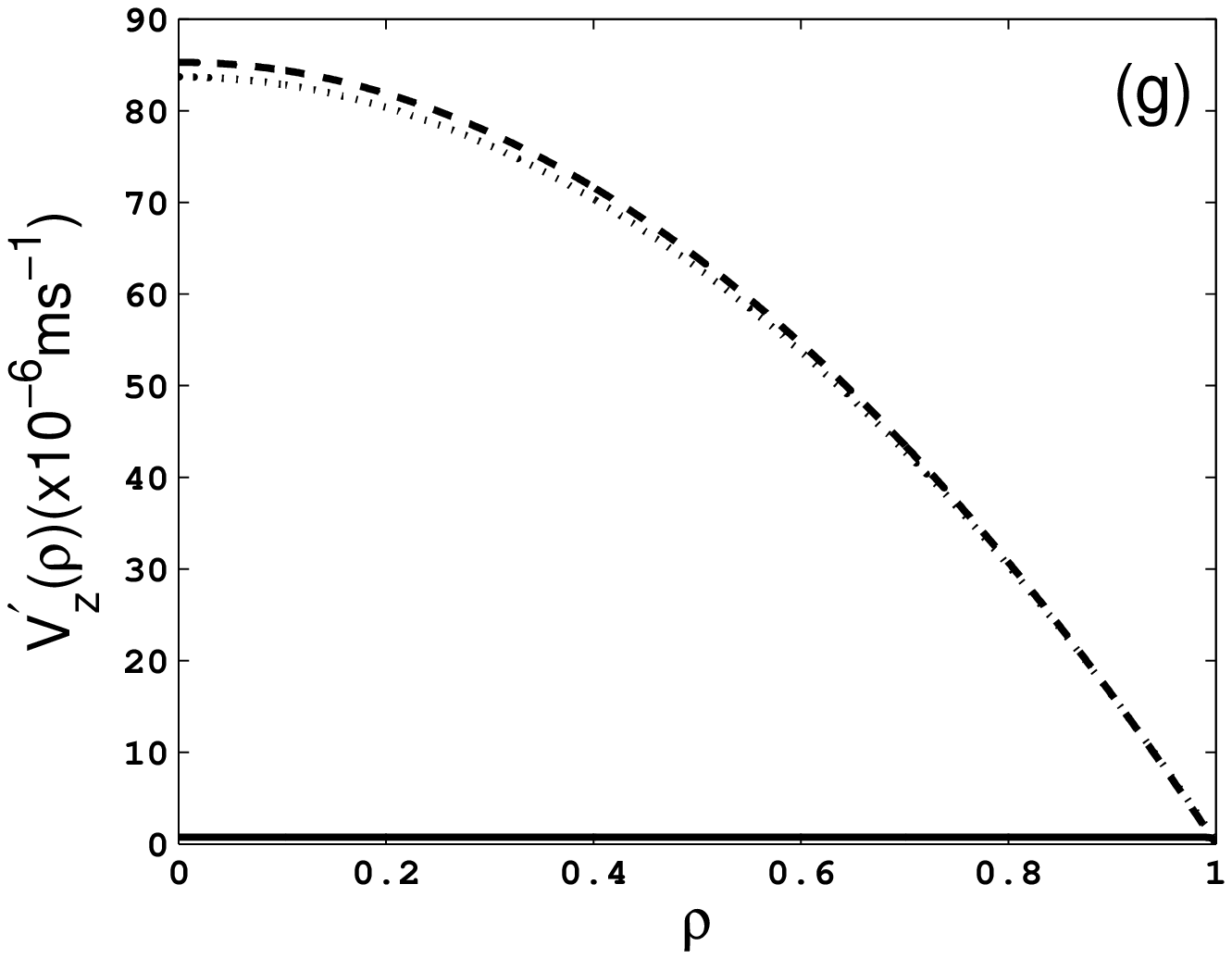,height=3.5cm}
\epsfig{file=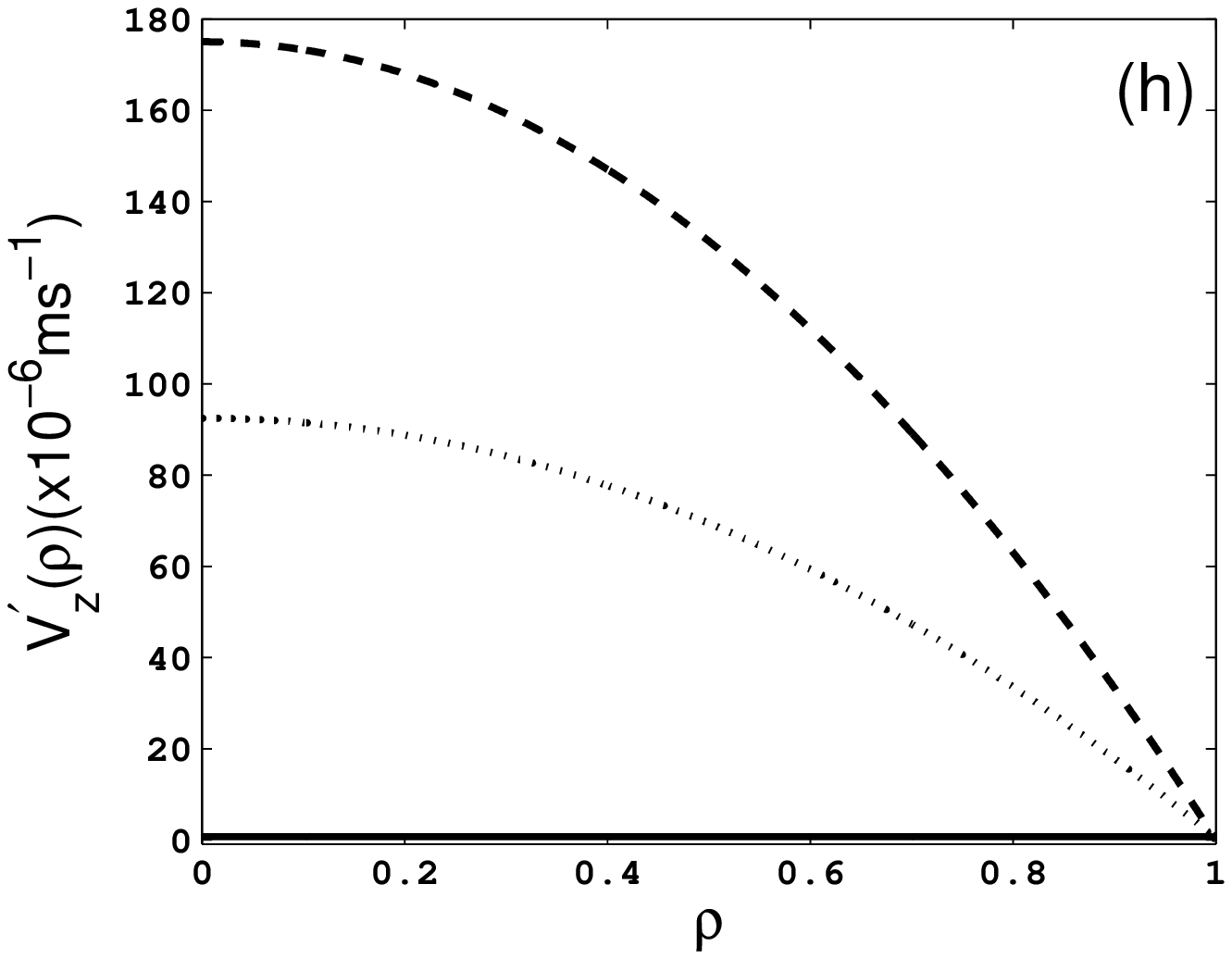,height=3.5cm}\\
\caption{Variation of $V_z^\prime(\rho)$ with $\rho$ when (a--d)
$m_o=50$ or (e--h) $m_o=500$, for $k_1=0$ (solid curves), $k_1=0.5$
(dotted curves), and $k_1=0.95$ (dashed curves). (a, e)
$\beta_o=-0.01$, (b, f) $\beta_o=-0.1$, (c, g) $\beta_o=-0.95$, (d,
h) $\beta_o=-1$. } \label{Vz_rfig1}
\end{figure}

Figure \ref{Vz_rfig1}  also shows that the speed of a micropolar
fluid ($k_1\ne 0$) decreases as $k_1$ increases {for all $m_o$ but
not for all $\beta_o$}. In addition to the fluid speed increasing
with $m_o$ (just as for a simple Newtonian fluid), the fluid speed
also increases with $\vert\beta_o\vert$ for all $m_o\gg1$ and
$k_1>0$.

The foregoing trends are in accord with the results of an analytical
investigation  in the central portion of the microcapillary. With
the assumption that $m_o\geq 10$, Eq.~\r{Vzrho} yields
\begin{equation}
\label{4.5} V^\prime_z(0) \simeq U\left(
1+k_1\left\{\frac{k_2}{k_0^2-m_o^2}-\frac{m_o}{k_0}\left(\frac{k_2}{k_0^2-m_o^2}
+\frac{\beta_o}{1+k_1\beta_o}\right)\left[\frac{I_0(k_0)-1}{I_1(k_0)}\right]\right\}\right)\,.
\end{equation}
For simple Newtonian fluids ($k_1=0$), Eq.~(\ref{4.5}) shows that
$V^\prime_z(0)=U$ is independent of  $m_o$ as well as of, as
expected, $\beta_o$. Plots of $V^\prime_z(0)$ vs. $m_o$ in
Figure~\ref{Vz_r0fig2} indicate that the fluid speed at the center
of the microcapillary increases with $m_o$, and therefore with $R$,
for micropolar fluids ($k_1\neq0$). When microrotation effects can
be neglected at the wall of the microcapillary---i.e., when
$\beta_o=0$---Eq.~(\ref{4.5}) simplifies to
\begin{equation}\label{4.6}
V^\prime_z(0)\Big\vert_{\beta_o=0}\simeq
U\left(1+\frac{k_1k_2}{k_0^2-m_o^2}
\left\{1-\left(\frac{m_o}{k_0}\right)\left[\frac{I_0(k_0)-1}{I_1(k_0)}\right]\right\}\right)\,,
\end{equation}
and {Figure}~\ref{Vz_r0fig2} shows that the effect of $m_o$ on
$V^\prime_z(0)$ is very weak but the effect of $k_1$ is
considerable. As $\vert\beta_o\vert$ increases, $V^\prime_z(0)$
acquires a stronger tendency in a micropolar fluid to increase with
$m_o$, which can be concluded from {Figure}~\ref{Vz_r0fig2}.

\begin{figure}[!htb]
\centering \psfull
\epsfig{file=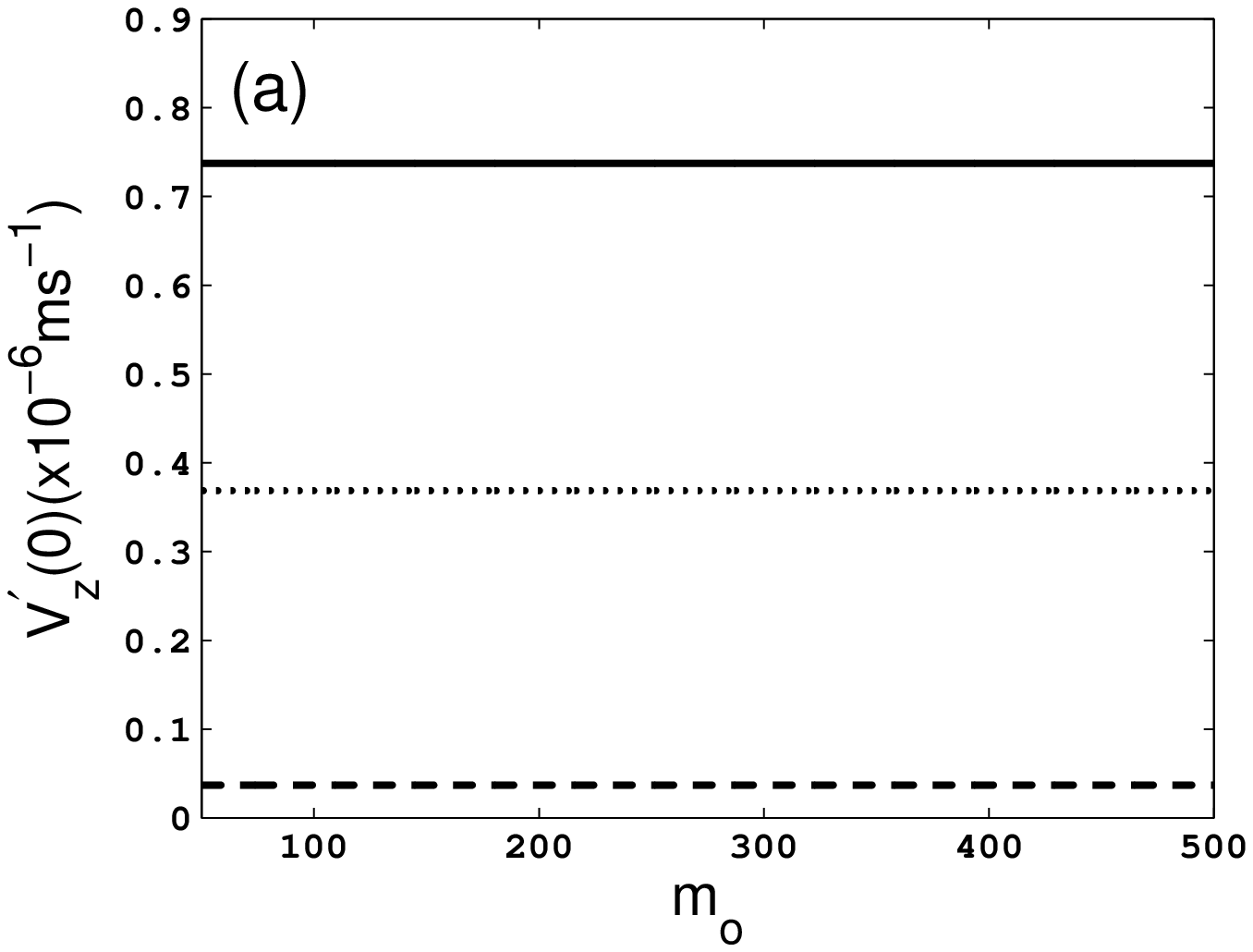,height=3.5cm}
\epsfig{file=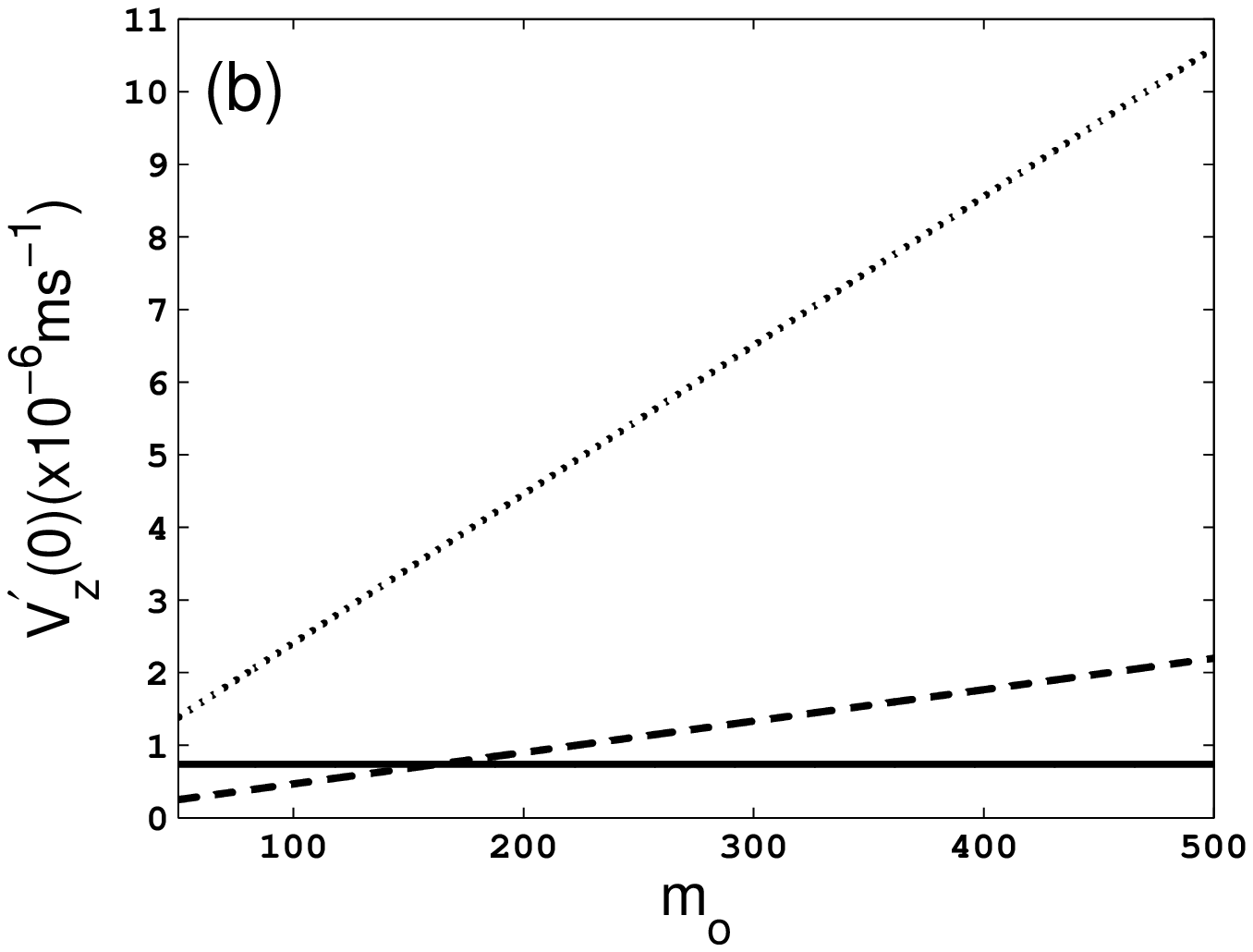,height=3.5cm}
\epsfig{file=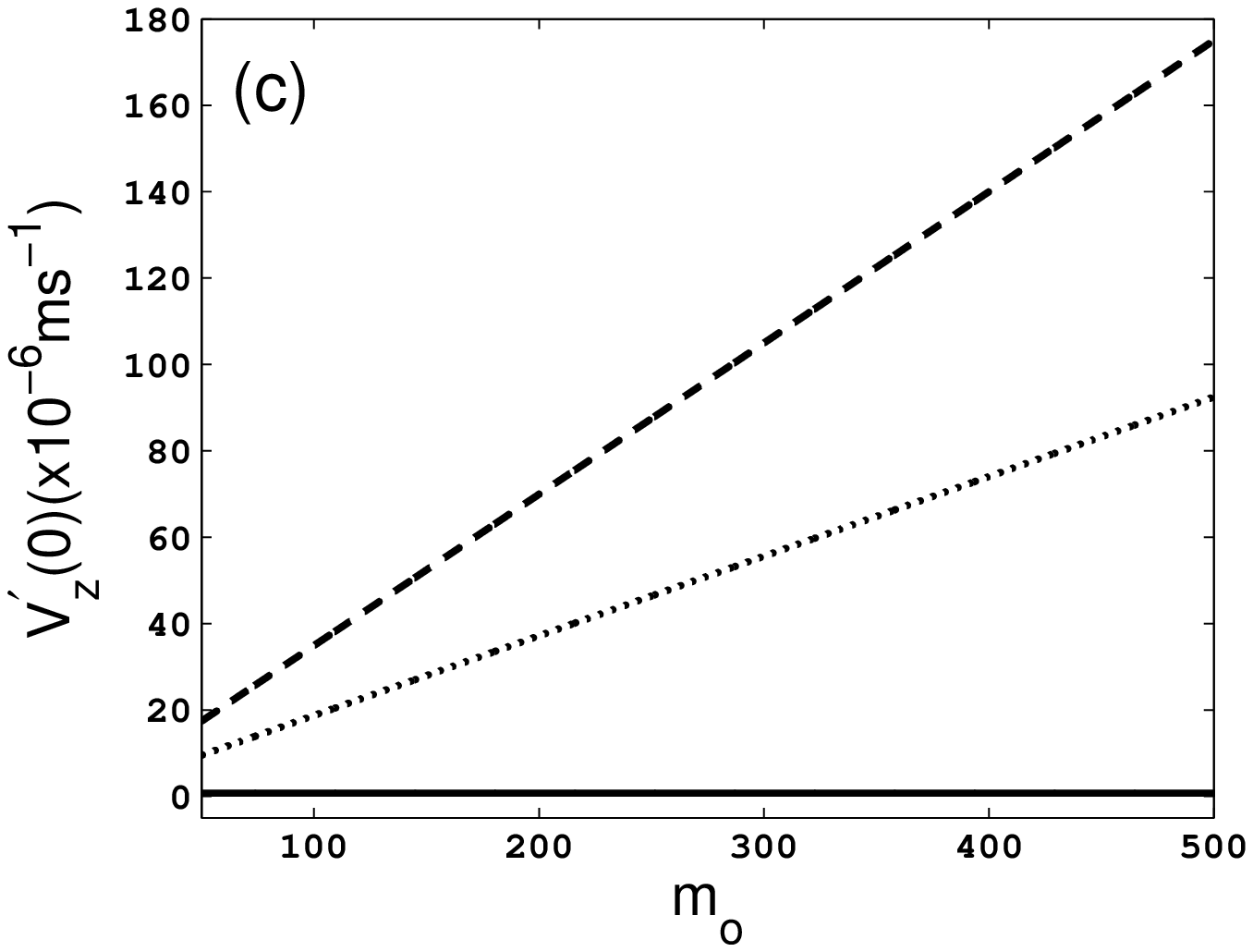,height=3.5cm}
\caption{Variation of $V_z^\prime(0)$ with $m_o$ for $k_1=0$ (solid
curves), $k_1=0.5$ (dotted curves), and $k_1=0.95$ (dashed curves).
(a)  $\beta_o=0$, (b) $\beta_o=-0.2$, (c) $\beta_o=-1$. }
\label{Vz_r0fig2}
\end{figure}

The fluid-speed gradient ${dV_z^\prime}/{d\rho}$ is not only an
important factor in microrotation per Eqs.~(\ref{2.22})$_1$ and
(\ref{3.5}), but it also influences several components of the stress
tensor identified in Eqs.~(\ref{4.1.3.1}). The Debye--H\"uckel
expression for ${dV_z^\prime}/{d\rho}$ follows from
Eqs.~(\ref{3.4}), (\ref{3.5}), and (\ref{3.8}) as
\begin{equation}
\label{4.3}  \frac{dV^\prime_z(\rho)}{d\rho}=-U\left[c_1 k_1
I_1(k_0\rho)+m_o\left(1+\frac{k_1
k_2}{k_0^2-m_o^2}\right)\frac{I_1(m_o\rho)}{I_0(m_o)}\right]\,.
\end{equation}
Now, this gradient vanishes as $\rho\to0$, i.e., at the centre of
the  microcapillary,   in conformity with boundary condition
(\ref{2.19-1}). But it has high magnitudes near the wall, as can be
gathered by setting $\rho =1$ in Eq.~(\ref{4.3}) to obtain
\begin{equation}
\frac{dV^\prime_z(\rho)}{d\rho}\Big\vert_{\rho=1}=
-\frac{Um_o}{1+k_1\beta_o}\,\frac{I_1(m_o)}{I_0(m_o)}\,.
\end{equation}
When $m_o\geq10$, the foregoing expression can be further
approximated as
\begin{equation}
\label{4.4}
 \frac{dV^\prime_z(\rho)}{d\rho}\Big\vert_{\rho=1} \simeq -\frac{Um_o}{1+k_1\beta_o}\,,
\end{equation}
evincing a direct proportionality with $m_o$ (or $R$) for all
$\vert\beta_o\vert$---for both simple Newtonian and micropolar
fluids.

\subsection{Fluid flux}\label{ff}

From an engineering perspective, the fluid flux   should be considered in addition to fluid speed. In the
present context, it is defined as
\begin{equation}\label{Qprime}
Q^\prime =2\pi \int^R_0 \rho^\prime\, V^\prime_z(\rho^\prime) \,
{d\rho^\prime}=2\pi{R^2U}\int^1_0 \rho\,V_z(\rho) \,d\rho\,.
\end{equation}
On substituting the Debye--H\"uckel expression (\ref{Vzrho}) in the
integrand on the right side of Eq.~(\ref{Qprime}), we get
\begin{equation}
\label{4.9} Q^\prime={\pi} R^2 U\left[\frac{c_1k_1}{k_0}I_2(k_0)+
\left(1+\frac{k_1k_2}{k^2_0-m^2_o}\right)\frac{I_2(m_o)}{I_0(m_o)}\right]\,,
\end{equation}
which simplifies to
\begin{equation}
\label{4.10} Q^\prime\simeq{\pi} R^2
U\left[\frac{c_1k_1}{k_0}I_2(k_0)+
\left(1+\frac{k_1k_2}{k^2_0-m^2_o}\right)\left(1-\frac{2}{m_o}\right)
\right]\,
\end{equation}
for $m_o\geq10$. Clearly then, the fluid flux depends on $k_1$,
$\beta_o$, and $m_o$.

{The dependence of $Q^\prime$ on $\beta_o$ can be identified using
Eqs.~(\ref{3.10}) and (\ref{4.9}). We can write $Q^\prime$ as the
sum of two parts, one of which is independent of $\beta_o$ and the
other depends linearly on $\beta_o/(1+k_1\beta_o)$.}

\begin{figure}[!htb]
\centering \psfull
\epsfig{file=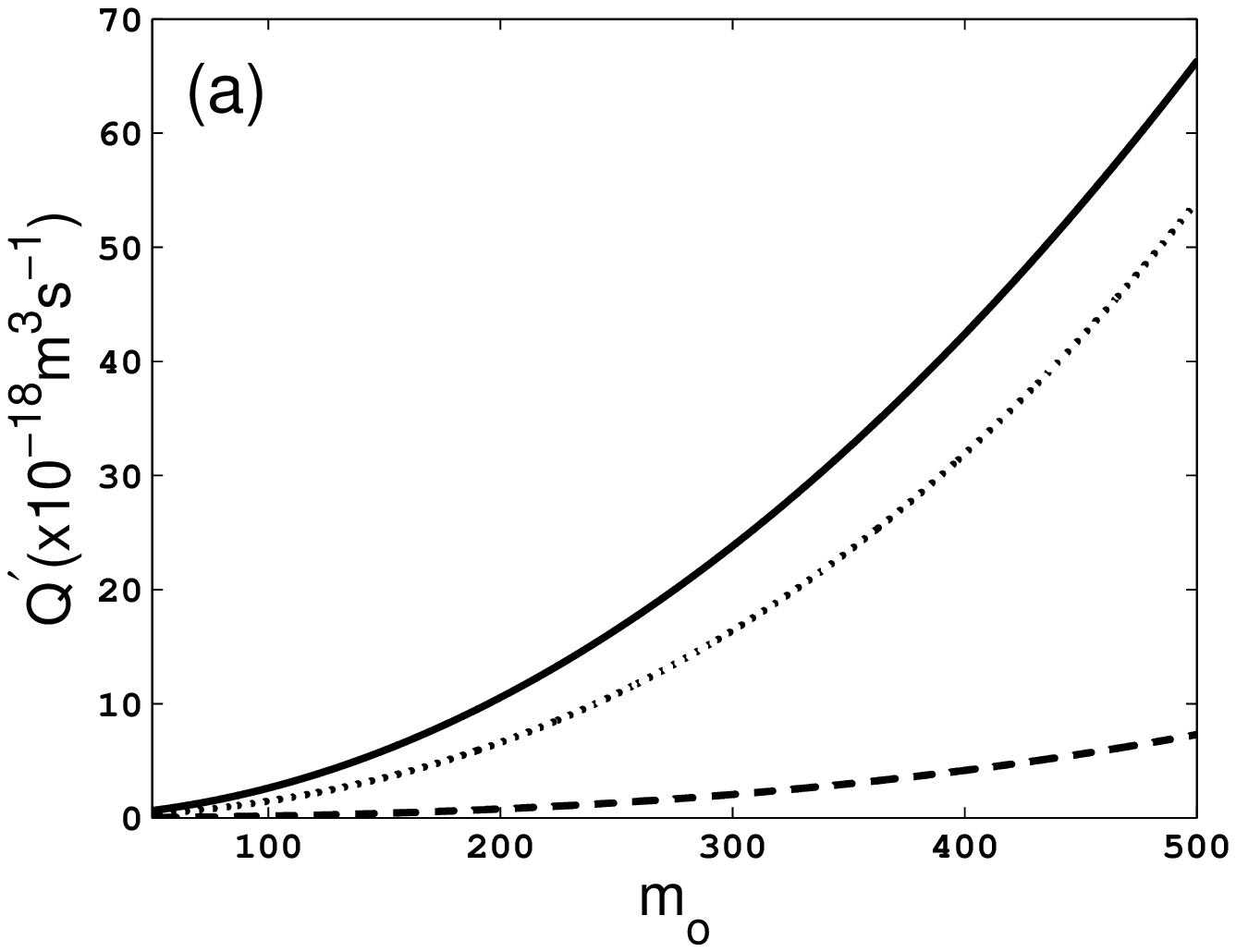,height=3.5cm}
\epsfig{file=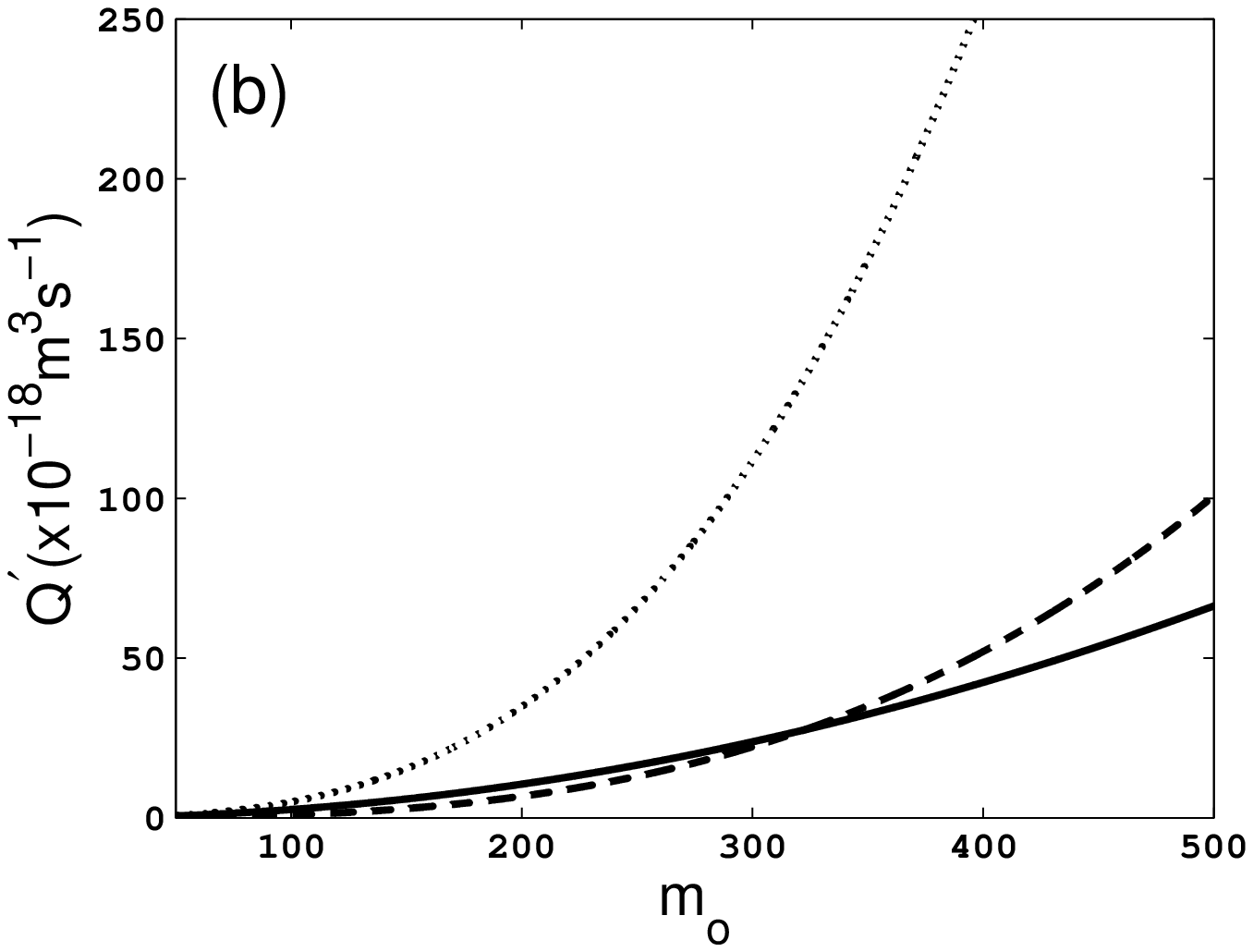,height=3.5cm}
\epsfig{file=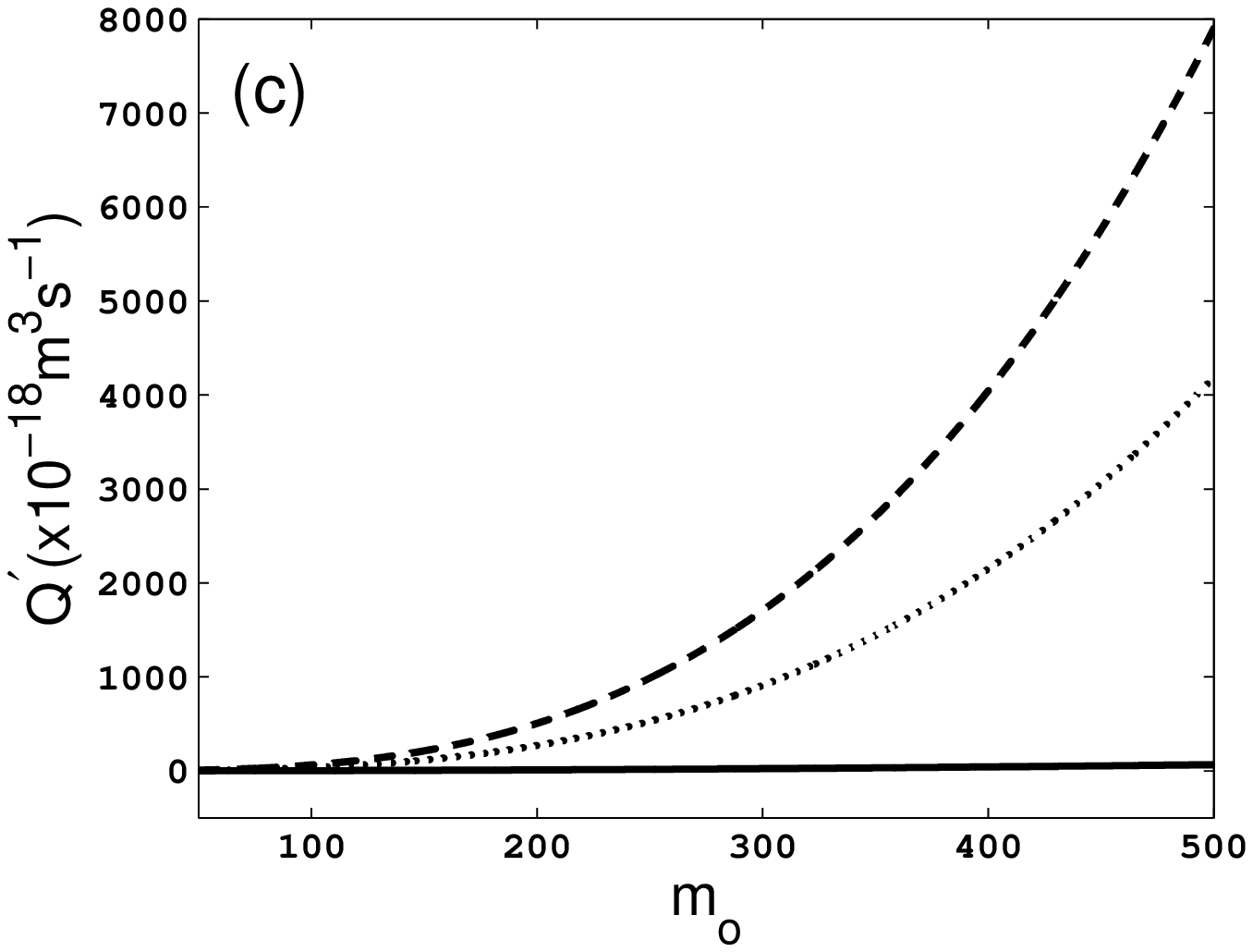,height=3.5cm}\\
\caption{Variation of $Q^\prime$ with $m_o$ for $k_1=0$ (solid
curves), $k_1=0.5$ (dotted curves), and $k_1=0.95$ (dashed curves).
(a)  $\beta_o=-0.01$, (b) $\beta_o=-0.2$, (c) $\beta_o=-1$. }
\label{Qfig3}
\end{figure}

{Although for simple Newtonian fluids ($k_1=0$), Eq.~(\ref{4.10})
yields
\begin{equation}\label{4.11}
1-\frac{Q^\prime}{{\pi}R^2U} \simeq \frac{2}{m_o}
\end{equation}
for $m_o\gg1$, the relationship of $Q^\prime$   and   $m_o$    is
more complicated when $k_1\ne0$.} The variation of ${Q^\prime}$ with
respect to $m_o$  is illustrated in  {Figure}~\ref{Qfig3} for nine
different combinations of $k_1$ and $\beta_o$. This figure shows
that $Q^\prime$ increases with $m_o$ (and, therefore, with $R$)
 for   {simple Newtonian fluids; the same trend
 exists for micropolar fluids, regardless of the value of $\beta_o\in[-1,0]$.}
Furthermore, $Q^\prime$ intensifies with  { increasing
$\vert\beta_o\vert$ for all $m_o\in[50,500]$ and $k_1\in(0,1)$, the
intensification rate $dQ^\prime/d\vert\beta_o\vert$ itself
increasing concurrently}.

{Figure~\ref{Qfig4} shows the dependency of $Q^\prime$ on $k_1$ for
12 different combinations of $\beta_o$ and $m_o$. Above a threshold
value of $\vert\beta_o\vert$ which is quite small, the following
trend is followed: from the value
$Q^\prime_0\simeq\pi{R^2}U(1-2/m_o)$ at $k_1=0$, the fluid flux
$Q^\prime$ increases linearly with $k_1$ for $k_1\ll 1$, then
increases nonlinearly with $k_1$ to a maximum value ${\tilde
Q}^\prime$ at $k_1={\tilde k}_1$, and finally drops monotonically to
$Q^\prime_1$ at $k_1=1$, for $\beta_o\in(-1,0]$ and $m_o\gg1$.
Whereas ${\tilde Q}^\prime$ increases with both $m_o$ and
$\vert\beta_o\vert$, ${\tilde k}_1$ is almost totally independent of
$m_o$ but increases with $\vert\beta_o\vert$. Calculations show that
$Q^\prime_1{\sim}m_o^n$, with the exponent $n$ lying between 2 and 3
and increasing with $\vert\beta_o\vert$. Furthermore, for
$\beta_o=-1$,  $Q^\prime$ increases linearly with $k_1$ for
$k_1\in[0,1]$ and ${\tilde k}_1=1$.}

\begin{figure}[!htb]
\centering \psfull
\epsfig{file=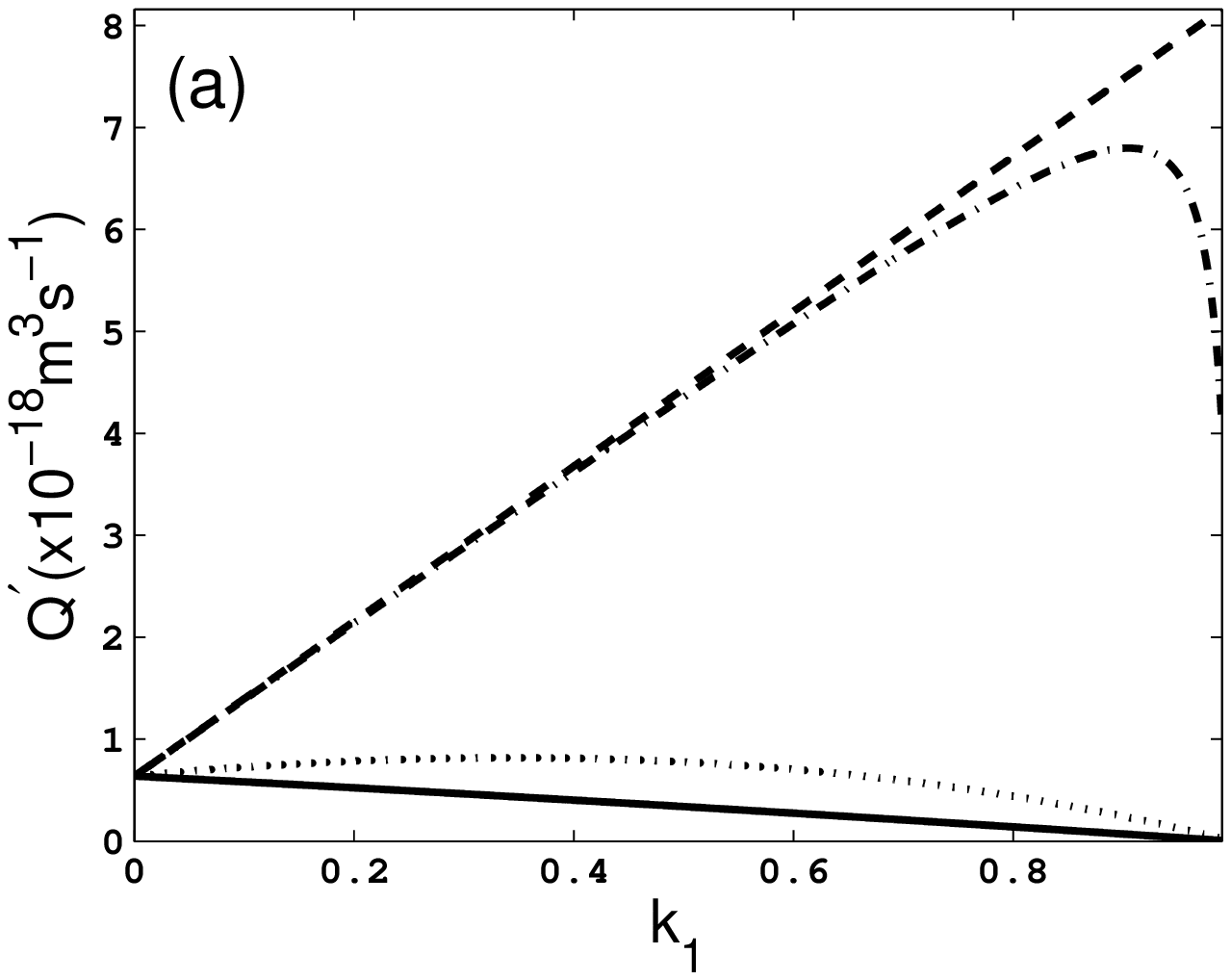,height=3.5cm}
\epsfig{file=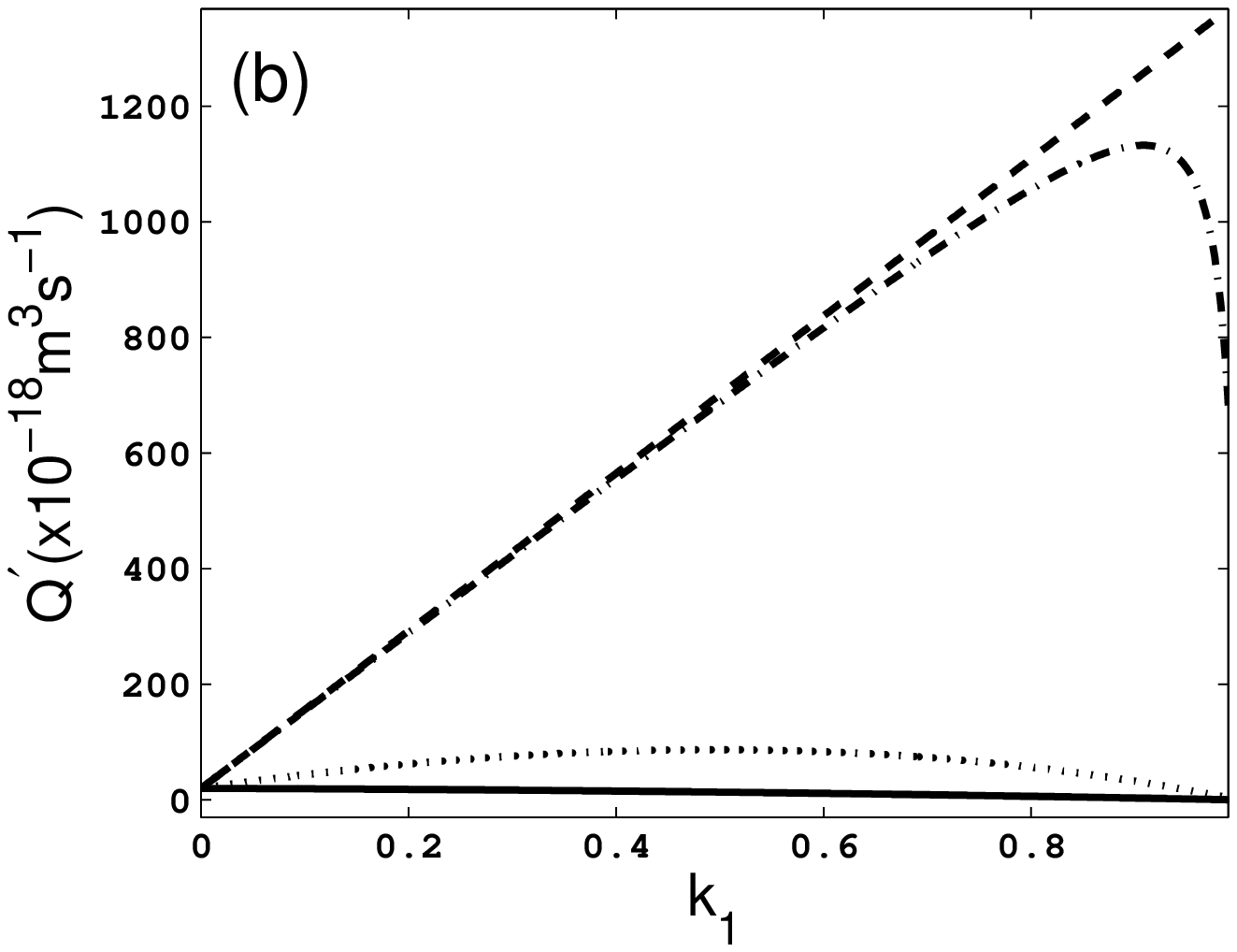,height=3.5cm}
\epsfig{file=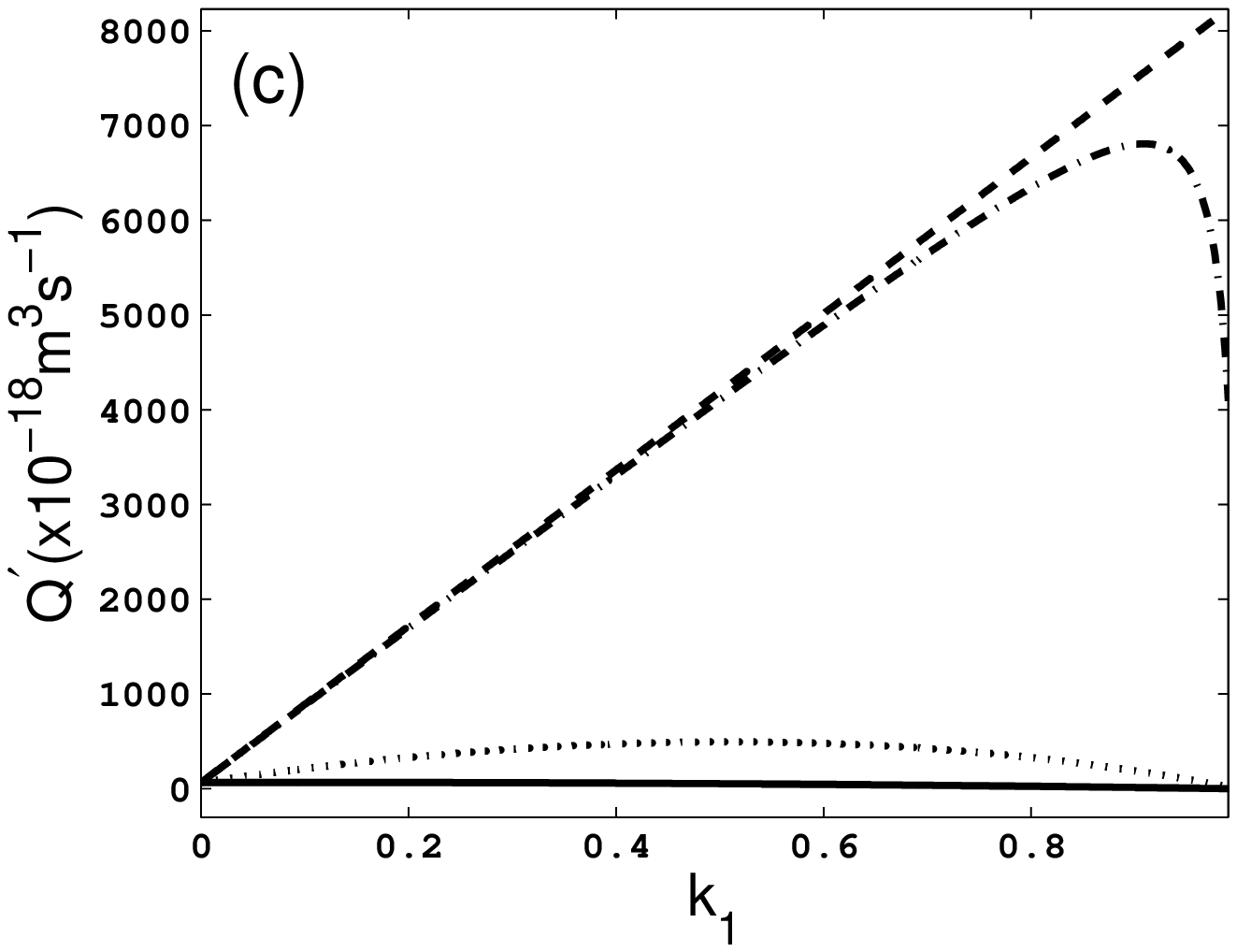,height=3.5cm}\\
\caption{Variation of $Q^\prime$ with $k_1$ for $\beta_o=-0.01$
(solid curves), $\beta_o=-0.2$ (dotted curves), $\beta_o=-0.99$
(dashed-dotted curves), and $\beta_o=-1.0$ (dashed curves).(a)
$m_o=50$, (b) $m_o=275$, and (c) $m_o=500$. } \label{Qfig4}
\end{figure}

\subsection{Microrotation}

{The microrotation $v^\prime_\theta(\rho)$  exists in micropolar
fluids but not in simple Newtonian fluids. Accordingly, representative plots of
$v^\prime_\theta$ vs. $\rho$, computed using the Debye--H\"uckel
expression (\ref{3.8}), are presented in Figure~\ref{Spin_r} for
$m_o\gg1$ and $k_1 > 0$. This figure indicates that
$v^\prime_\theta(\rho)$
\begin{itemize}
\item[(i)] increases linearly  with $\rho$,
\item[(ii)] increases with $|\beta_o|$ for
all $k_1\in(0,1)$ and $\rho\in(0,1]$, and
\item[(iii)]  decreases as $k_1$
increases for all $m_o\in[50,500]$ and $\beta_o\in[-1,0)$.
\end{itemize}}

{By virtue of the boundary condition (\ref{2.22})$_2$, the
microrotation is weak in the the central part of the microcapillary,
but it is maximum at the wall $\rho =1$ for all $m_o\gg1$,
$\beta_o\in[-1,0)$, and $k_1>0$. Indeed, Eq.~(\ref{3.8}) yields
\begin{equation}
\label{4.13}
v^\prime_\theta(\rho)\Big\vert_{\rho=1}\approx-\frac{U\beta_o}{\lambda_D(1+k_1\beta_o)}\,.
\end{equation}
This equation shows that the microrotation at the wall of the
microcapillary is independent of  $R$f, which is in agreement with
Figure~\ref{Spin_r}. This figure also shows that, for $\beta_o=-1$,
$v^\prime_\theta$ at $\rho=1$ is independent of $k_1$, which agrees
with the result
\begin{equation}
\label{4.14}
v^\prime_\theta(\rho)\Big\vert_{\rho=1,\beta_o=-1}\approx-\frac{\epsilon\psi_oE_o}{\mu\lambda_D}\,.
\end{equation}
obtained after using the definitions (\ref{2.5}) and (\ref{k-defs})
in Eq.~(\ref{4.13}).}

\begin{figure}[!htb]
\centering \psfull
\epsfig{file=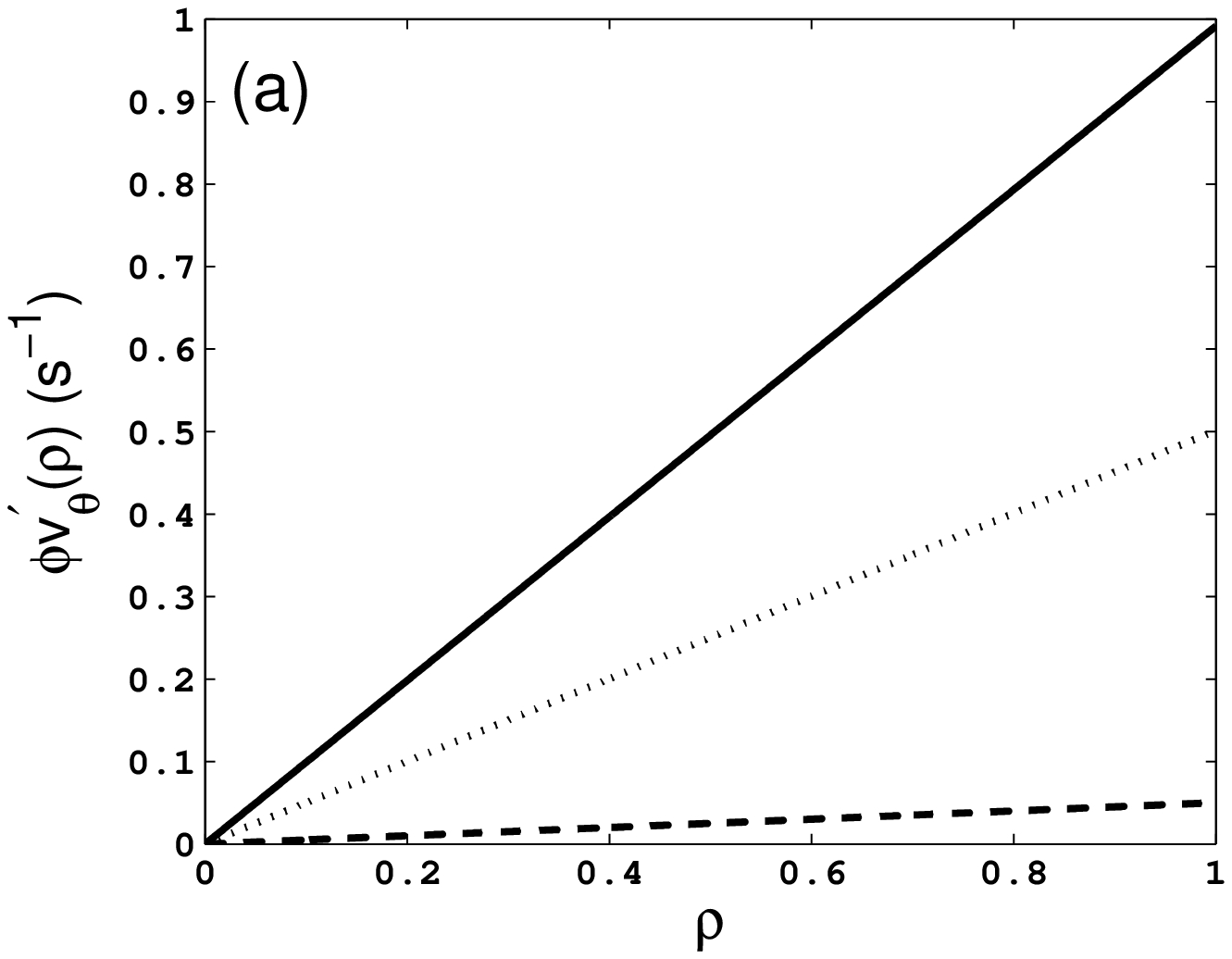,height=3.5cm}\,\epsfig{file=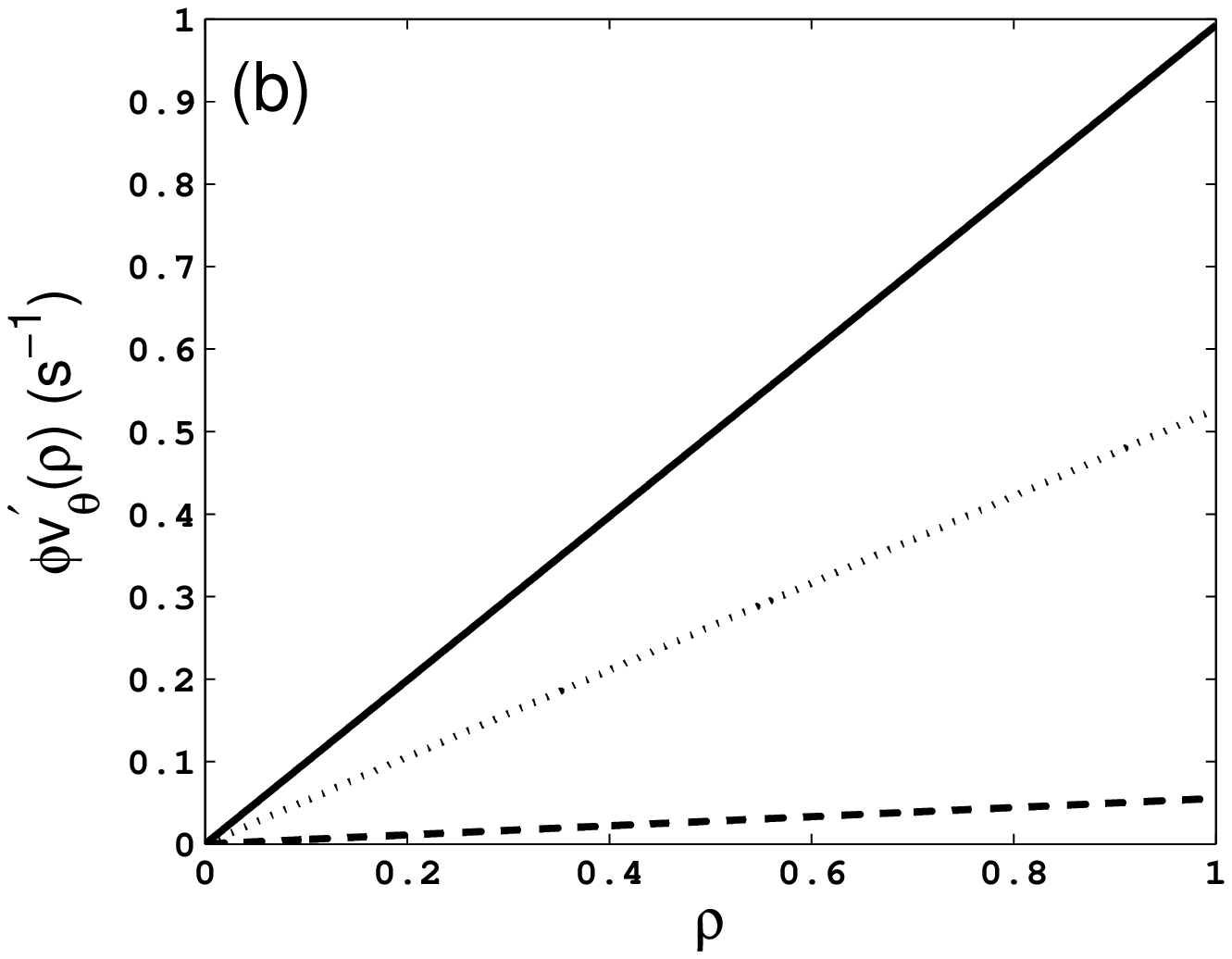,height=3.5cm}
\epsfig{file=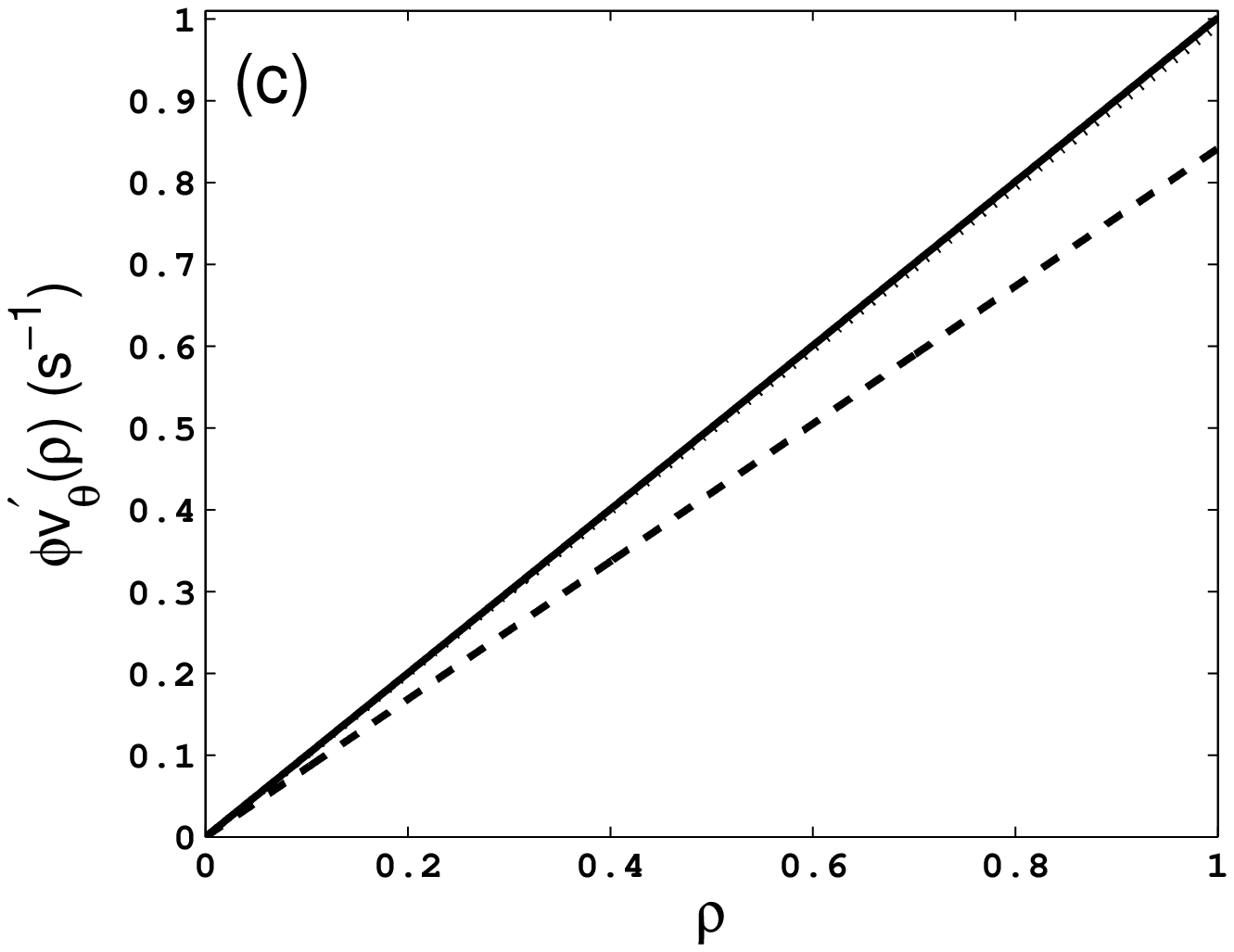,height=3.5cm}\,\epsfig{file=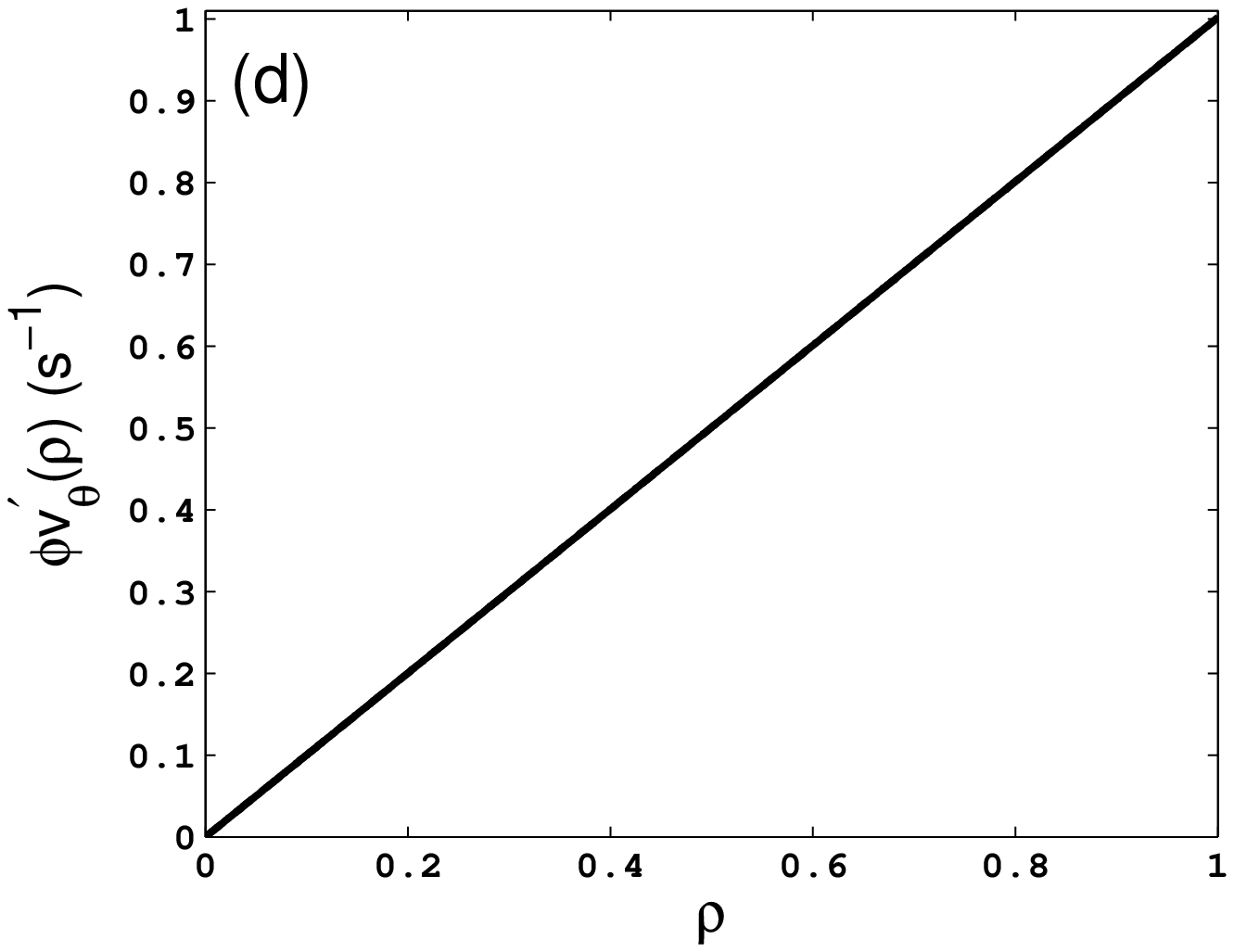,height=3.5cm}
\epsfig{file=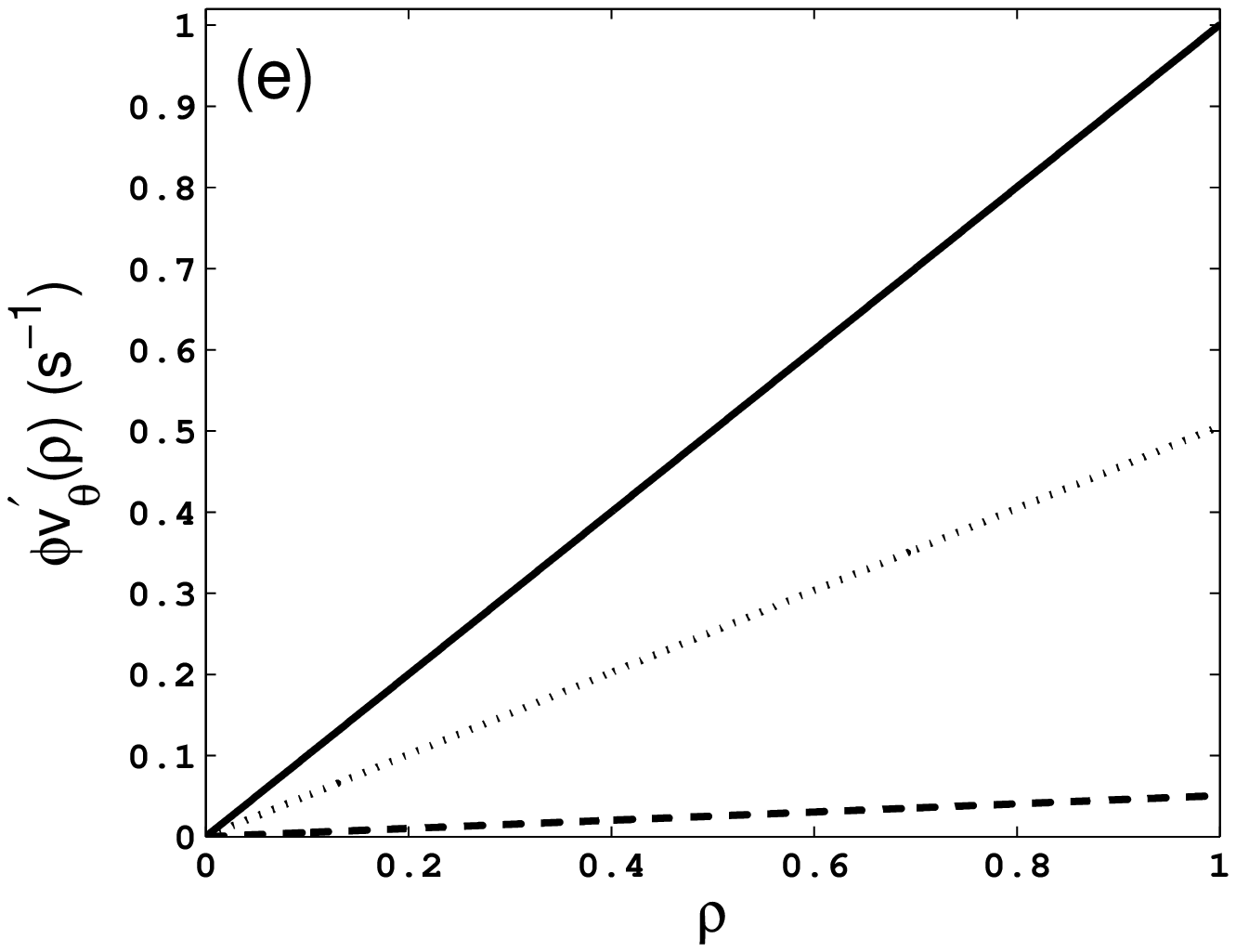,height=3.5cm}\,\epsfig{file=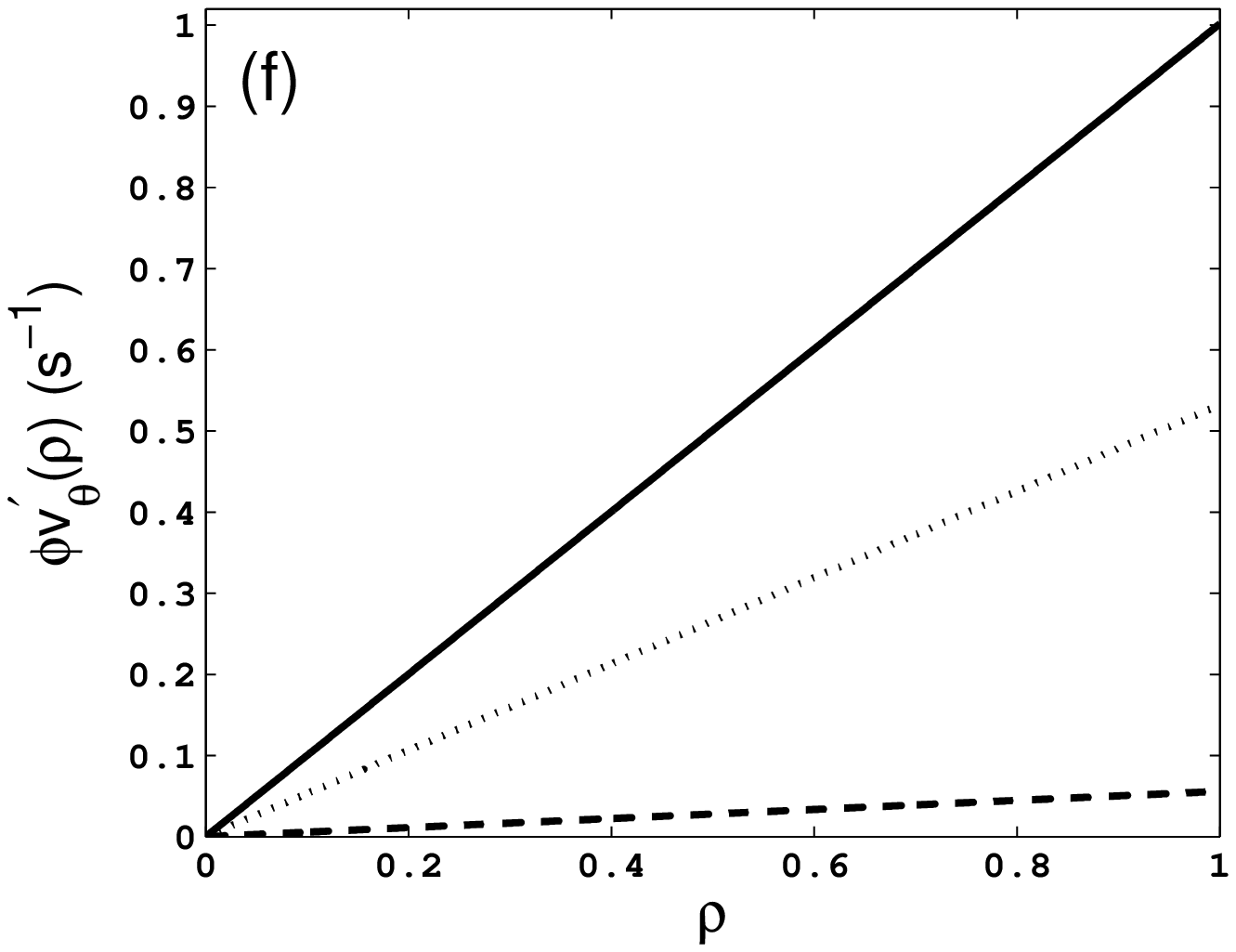,height=3.5cm}
\epsfig{file=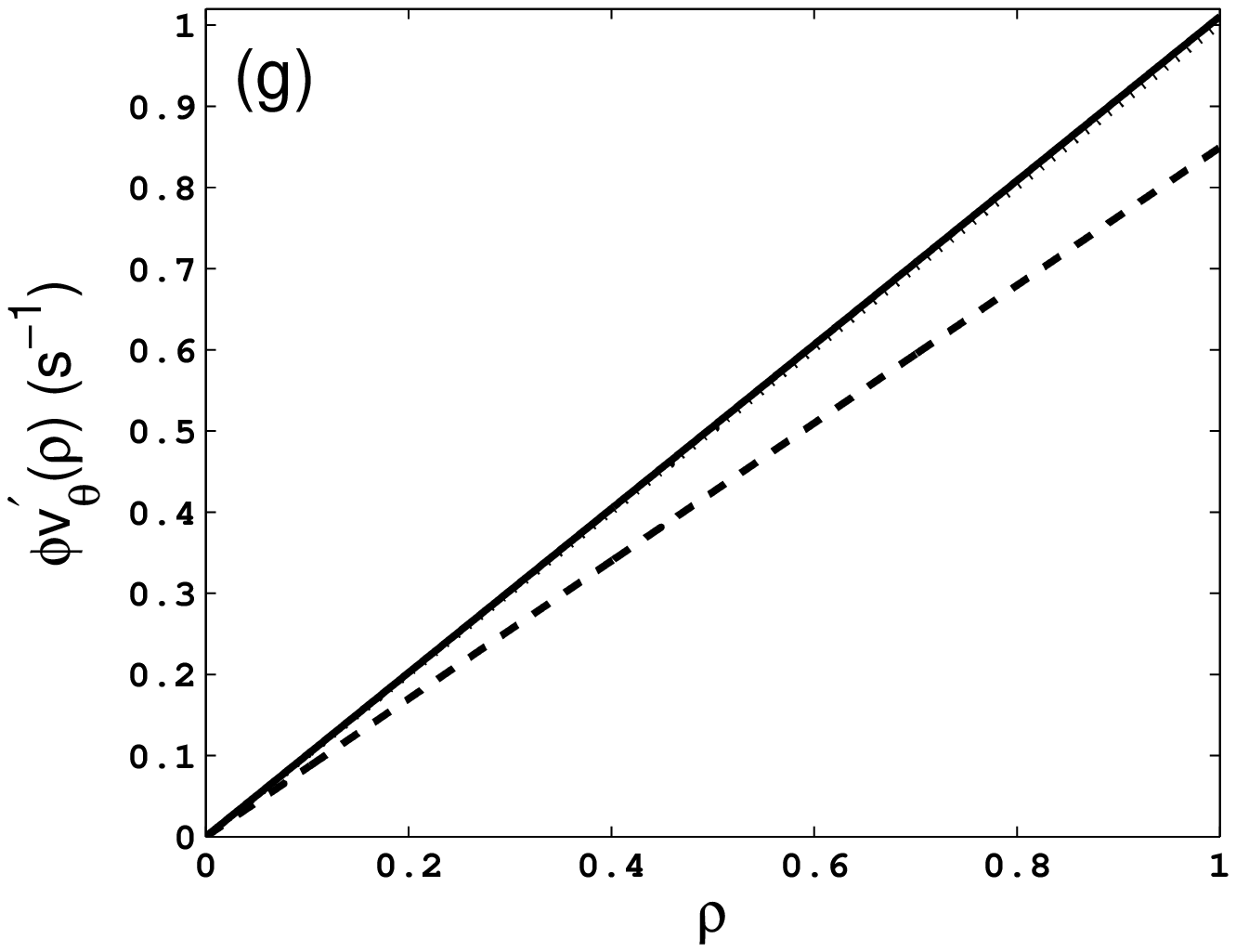,height=3.5cm}\,\epsfig{file=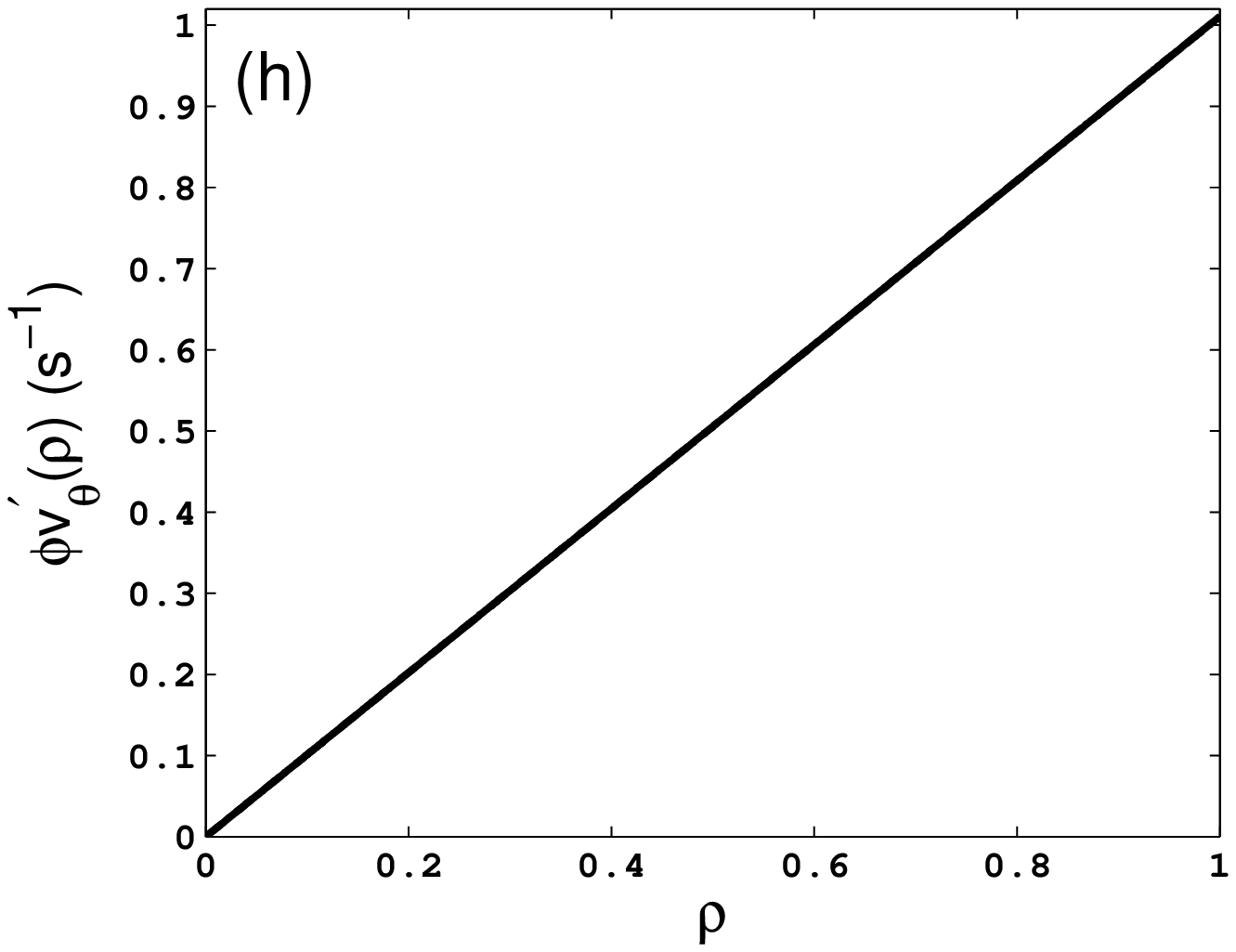,height=3.5cm}
\caption{Variation of $\phi v^\prime_\theta(\rho)$, where the
normalization factor $\phi=(68|\beta_o|)^{-1}$, with $\rho$ when
(a--d) $m_o=50$ or (e--h) $m_o=500$, for $k_1=0.01$ (solid curves),
$k_1=0.5$ (dotted curves), and $k_1=0.95$ (dashed curves). (a, e)
$\beta_o=-0.001$, (b, f) $\beta_o=-0.1$, (c, g) $\beta_o=-0.99$, (d,
h) $\beta_o=-1$. } \label{Spin_r}
\end{figure}


All of the foregoing conclusions about the microrotation can be
encapsulated as the approximation
\begin{equation}
\label{vthetaprime-simple} v^\prime_\theta(\rho) \simeq
-\frac{U}{R}\,\frac{\beta_o}{1+k_1\beta_o}\,m_o\rho\,,
\end{equation}
which is valid for $m_o\gg1$. Its derivation proceeds as follows: A
comparison of magnitudes for large $m_o$ suggests that
${k_2}/\left(k_0^2-m_o^2\right)$ can be neglected in favor of
${\beta_o}/\left(1+k_1\beta_o\right)$ while $I_1(m_o)/I_0(m_o)\simeq
1$ in Eq.~(\ref{3.10}), leading to
\begin{equation}
c_1\simeq - \frac{\beta_o}{1+k_1\beta_o}\,m_o\frac{1} {I_1(k_0)}\,.
\end{equation}
A similar argument leads to the neglect of the second term on the
right side of Eq.~(\ref{3.8})\, yielding
\begin{equation}
v_\theta(\rho)\simeq -
\frac{\beta_o}{1+k_1\beta_o}\,m_o\frac{I_1(k_0\rho)}{I_1(k_0)}\,.
\end{equation}
Even for very large $m_o$, $k_0$ is very small so that
$I_1(k_0)\simeq k_0/2$ and $I_1(k_0\rho)\simeq k_0\rho/2$;
Eq.~(\ref{vthetaprime-simple}) then follows for $m_o\gg1$.

\subsection{Stress  tensor}
After the use of Eqs.~(\ref{3.8}) and (\ref{Vzrho}) in
Eqs.~(\ref{4.1.3.0})$_1$ and (\ref{4.1.3.0})$_2$, we obtain the
non-zero components of the stress tensor as
\begin{equation}
\label{4.14} \sigma^\prime_{\rho{z}}(\rho)
=\frac{\epsilon\psi_oE_o}{\lambda_D}\frac{I_1(m_o\rho)}{I_0(m_o)}
\end{equation}
and
\begin{equation}
\label{4.15} \sigma^\prime_{z\rho
}(\rho)=\frac{\epsilon\psi_oE_o}{\lambda_D} \left\{
\frac{c_1k_1\left(2-k_1\right)}{m_o}
 I_1(k_0\rho)+ \left[1 +
\frac{k_1m_o^2}{k^2_0-m^2_o}\right] \frac{I_1(m_o\rho)}{I_0(m_o)}
\right\}\,,
\end{equation}
consistently with the Debye--H\"uckel approximation. For a simple
Newtonian fluid ($k_1=0$), from the foregoing expressions we get
$\sigma^\prime_{\rho{z}}(\rho) = \sigma^\prime_{{z}\rho}(\rho)$,
which conforms to the symmetry of the stress tensor in the absence
of micropolarity. In a micropolar fluid ($k_1\ne0$), the stress
tensor of Eq.~(\ref{stress-def}) has both symmetric and
skew-symmetric components, which explains why $
\sigma^\prime_{\rho{z}}(\rho) \ne \sigma^\prime_{{z}\rho}(\rho)$.

What is really surprising is that $\sigma^\prime_{\rho{z}}(\rho)$ is
the same for simple Newtonian as well as micropolar fluids. This
component of the stress tensor depends on $m_o$, but  neither on
$k_1$ nor on  $\beta_o$. Furthermore, per Eq.~(\ref{4.14}), it is
absent on the axis of the microcapillary because $I_1(0)=0$. Indeed,
as can be deduced from Fig.~\ref{StressRZ_r},
$\sigma^\prime_{\rho{z}}(\rho)$ is negligible in most of the
microcapillary and becomes significant in magnitude very near to and
on the wall, with
\begin{equation}
\label{4.14add} \sigma^\prime_{\rho{z}}(\rho)\Big\vert_{\rho=1}
=\frac{\epsilon\psi_oE_o}{\lambda_D}\,.
\end{equation}
 As $m_o$ increases, the region in which the 
 magnitude of $\sigma^\prime_{\rho{z}}(\rho)$ is significant
 shrinks.

\begin{figure}[!htb]
\centering \psfull
\epsfig{file=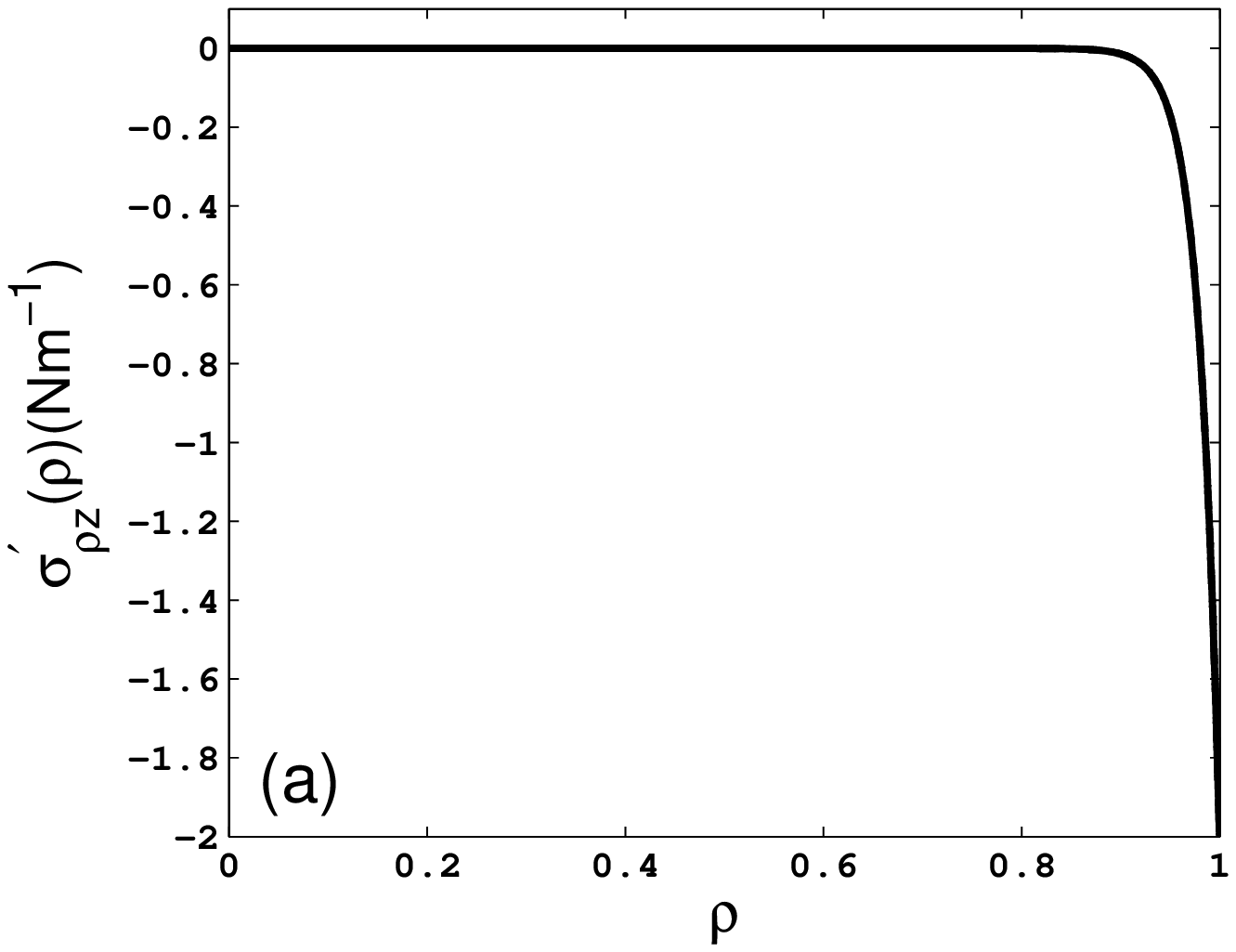,height=3.5cm}\,
\epsfig{file=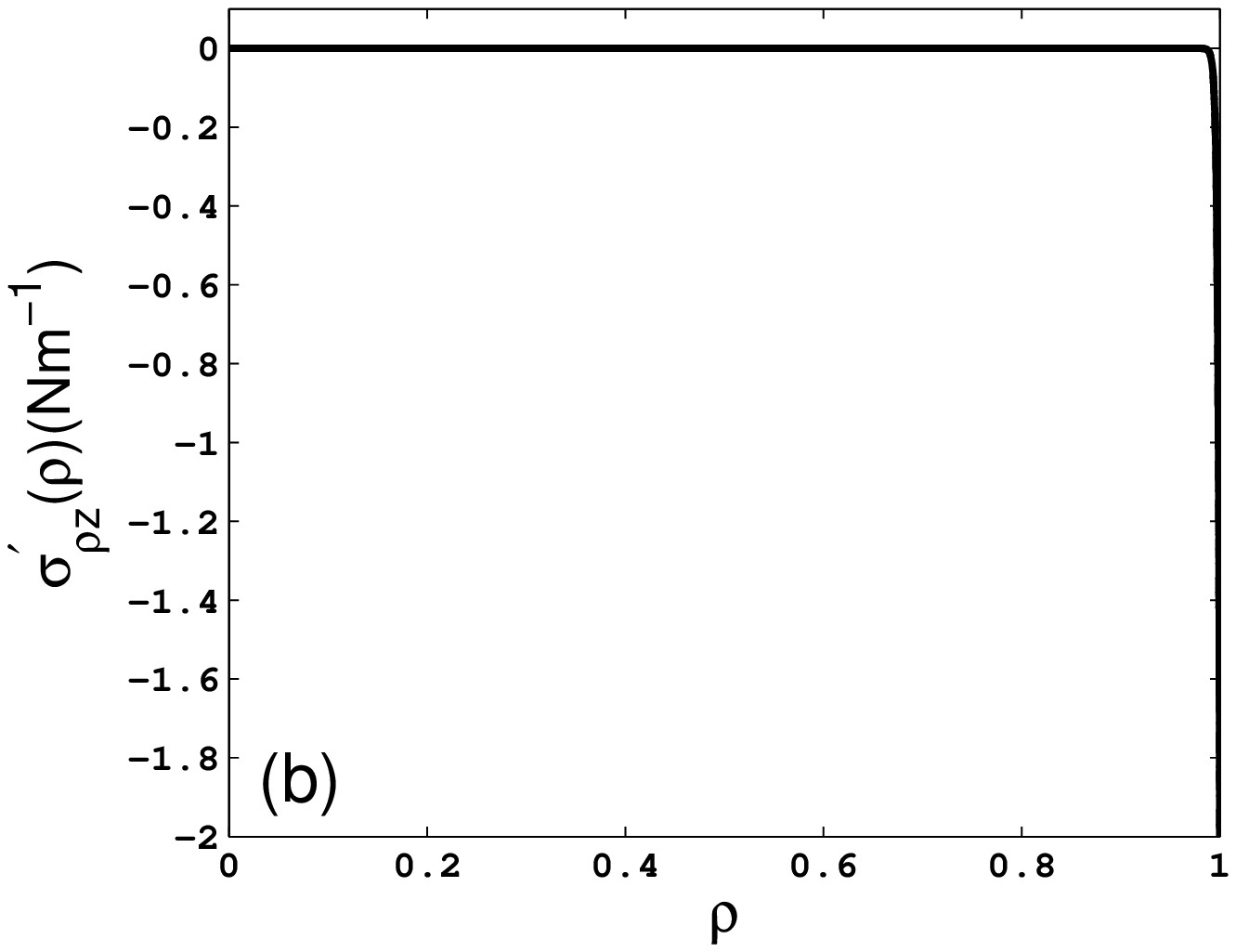,height=3.5cm}
\caption{Variation of $\sigma^\prime_{\rho z}(\rho)$ with $\rho$
when  (a) $m_o=50$ or (b) $m_o=500$. There is no dependence on
either $k_1$ or $\beta_o$.} \label{StressRZ_r}
\end{figure}

The difference $
\sigma^\prime_{\rho{z}}(\rho)-\sigma^\prime_{z\rho}(\rho)$ is
entirely due to  micropolaity, as becomes evident on comparing the
spatial profiles of $\sigma^\prime_{z\rho}(\rho)$ in
Fig.~\ref{StressZR_r} with the spatial profiles of
$\sigma^\prime_{\rho{z}}(\rho)$ in Fig.~\ref{StressRZ_r}. Clearly,
$\sigma^\prime_{z\rho}(\rho)$ depends on $m_o$, $k_1$, and
$\beta_o$. Now, $\sigma^\prime_{z\rho}(0)=0$ but
\begin{equation}
\label{4.16}
\sigma^\prime_{z\rho}(\rho)\Big\vert_{\rho=1}\approx\frac{\eps\psi_oE_o
}{\lambda_D}\,\frac{(1-k_1-k_1\beta_o)}{(1+k_1\beta_o)}\left(1-\frac{1}{2m_o}\right)
\,
\end{equation}
for $m_o\gg1$, and Fig.~\ref{StressZR_r} shows that
$\sigma^\prime_{z\rho}(0)>\sigma^\prime_{z\rho}(\rho)
>\sigma^\prime_{z\rho}(1)$ for $\rho\in(0,1)$. The region around the
central axis of the microcapillary can be called a
$\sigma^\prime_{z\rho}$--free zone, which shrinks rapidly as
 $|\beta_o|$ increases.

Figure~\ref{StressZR_r} also shows that $|\sigma^\prime_{z\rho}|$
not only increases with $k_1$ for $m_o\gg1$ but it also  increases
with $m_o\gg1$ for all $k_1\in[0,1]$ and $\beta_o\in(-1,0)$. As
$\beta_o\rightarrow 0$, the distinction between the micropolar and
the simple Newtonian fluids disappears for
$\sigma^\prime_{z\rho}(\rho)$.

\begin{figure}[!htb]
\centering \psfull
\epsfig{file=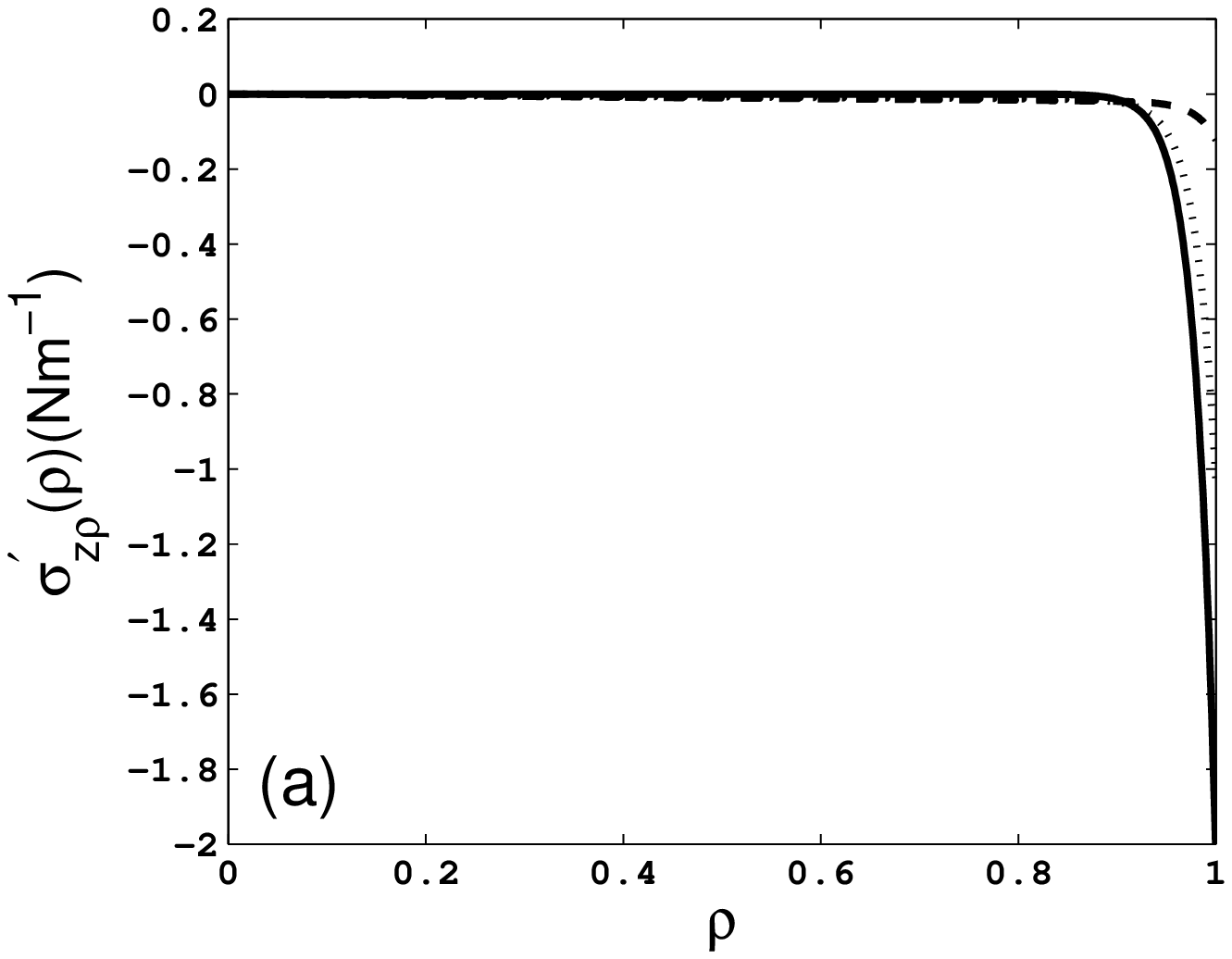,height=3.5cm}
\epsfig{file=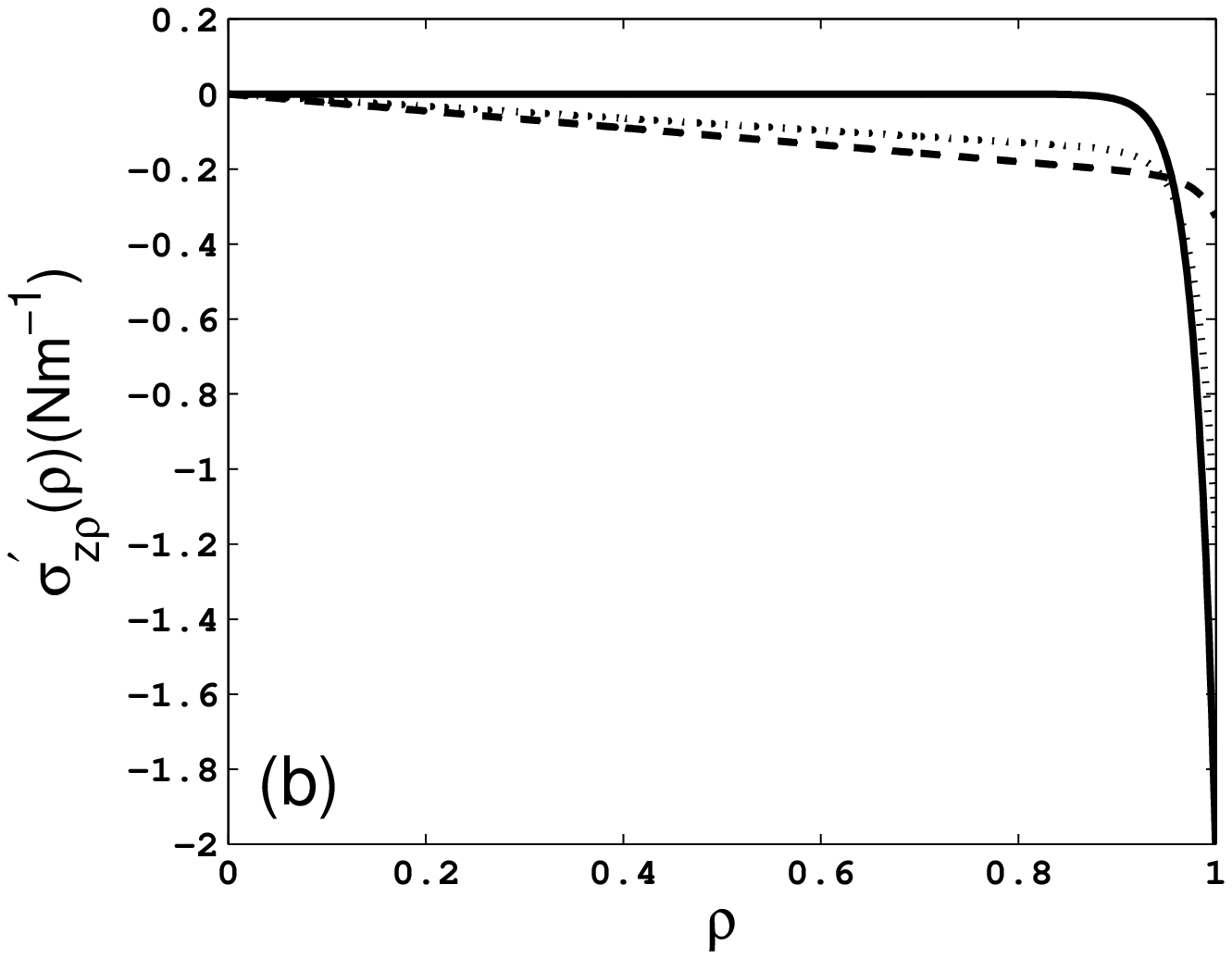,height=3.5cm}\\
\epsfig{file=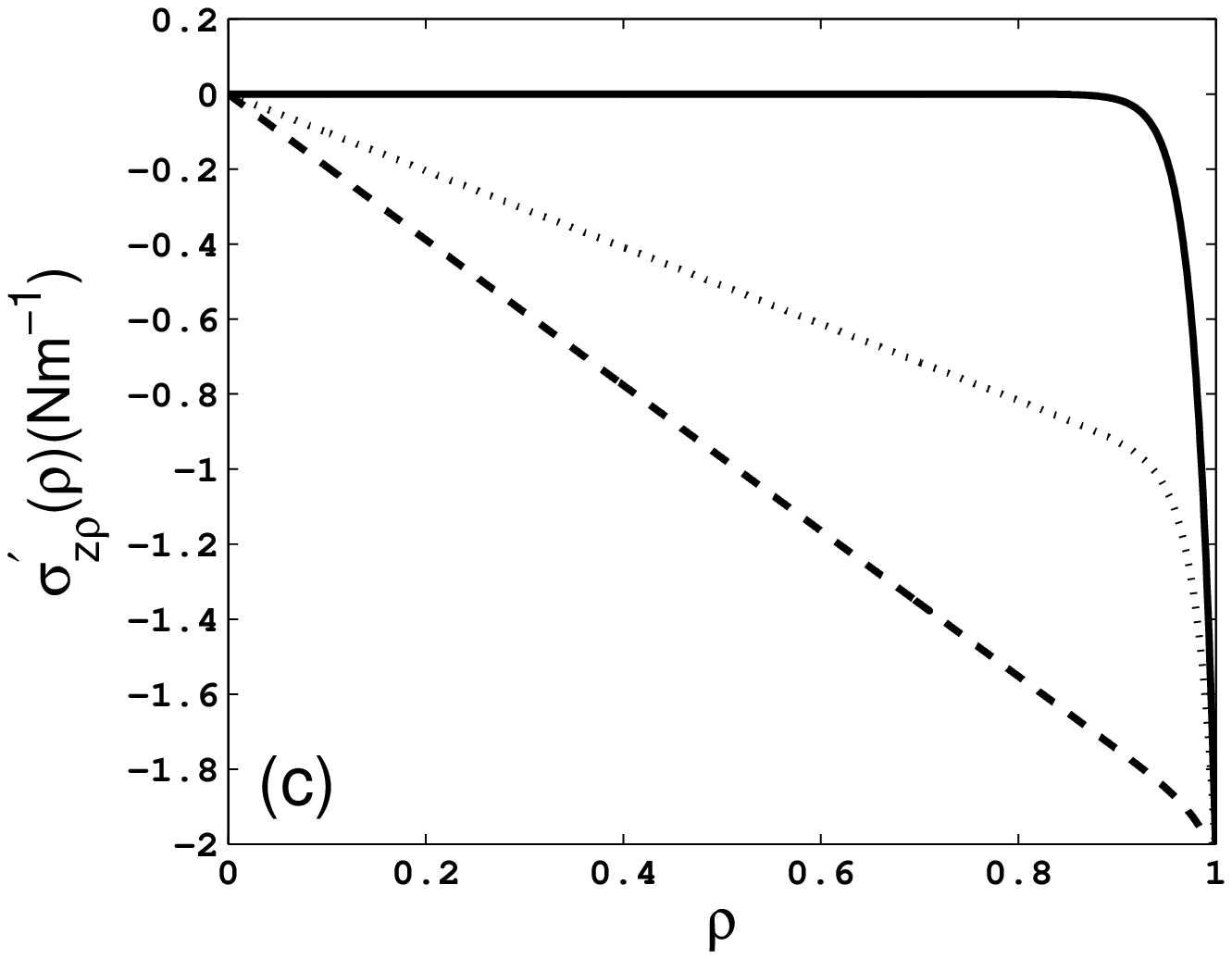,height=3.5cm}
\epsfig{file=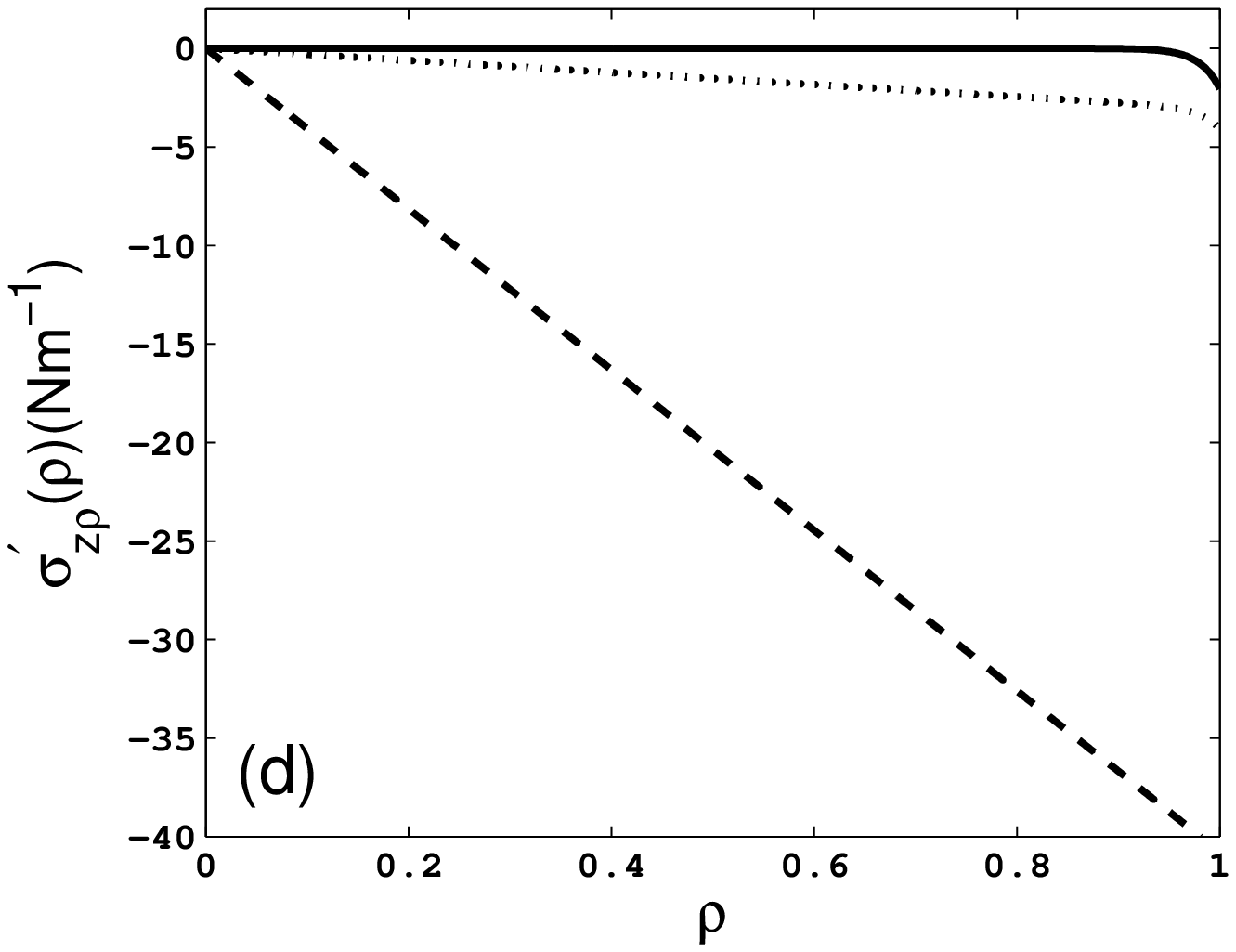,height=3.5cm}\\
\epsfig{file=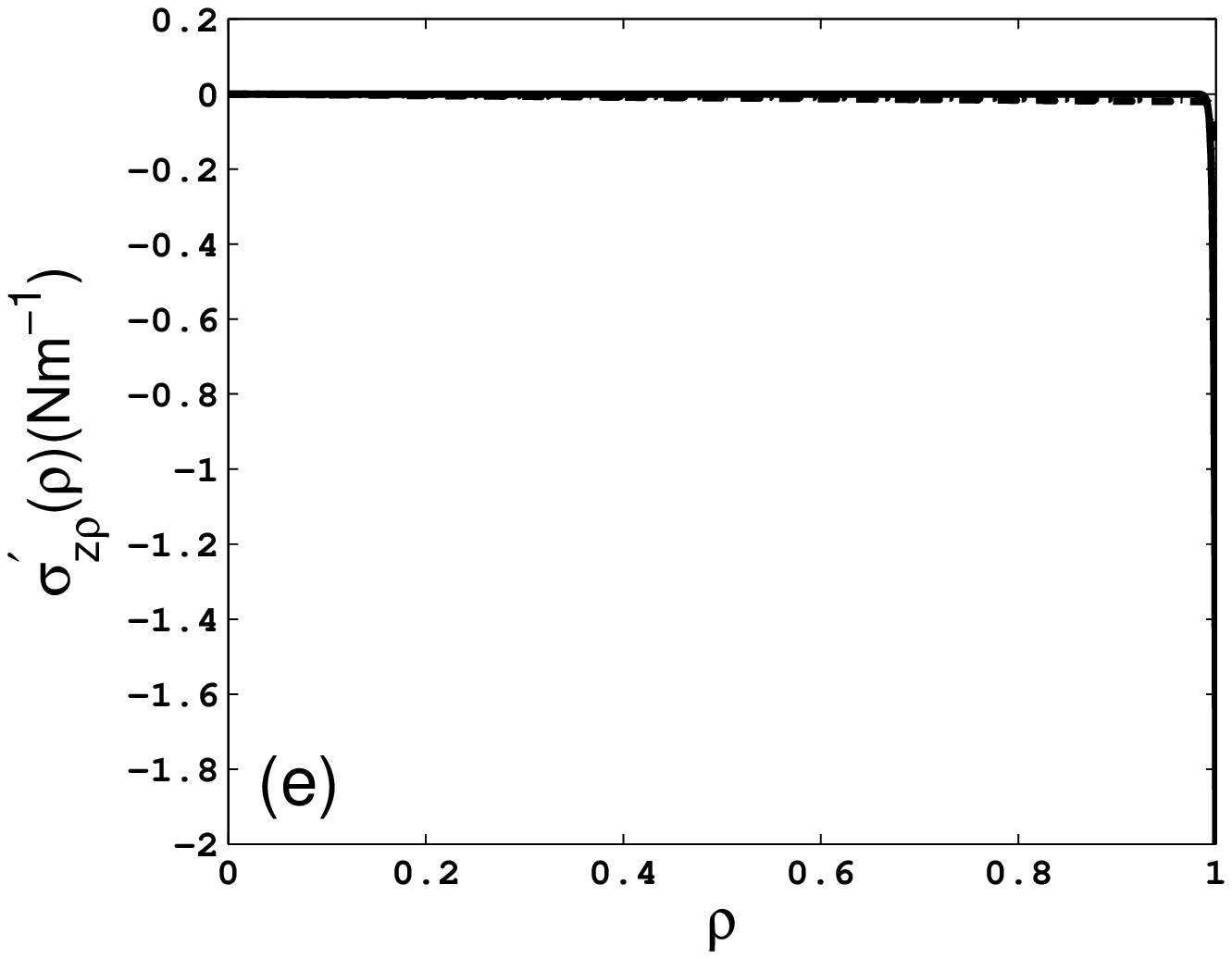,height=3.5cm}
\epsfig{file=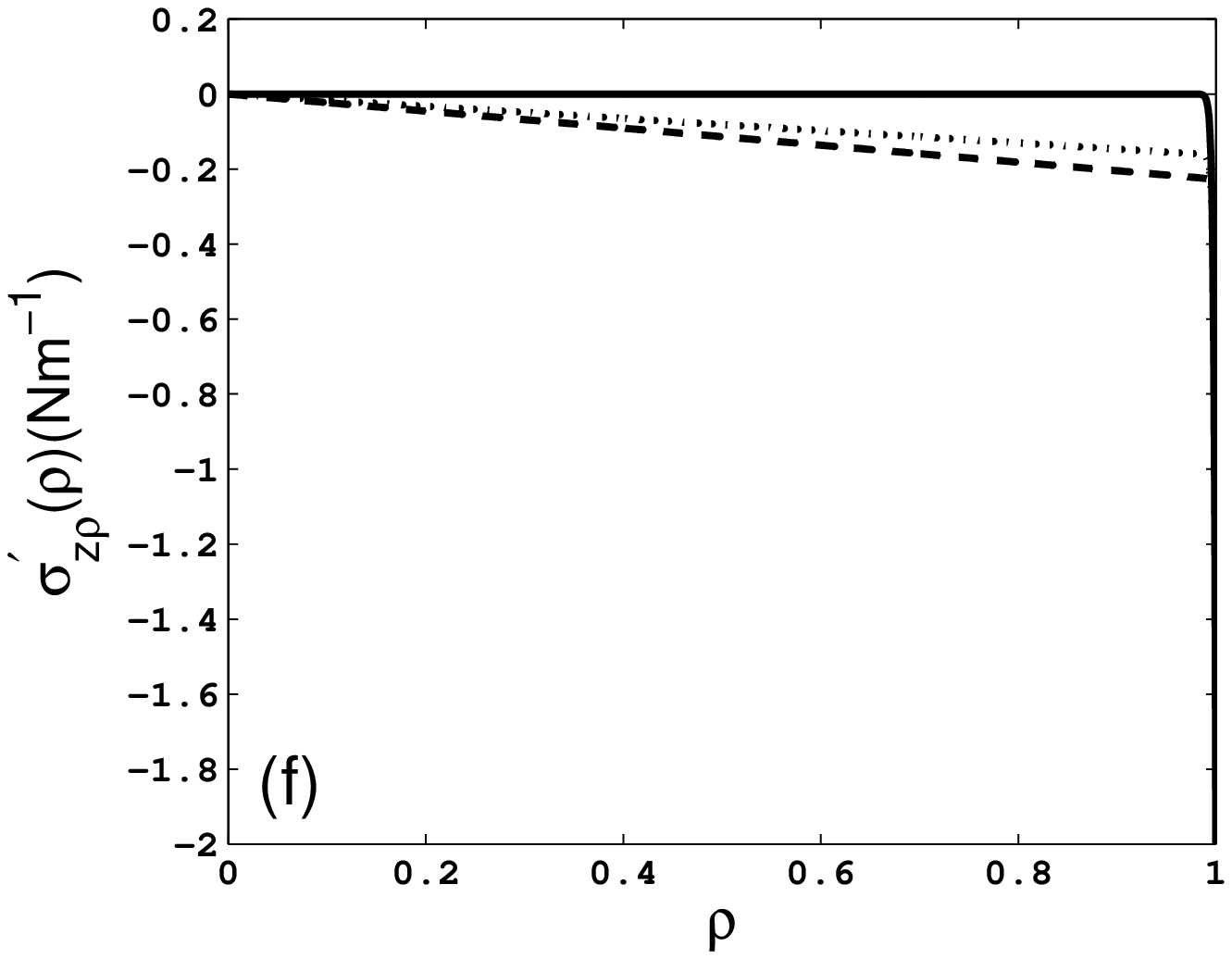,height=3.5cm}\\
\epsfig{file=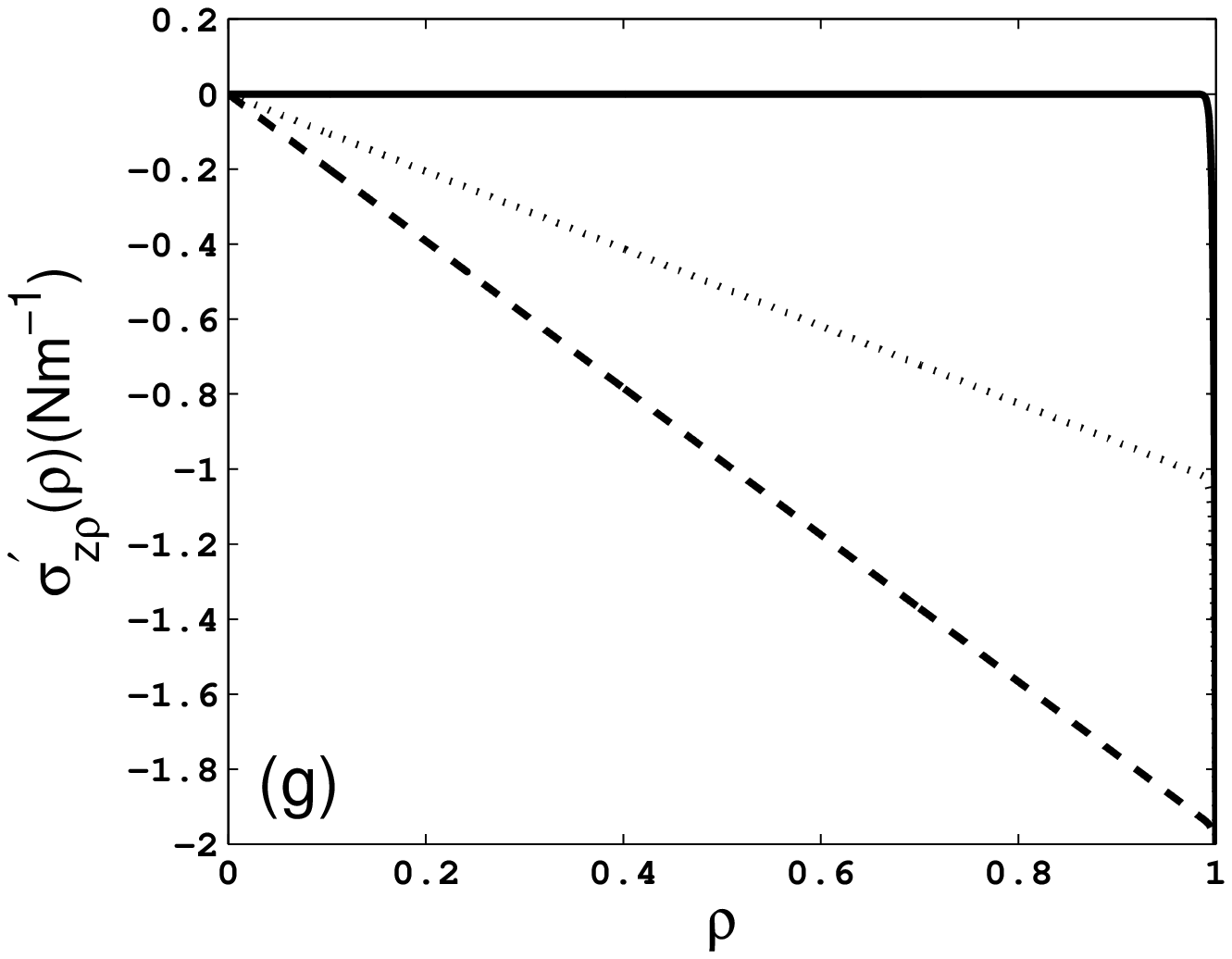,height=3.5cm}
\epsfig{file=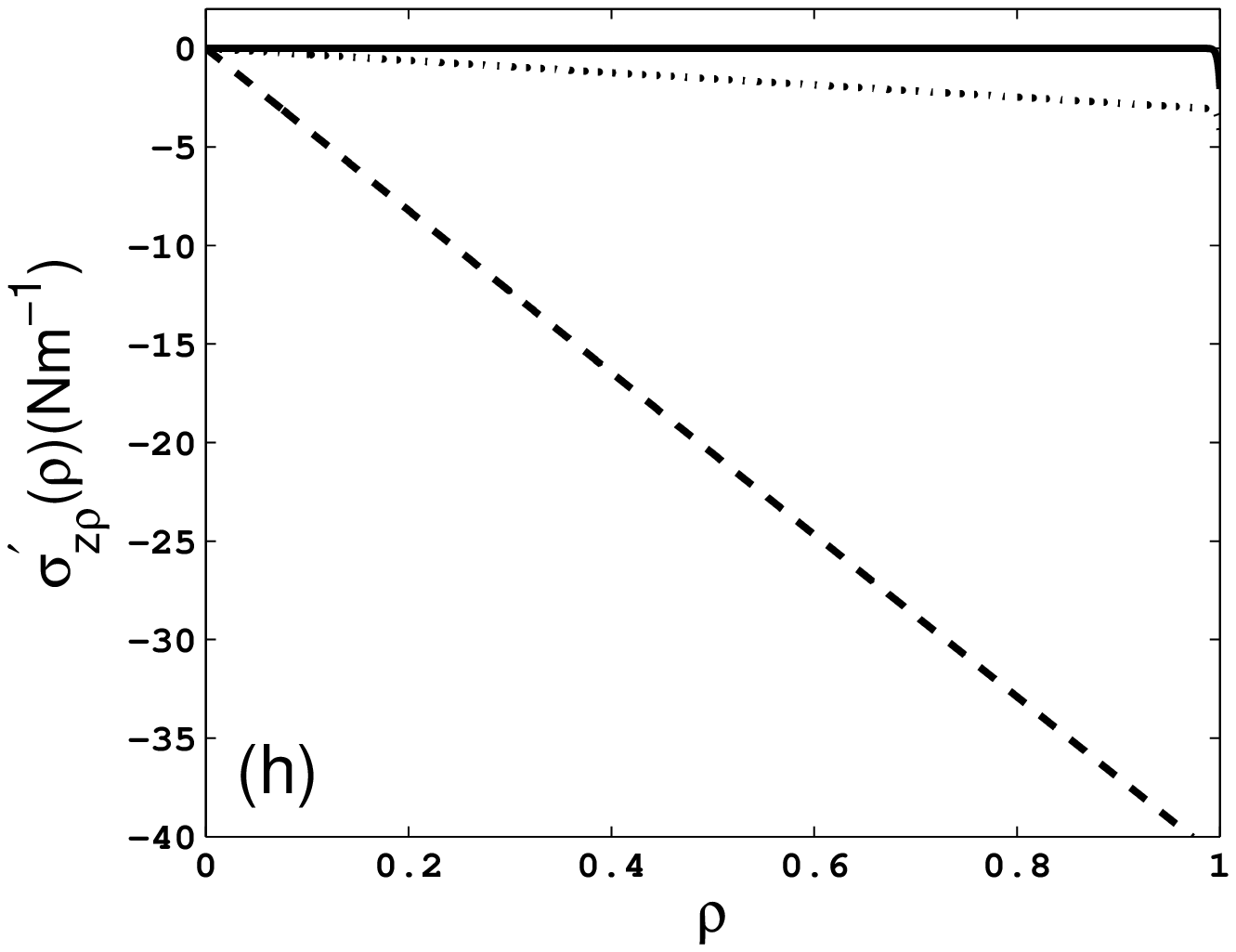,height=3.5cm}\\
\caption{Variation of $\sigma^\prime_{z\rho}(\rho)$ with $\rho$ when
(a--d) $m_o=50$ or (e--h) $m_o=500$, for $k_1=0$ (solid curves),
$k_1=0.5$ (dotted curves), and $k_1=0.95$ (dashed curves). (a, e)
$\beta_o=-0.001$, (b, f) $\beta_o=-0.1$, (c, g) $\beta_o=-0.5$, (d,
h) $\beta_o=-1$. } \label{StressZR_r}
\end{figure}

\subsection{Couple stress tensor}

The couple stress arises in conjunction with the microrotation, as
delineated by Eq.~(\ref{couplestress-def}), and therefore cannot
exist in a simple Newtonian fluid. Consistently with the
approximations made for a microcapillary, the only two non-zero
components of the couple stress tensor are given by
Eqs.~(\ref{4.1.3.0})$_{3,4}$ or, equivalently,
Eqs.~(\ref{4.1.3.1})$_{3,4}$. Substitution of the Debye--H\"uckel
approximation~(\ref{3.8}) for $v_\theta$ therein yields
\begin{eqnarray}
\nonumber &&
m^\prime_{\rho\theta}(\rho)=\frac{\gamma{U}}{R^2}\left\{
c_1\left[k_0k_7I_0(k_0\rho)-\frac{k_7+1}{\rho}I_1(k_0\rho)\right]\right.
\\
\label{4.18} &&\qquad\qquad\left.
+k_8\left[m_ok_7I_0(m_o\rho)-\frac{k_7+1}{\rho}I_1(m_o\rho)\right]\right\}\,
\end{eqnarray}
and
\begin{eqnarray}
\nonumber &&
m^\prime_{\theta\rho}(\rho)=\frac{\gamma{U}}{R^2}\left\{
c_1\left[k_0I_0(k_0\rho)-\frac{k_7+1}{\rho}I_1(k_0\rho)\right]\right.
\\
\label{4.19} &&\qquad\qquad\left.
+k_8\left[m_oI_0(m_o\rho)-\frac{k_7+1}{\rho}I_1(m_o\rho)\right]\right\}\,,
\end{eqnarray}
where
\begin{equation}
\label{4.20} k_8=\frac{k_2}{k_0^2-m_o^2}\,\frac{m_o}{I_0(m_o)}\,.
\end{equation}

Equations (\ref{4.18}) and (\ref{4.19}) yield the difference
\begin{equation}
m^\prime_{\rho\theta}(\rho)-m^\prime_{\theta\rho}(\rho)
=\frac{\gamma{U}}{R^2}\left(k_7-1\right)\left[c_1k_0I_0(k_0\rho)+k_8m_oI_0(m_o\rho)\right]\,
\end{equation}
between the two non-zero components of $\=m^\prime$. This difference
vanishes for all $\rho\in\left[0,1\right]$ when $\beta=\gamma$
(i.e., $k_7=1$), in agreement with Eq.~(\ref{couplestress-def}).
Furthermore, as
\begin{equation}
m^\prime_{\rho\theta}(0)=-m^\prime_{\theta\rho}(0)
=\frac{\gamma{U}}{2R^2}\left(k_7-1\right)\left(c_1k_0+k_8m_o\right)\,,
\end{equation}
the condition $\beta=\gamma$ also implies that the couple stress
tensor is   nonexistent on the axis of the microcapillary.

A more general but slightly approximate conclusion can be drawn from
Eq.~(\ref{vthetaprime-simple}) for $m_o\gg1$ as follows. That
equation yields $v_\theta(\rho) \simeq -
{\beta_o}\left({1+k_1\beta_o}\right)^{-1}m_o\rho$, whose use in
Eqs.~(\ref{4.1.3.1})$_{3,4}$ and (\ref{1-dim}) provides
\begin{equation}
\label{m-simple}
m^\prime_{\rho\theta}(\rho)=-m^\prime_{\theta\rho}(\rho) \simeq -
\frac{\gamma{U}}{R^{2}} (k_7-1) \frac{m_o\beta_o}{1+k_1\beta_o}
\end{equation}
for $m_o\gg1$. In other words, the couple stress tensor is
skew-symmetric as well as uniform throughout the cross-section of
the microcapillary. The approximate equations (\ref{m-simple}) were
verified by directly using Eqs.~(\ref{4.18}) and (\ref{4.19}) to
plot the variations of $m^\prime_{\rho\theta}(\rho)$ and
$m^\prime_{\theta\rho}(\rho)$ with $\rho$ in Figs.~\ref{CStress12_r}
and \ref{CStress21_r}, respectively.

\begin{figure}[!htb]
\centering \psfull
\epsfig{file=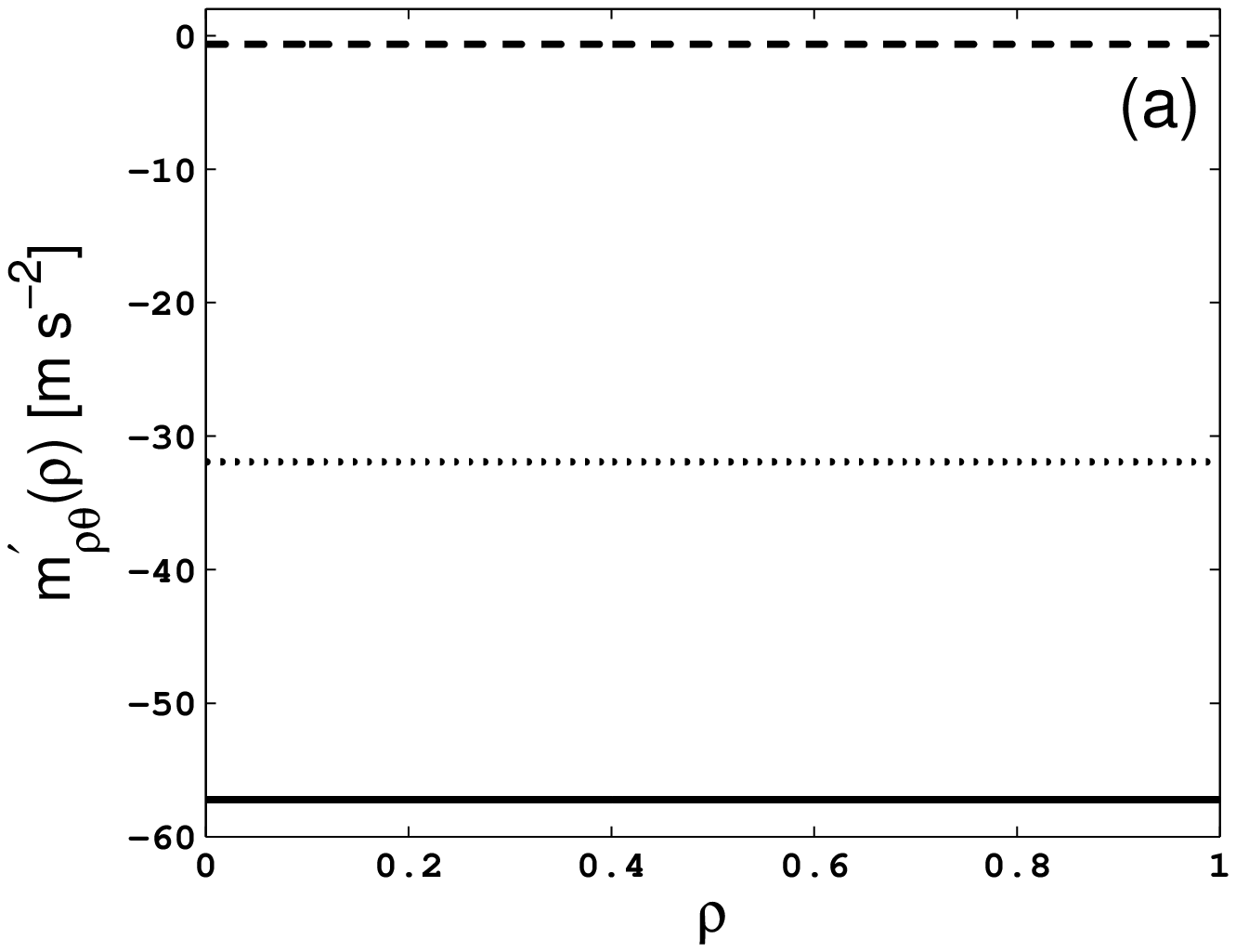,height=3.5cm}
\epsfig{file=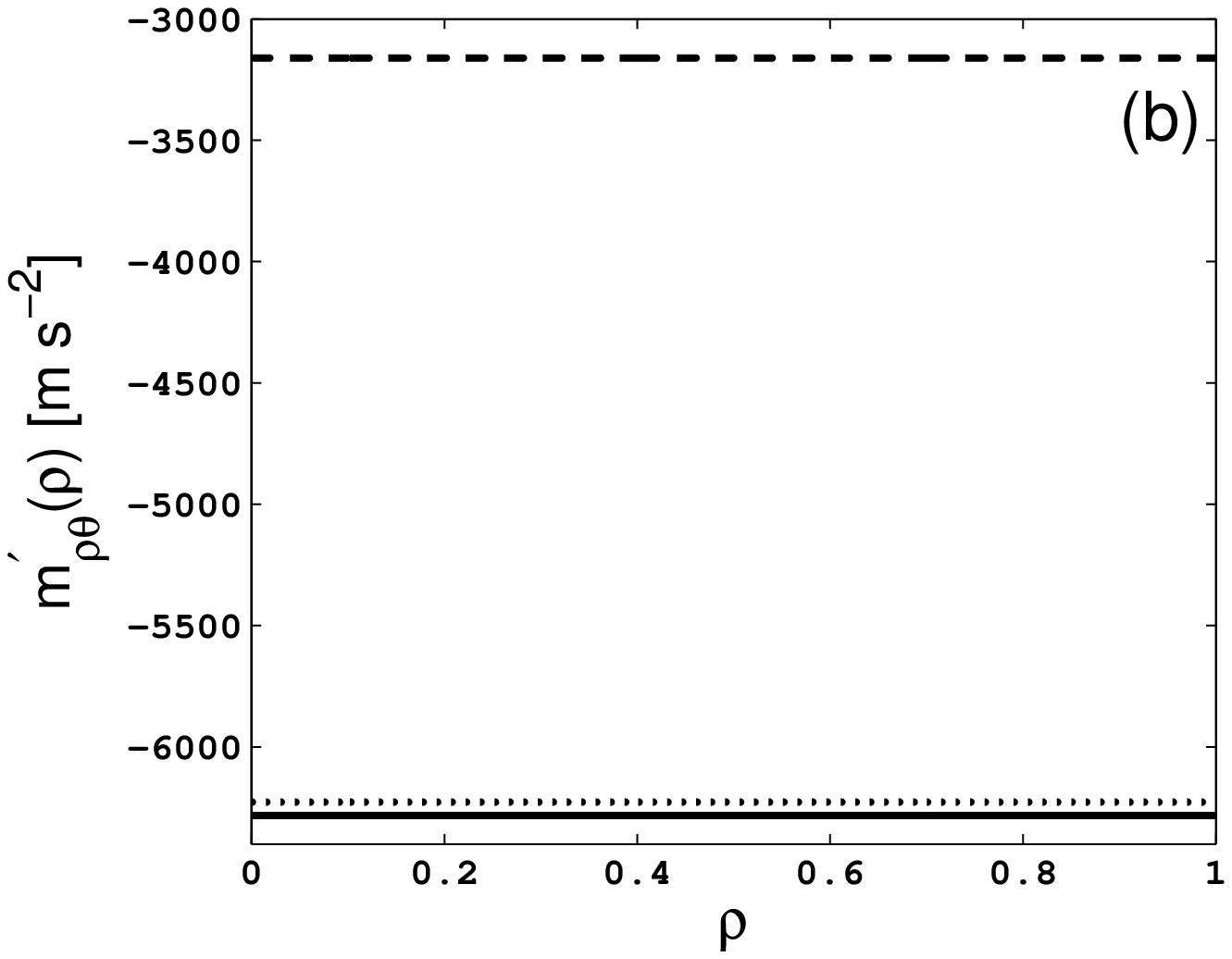,height=3.5cm}\\
\epsfig{file=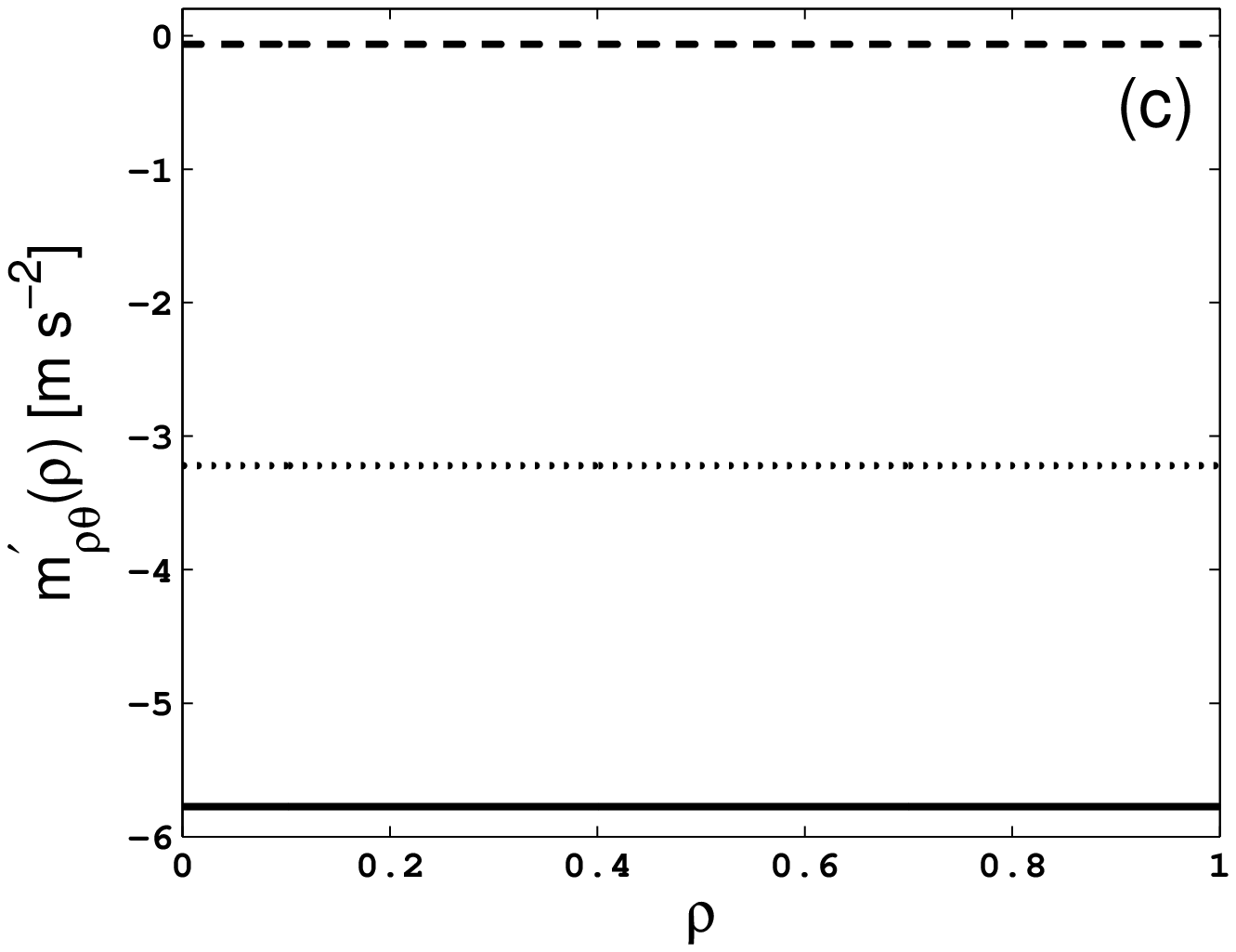,height=3.5cm}
\epsfig{file=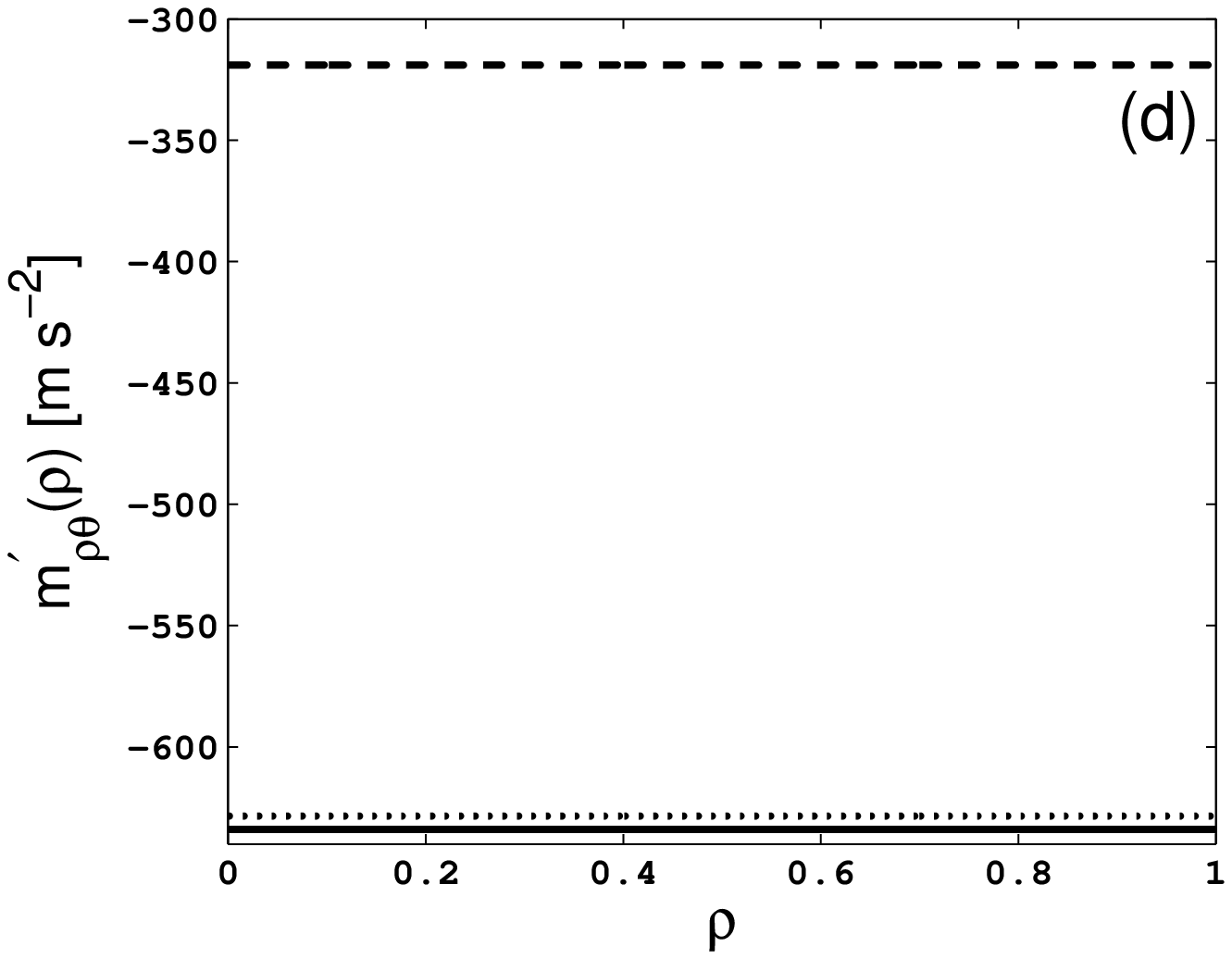,height=3.5cm}\\
\caption{Variation of $m^\prime_{\rho\theta}(\rho)$ with $\rho$ when
$k_7=0.5$ and (a,b) $m_o=50$ or (c,d) $m_o=500$, for $k_1=0.1$
(solid curves), $k_1=0.5$ (dotted curves), and $k_1=0.99$ (dashed
curves). (a, c) $\beta_o=-0.01$, (b, d) $\beta_o=-0.99$. }
\label{CStress12_r}
\end{figure}

\begin{figure}[!htb]
\centering \psfull
\epsfig{file=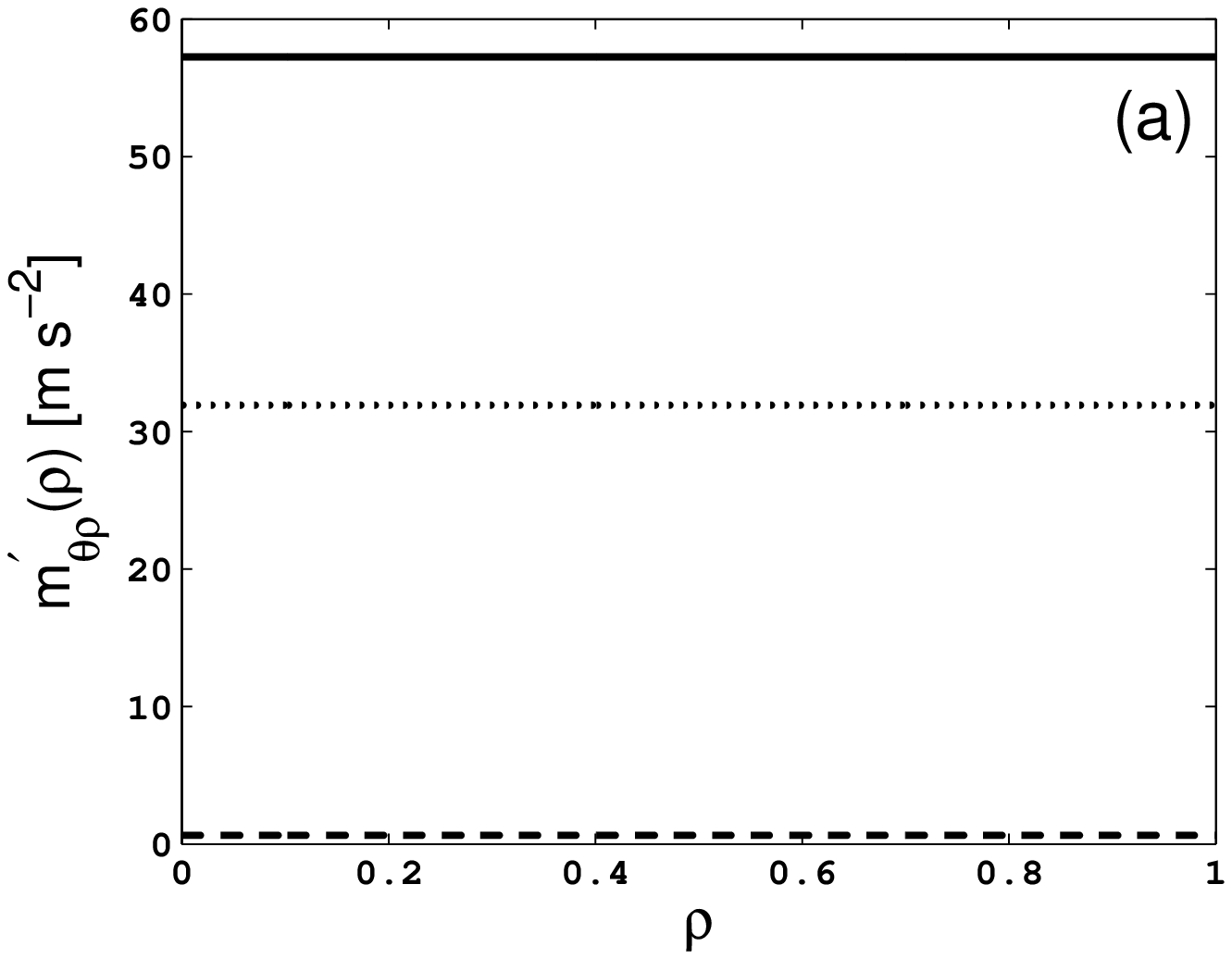,height=3.5cm}
\epsfig{file=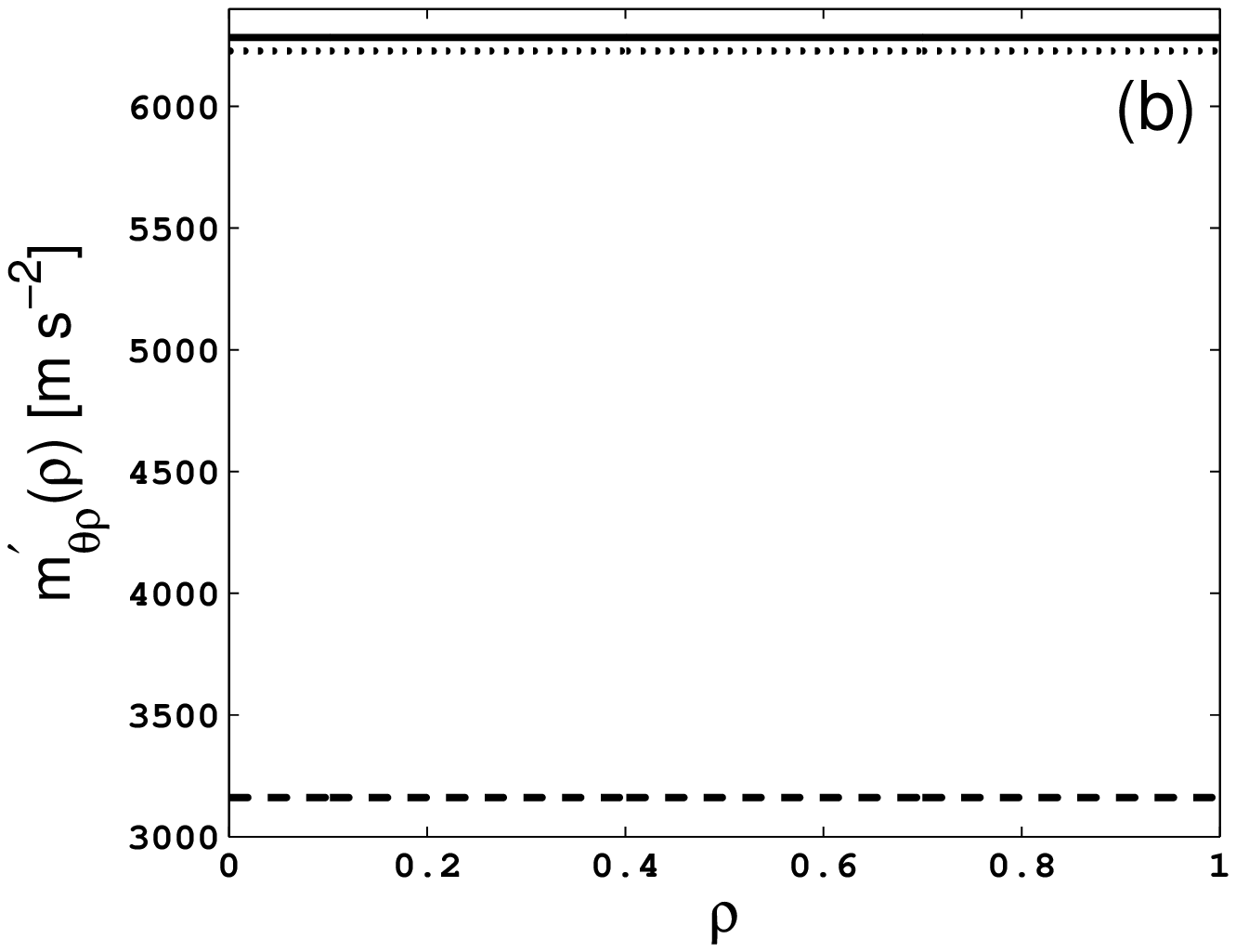,height=3.5cm}\\
\epsfig{file=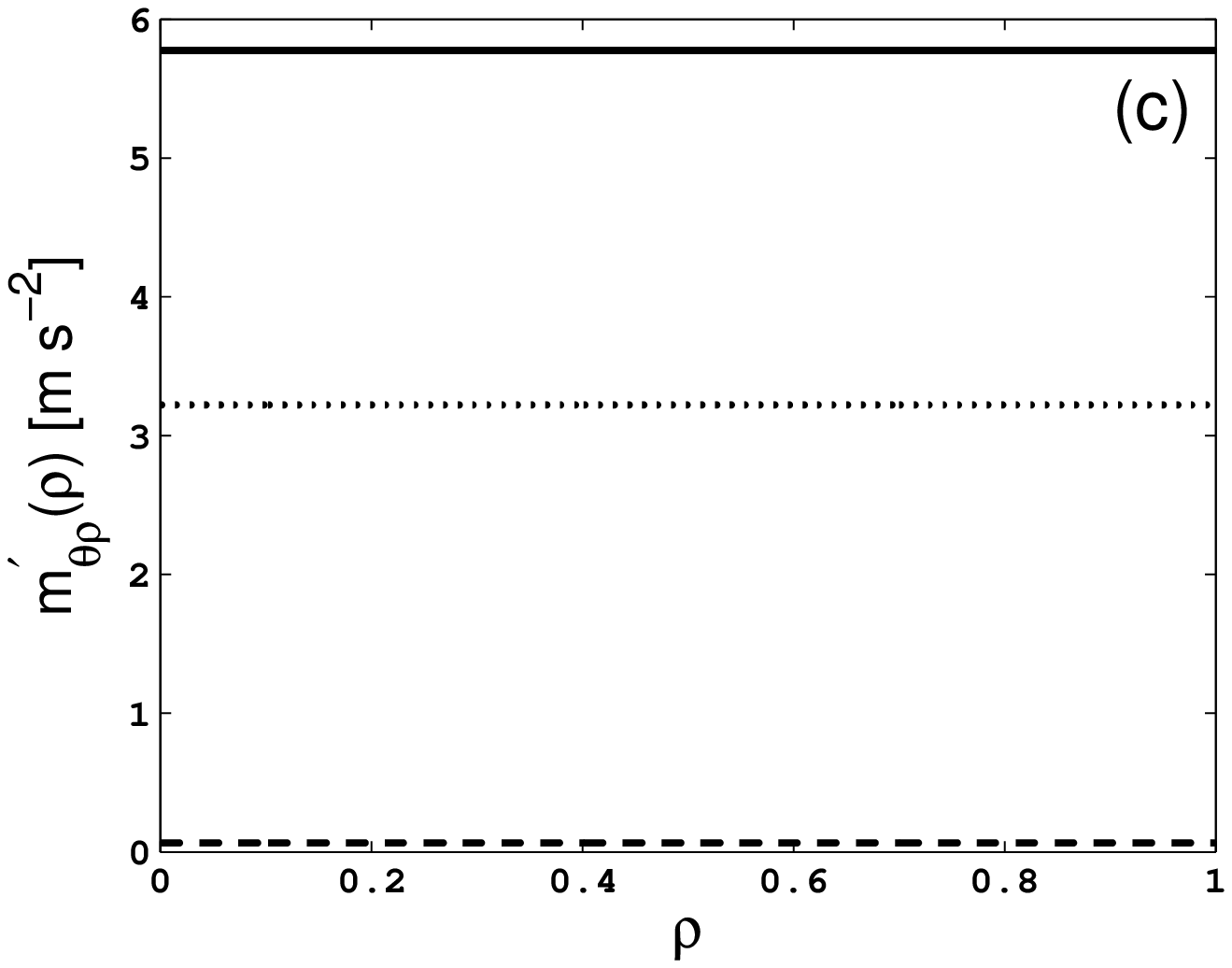,height=3.5cm}
\epsfig{file=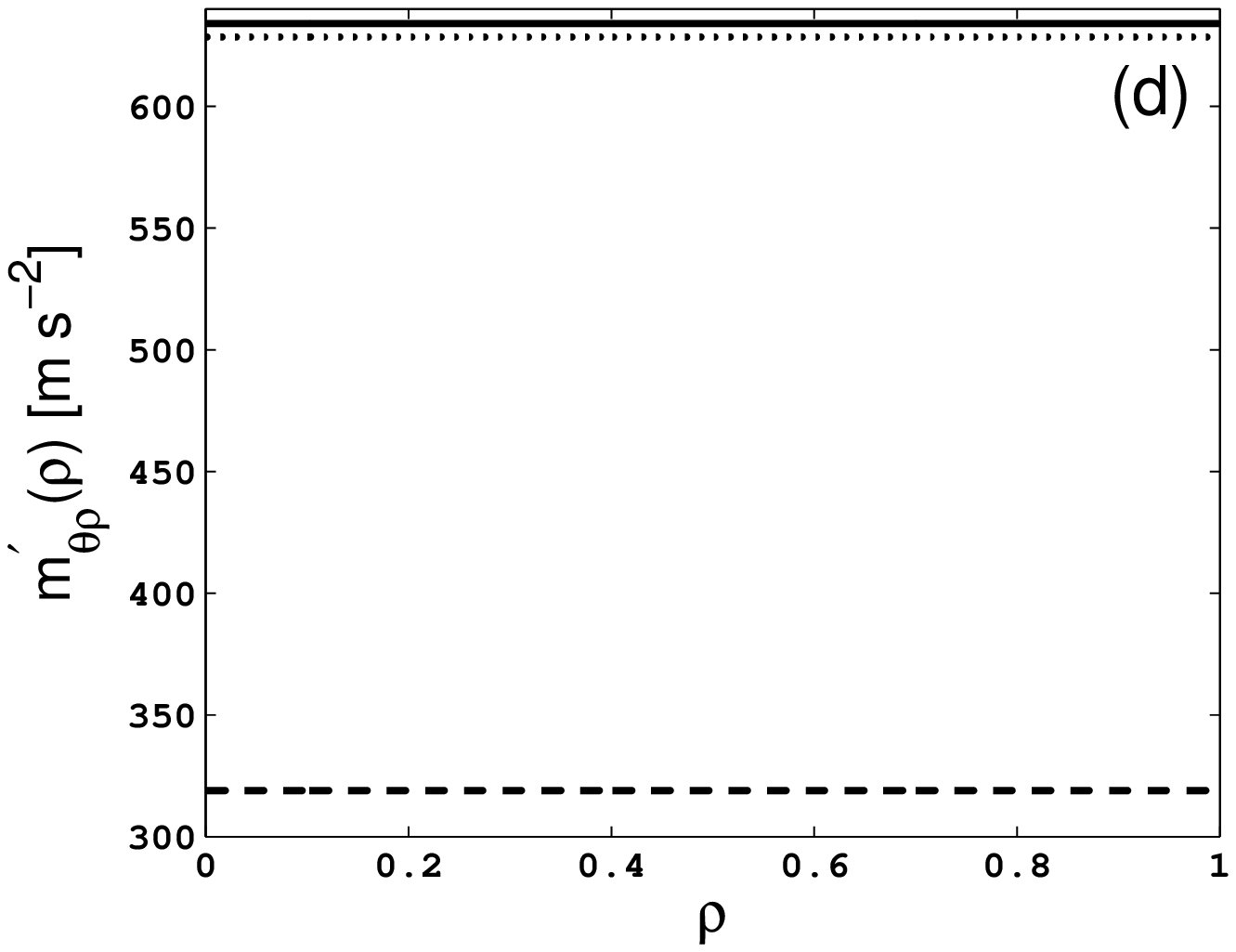,height=3.5cm}\\
\caption{Same as Fig.~\ref{CStress12_r}, except that
$m^\prime_{\theta\rho}(\rho)$ is plotted against $\rho$. }
\label{CStress21_r}
\end{figure}
\section{CONCLUDING REMARKS}\label{CR}

We formulated the boundary-value problem of steady electro-osmotic
flow of a micropolar fluid in a microcapillary whose length is much
greater than its cross-sectional radius. Analytical solution of this
boundary-value problem was obtained under the assumption that both
the Debye length and the zeta potential are sufficiently small in
magntitude so that the  Debye--H\"uckel approximation could be used.

As the aciculate particles in a micropolar fluid can rotate without
translation, micropolarity must influence fluid speed. The derived
expressions and the calculated data indicate that the  fluid speed
depends significantly on the viscosity coupling parameter $k_1$,
which mediates the (micropolar) viscosity coefficient and Newtonian
shear viscosity coefficient. The axial   speed of a micropolar fluid
is below the speed of a simple Newtonian fluid for small values of
the micropolar boundary parameter $\beta_o$---which relates the
velocity gradient and microrotation at the wall of the
microcapillary---but exceeds the latter for $\beta_o$ close to $-1$.
In addition, the axial speed is independent of the radius of the
microcapillary when the fluid if of simple Newtonian type
 but not when it is micropolar, provided the  Debye length is fixed; indeed, the magnitude of the
 axial speed in a micropolar fluid
 intensifies  as the radius increases.

The fluid flux increases as either the microcapillary radius
increases and/or the Debye length
 decreases, whether the fluid is simple Newtonian or micropolar---in the latter case, regardless of the value
 of the micropolar boundary parameter $\beta_o$. The flux of a micropolar fluid increases
 as the magnitude of that boundary parameter intensifies.

Although microrotation greatly influences the speed and the flux, it
vanishes on the axis of the microcapillary and increases linearly in
the radial direction. Microtation increases with the magnitude of
the micropolar boundary parameter $\beta_o$, but it decreases as the
viscosity coupling parameter $k_1$ increases. Quite surprisingly,
microrotation at the wall is independent of the cross-sectional
radius of the microcapillary. Moreover, when the boundary layer is
turbulent (i.e., $\beta_o=-1$), microrotation at the wall is also
independent of $k_1$.

The stress tensor in the fluid has just two non-zero components, one
of which is totally unaffected by the micropolarity of the fluid.
That component does not exist on the axis and is largely confined to
the region close to the wall. The other component is also absent on
the axis and it gets progressively concentrated on the region close
to the wall as $\beta_o\to-1$.

Unlike all foregoing physical parameters, both non-zero components
of the skew-symmetric couple stress tensor are uniform in a
micropolar fluid throughout the cross-section of the microcapillary.
The couple stress tensor does not exist in a simple Newtonian fluid.

Our conclusions are significant for the design of microcapillaries
are the selection of materials for labs-on-a-chip. For instance,
turbulence caused by mixing of a (micropolar) body fluid with a
(simple Newtonian) reagent fluid is likely to result in higher
electro-osmotically induced  flux and axial speed in a
microcapillary than if both fluids are of the simple Newtonian type,
suggesting that the microcapillary be designed with a larger
cross-sectional diameter. Higher stress at the wall of a
microcapillary transporting a micropolar fluid suggests that stiffer
materials be used for the construction of labs-in-a-chip than if all
fluids were to be simple Netwonian. As all of our conclusions apply
when the zeta potential is sufficiently small in magnitude and the
the cross-sectional radius of the microcapillary  exceeds the Debye
length, numerical solution of Eqs.~(\ref{2.1})--(\ref{2.3}) is
required for more general situations. We plan to take up that
investigation next.

\end{document}